# PHYSICAL VACUUM AS A SYSTEM MANIFESTING ITSELF ON VARIOUS SCALES – FROM NUCLEAR PHYSICS TO COSMOLOGY

Timashev Serge F.[1, 2]
[1] National Research Nuclear University MEPhI, Moscow, Russia
[2] Semenov Institute of Chemical Physics, Moscow, Russia

Rising to Mach, the problem of establishing the unity of the physical essence of the Universe on all space-time scales of its evolution is, apparently, one of the main problems in contemporary epistemology. The issues are primarily related to the problems of the dynamics of the Universe, namely, the need to postulate hypothetical entities - dark energy and dark matter, whose nature is unknown, but whose contribution to the energy content of the Universe is about 95%. A "severe trial for the entire fundamental theory" turned out to be a discrepancy of 120 orders of magnitude between the experimentally established value of the cosmological constant and the calculated value of the energy density of the physical vacuum. Sharp questions arise also regarding the physical essence of gravitation and the nature of its unique smallness as compared with other interactions – nuclear ones, strong and weak, as well as electromagnetic ones. Sufficiently acute problems in the perception of the Universe as a single holistic system arise also at the level of microscales. First of all, we mean here the so called low-energy nuclear reactions, which are realized in the conditions of nonequilibrium low temperature plasma. The fundamental problems of such processes are usually not widely discussed, and many physicists are very critical of the very possibility of implementing such processes. Our main hypothesis for understanding the key problems of contemporary fundamental physics, including the indicated cosmological problems as well as a new class of electron-initiated nuclear chemical processes, is to introduce the basic reference system associated with the electromagnetic component of physical vacuum - EM vacuum of the expanding Universe, with a selection of global time t, being common for all points of the expanding Universe and measured starting from the time of t = 0, which corresponds to the Big Bang.

TABLE OF CONTENTS

I. INTRODUCTION. ON PROBLEMS OF CONTEMPORARY FUNDAMENTAL PHYSICS: QUO VADIS?

The problem of establishing the unity of the physical essence of the Universe on all the space-time scales of its evolution is, apparently, one of the main problems in contemporary epistemology. The idea of such unity is contained in the statement of E. Mach [1], according to which "… even in the simplest case, in which apparently we deal with the mutual action of only *two* masses, the neglecting of the rest of the world is *impossible*". The expression of such a unity of the Universe is usually associated with Planck's numbers, dimensionless characteristics of length $a_{Pl}$, time $t_{Pl}$ and mass $m_{Pl}$ introduced in 1899 on the basis of the fundamental constants $\hbar$, $c$ and $G$, that are the Planck's constant, the velocity of light in vacuum and the gravitational constant, respectively [2, 3]:

$$a_{Pl} = 2^{3/4}\sqrt{\frac{G\hbar}{c^3}} = 2^{1/2}\frac{\hbar}{m_{Pl}c} \approx 2.64 \cdot 10^{-33}\, cm; \quad t_{Pl} = \frac{a_{Pl}}{c} = 2^{3/4}\sqrt{\frac{G\hbar}{c^5}} \approx 0.88 \cdot 10^{-43}\, s;$$

$$m_{Pl} = 2^{-1/4}\sqrt{\frac{\hbar c}{G}} \approx 1.78 \cdot 10^{-5}\, g \quad \text{or} \quad w_{Pl} = \frac{m_{Pl}c^2}{t_{Pl}} = \frac{c^5}{2G} \approx 1.8 \cdot 10^{59}\, erg/s. \qquad (1)$$

Here the Planck numbers $a_{Pl}$ and $t_{Pl}$ could be considered as parameters of the microworld and the Planck number $w_{Pl}$ as a parameter of the Universe. Unfortunately, the collection of Planck numbers is insufficient to understand, which parameter, hidden from the eye, determined the Planck numbers ($a_{Pl}$) and ($t_{Pl}$) as parameters of the microworld, and other ($w_{Pl}$) as a parameter of the Universe? Is it possible to introduce such dimensionless parameter? Another question arises: to what extent can some physical images, physical processes be associated with the Planck numbers, or the specified length, time and power parameters are simply physical abstractions, and the mass parameter is something incomprehensible?

Another question arises: is it possible to give any physical meaning to the Planck numbers, or to the specified parameters of length, time and power, or are these numbers physical abstractions, and the mass parameter is something completely incomprehensible?

Questions arise also in connection with the physical essence of gravitation and the nature of its unique smallness as compared with other interactions – nuclear ones, strong and weak, as well as electromagnetic ones. The "internal mechanism" of the gravitational interaction remains unknown in contemporary physics. Here are two quotes. In 1964, in the lectures at Cornell University on the nature of fundamental interactions, R. Feynman emphasized ([4], p. 39) that "… up to today, from the time of Newton, no one has invented another theoretical description of the mathematical machinery behind this law which does not either say the same thing over again, or make the mathematics harder, or predict some wrong phenomena. So, there is no model of the theory of gravitation today, other than the mathematical form".

Creating General theory of relativity (GRT, 1921), A. Einstein, being aware of the limitation of the available information for solving this kind of problems, wrote ([5], p. 87): "We have seen, indeed, that in a more complete analysis the energy tensor can be regarded only as a provisional means of representing matter. In reality, matter consists of electrically charged particles, and is to be regarded itself as a part, in fact, the principal part, of the electromagnetic field. It is only the circumstance that we have not sufficient knowledge of the electromagnetic field of the concentrated charges that compels us, provisionally, to leave undetermined in presenting the theory, the true form of tensor" (with using G as a fundamental constant).

In connection with the such remark, a question appears. Is the gravitational constant $G$ a fundamental one? In fact, Einstein's quote means that the question of whether the gravitational constant $G$ is a carrier of information about gravity as a fundamental interaction and can be regarded as a universal constant together with $\hbar$ and $c$, remained open for Einstein and still remains open even after nearly 100 years since the creation of the GTR. In fact, it turned out to be impracticable and the so-called $\hbar c G$-disclosure plan, which began to be developed by M. Bronshtein in 1933 [6], and which meant to achieve an epistemological unity of the quantum theory (with its constant $\hbar$), special relativity theory (with its constant $c$) and the theory of gravitation (with its constant $G$ as a fundamental physical value).

The difficulties of perceiving the Universe as a holistic integral system increase significantly when developing dynamic models of the Universe dynamics. So, in the frame of the existing "standard model" of the dynamics of the Universe [7-10], it is not possible to understand the physical essence of 96% of the Universe energy content, which is connected with the so-called "dark energy" (73%) and "dark matter" (23%). In the standard model, the dark energy density $\varepsilon_V$ manifests itself as the factor governing the expansion of the Universe and related to the cosmological constant $\Lambda$, which is determined by the fluctuations of the physical vacuum introduced in particle physics by relation:

$$\varepsilon_V = \Lambda c^4 / 8\pi G. \qquad (2)$$

The questions as to the physical essence of such repulsive interactions remaining open at that. According to the present-day data, $\varepsilon_V \approx 0.66 \cdot 10^{-8}$ erg/cm$^3$ [11]. However, the attempts to link the quantity $\varepsilon_V$, determined on the basis of the relation (2) from the experimentally determined value of $\Lambda \approx 1,37 \cdot 10^{-56}$ cm$^{-2}$, with the parameters of the physical vacuum, were unsuccessful [7-10]. The differences were 120 orders of magnitude if one were focused on the vacuum of physical fields! Such catastrophic differences are considered to be a "severe trial for the entire fundamental theory" [9]. The "wanted" relations discussed in a number of works [9, 12, 13], which are being suggested purely formally, without resort to comprehensible physical models linking up the energy density $\varepsilon_V$ with the fundamental physical constants, surely cannot help solve the problem.

Sufficiently acute problems in the perception of the Universe as a single holistic system arise also at the level of microscales. First of all, we mean here the so-called low-energy nuclear reactions, which are realized in the conditions of nonequilibrium low-temperature plasma and in the conditions of mechanical activation of solid mixtures. Fundamental problems of such processes do not so extensively discussed, however extensively disapproved over the past almost 30 years. In the author's opinion, the misunderstanding and denial of the existence of the discussed phenomena by some physicists is one of the constituents of the general problems of contemporary fundamental physics.

It is commonly supposed that the processes in nuclear physics and chemistry are characterized by the energies, the values of which differ by 4 and more orders of magnitude – from tens of *keV* and more in nuclear physics to *eV* and their fractions in chemistry. However, even before the creation of quantum mechanics, there appeared information, in which it was stated that the nuclear transformation may occur at the energies of "chemical values". In 1922, Wendt and Irion [14] published the results, pursuant to which helium formation took place at the electric explosion of tungsten wire in a vacuum (temperature $T \approx 20{,}000$ K; helium was detected by a spectrograph). The explosion was initiated by the discharge current of the capacitor. However, Rutherford [15] considered this result as erroneous; in his experiment, in which the electron beam characterized by the energy of $E_e = 100$ keV was directed on the tungsten plate, helium did not appear. Though Wendt [16] did not agree with Rutherford, specifying that in this experiment the generated power was sufficiently lower, the scientists all around the world accepted the result obtained by Wendt and Irion as an erroneous one.

Another information on the possibility of nuclear transformations at low energies was connected with the discovery (in 1952) of neutron emission at high-current electric discharges in the low-temperature plasma, which was generated in tubes containing mixtures of deuterium, ~10 Torr, with the inert gas [17, 18]. In these experiments with electric discharges (10 kV on electrodes) in 2-m long tubes, neutrons ($10^8$/pulse) and hard X-ray quanta (300-400 keV) were detected. It should be noted that in the subsequent theoretical calculations [19, 20] it was shown that the acceleration of deuterons up to the energies in excess of 100 keV, which are sufficient for the realization of the $d + d \rightarrow {}^3He + n$ process, is quite possible because of the pinch instability development. Nevertheless, a possibility of the realization of such thermonuclear reaction in the conditions of the performed experiments still open to questions. The primary questions are connected with rather short periods of plasma containment in the conditions of the performed experiments, which can not assure the establishment of the Maxwell distribution of the energy of ions of the specified high values. Direct experimental data, which could confirm the formation of the helium-3 nuclei during the high-temperature synthesis, are not available as well.

About the same time, in 1954, Professor Cherdyntsev and the postgraduate student Chalov discovered the phenomenon of the separation of uranium even isotopes at transfer from the solid phase to the liquid phase (Cherdyntsev-Chalov effect) [21]. Later on, it was established that almost in the whole hydrosphere of the Earth the activity of $^{234}$U isotope exceeds the activity of $^{238}$U isotope [22-24], notwithstanding that both isotopes belong to the same series of radioactive transformations ("Uranium-Radium series"), and in the natural uranium ores in the conditions of the secular radioactive equilibrium, the activity of these isotopes should be the same. In the World Ocean, the value of $^{234}$U and $^{238}$U activity ratio, which is determined as follows:

$\eta_{234/238} \equiv {}^{234}U/{}^{238}U\ AR = \dfrac{T_{1/2}({}^{238}U)\theta_{234}}{T_{1/2}({}^{234}U)\theta_{238}}$, where $\theta_i$ and $T_{1/2}({}^iU)$ are the fraction of the $i$-th uranium isotope and its half-life period, respectively, amounts to 1.14 Bq/Bq. For the atmospheric precipitates, surface waters, ground waters of the platform and geosynclinal areas, unessential deviations of the isotope ratio in the range of 1.10-1.80 Bq/Bq exist. However, the ground waters of the Earth crust fractures in the seismically active zones are of great importance, and in the ground waters within the areas of kimberlitic magmatism the values up to 14 Bq/Bq have been detected. Near the Yucca Mountain adjacent to the proving ground in Nevada, where seismic activity exists and where an earthquake of 5.6 intensity on the Richter scale happened in 1992, the uranium isotopes concentration variations amount to 0.6-10 mcg/$l$, $\eta_{234/238}$ ratio varies in a broad range, in some cases reaching the values of 7-8 [25].

The beginning of the new stage in investigations of low-energy nuclear reactions (LENR) should be connected with the works of Deryagin *et al.* [26, 27], who have found emission of high-energy electrons from the freshly formed surfaces of solid bodies – fast electrons emission (up to 100 keV) in case of polymeric films exfoliation from the glass surface in high vacuum caused by microcracks generation in the near-surface layer. The analysis of the electrophysical processes occurring in the propagating crack, which was fulfilled in the 1980s-1990s in the works of Deryagin (together with Anisimova, Klyuyev, Lipson, Toporov) with using X-ray characteristics of the crack zone for determining quantitatively the energy of those electrons, which are emitted by freshly formed surfaces of the crack walls, allowed to establish the formation of the plasma state of the material in the propagating crack. The final stage of these investigations became the discovery of neutrons generation [14] in case of heavy ice ($D_2O$) targets destruction by the metallic bullet, moving with the initial velocity of 100-200 m/sec. In this case, the quantity of the generated neutrons exceeded the background level by several times. In case of similar effect on the usual ice ($H_2O$), neutrons generation has not been detected. Note, that it is the role of mechanically activated factors in the areas with the high seismic activity, where high shear stresses and cracks in the uranium ore are generated, is considered as the main factor of the Cherdyntsev-Chalov effect manifestation in the acceleration of the process of the radionuclides decay.

However, the strongest response of scientists was evoked by the results published by Fleischmann, Pons and Hawkins [28], who have demonstrated that during the heavy water ($D_2O$) electrolysis with a Pd cathode, excessive heat is produced, neutrons are generated and tritium is formed. The results of the works devoted to such "cold fusion", namely, during the heavy water electrolysis, inspired a great number of similar investigations, though some difficulties in reproducing the results in different laboratories have been noted. It became quite clear that in order to solve complex problems arising in connection with the phenomenon of low-energy nuclear reactions, the problem of determining the mechanism of nuclear transformations may become the key one. Since the excessive heat, which is emitted in the result of nuclear transformations, may lead to an increase in the efficiency of power plants and the coefficient of performance (COP) may exceed one, the studies of LENR may become very important, especially if the "cold processes" is not accompanied by dangerous ionizing radiation or neutron flows. The latter is especially hardly accepted by nuclear physics experts since their solid foundations are crushed.

The situation connected with the initiation of the LENR appeared to be more unambiguous and easily reproducible in the conditions of low-temperature plasma, in particular, under of a high-current glow discharge adjacent to the metallic cathode (investigations performed by Savvatimova at NPO "Luch", Podolsk [29-32]); within the vapor phase adjacent to the metal surface during the laser ablation of metals in the water environment (investigations performed by Shafeev's group at IOFAN [33-27]); during burning of the thermite powder mixtures in the air environment [38]. The specific examples will be discussed below.

A qualitative shift in large-scale investigations of low-energy nuclear reactions is connected with the pioneer research works of Rossi-Foccardi and the creation of the pilot nuclear power facilities E-CAT by Rossi [39, 40], characterized by the coefficient of performance (COP factor) substantially exceeding 1. Fine nickel powder of the natural isotopic content mixed with a relatively small fraction (~ 10%) of lithium aluminum hydrate is used as the fuel for these facilities. Impressive results were obtained during the test of one element of the E-CAT Rossi facility, the report of which is presented in [40]. This element was represented by an alumina cylinder, 2 cm in diameter and 20 cm in length, in which 1 g of "fuel" (nickel powder + $LiAlH_4$) was loaded. The reactor operating point was set to about 1260 °C in the first half of the run, and at about 1400 °C in the second half. The measured energy balance between input and output heat yielded a COP factor of about 3.2 and 3.6 for the 1260 °C and 1400 °C runs, respectively. The total net energy obtained during the 32 days run was about 1.5 MWh. This amount of energy is far more than can be obtained from any known chemical sources in the small reactor volume. During the tests period, the isotopic content of the "fuel" experienced qualitative changes: among 5 initially stable nickel isotopes in "ash", the main fraction belonged to $^{62}Ni$, the initial fraction of which amounted to near 4%, while the fraction of the initially main isotope $^7Li$ decreased to 8%. It is necessary to point out that Rossi facilities, like any other nuclear power facilities, which are developed based on low-energy nuclear reactions, in practice are safe during their operation. This is evidenced by the fact that dangerous ionizing radiations, neutron flows, radioactive products are practically absent.

These latter factors, which are formally not consistent with those traditional assumptions that have been developed in nuclear physics during many years, are considered by many experts as the foundation for the denial of the LENR phenomenon itself. Here one may additionally note that one year later, in 2016, relatively successful annual tests of the MW E-CAT facility have been performed (with several short stops, at total COP factor of near 7) [41].

The author believes that the declared lack of understanding of the discussed LENR phenomenon itself, as well as the above, seemingly intractable problems of cosmology and the dynamics of the Universe, are not particular problems. Actually, it reflects a number of gnoseological problems in the understanding of the physical essence of the Universe, which, in the author's opinion, point to the general crisis of contemporary fundamental physics, and first of all, of its basic principles.

In accordance with Nicholas of Cusa, the stages of cognition of Nature and the World Order in general are based on the understanding of the new boundaries of "non-acquaintance". Such an approach was clearly demonstrated by the crisis of physics, which happened at the turn of the past century, connected with the discovery of radioactivity, line spectra of atoms, mysterious aspects of solid body radiation, the negative results of the Michelson-Morley experiment to detect the Universe base medium (ether). It was this non-acquaintance exactly that served as the basis for the new paradigms – quantum mechanics, special and general theories of relativity were created, which have become the fundamental sciences for natural science taken as a whole. Furthermore, quantum mechanics application for nuclear physics and chemistry not only had an effect on solving many problems in energetics and engineering but also to a certain extent determined the global character of contemporary civilization.

The author realizes that his opinion on the crisis of contemporary fundamental physics may be accepted by not all experts. Such unanimity of opinions, assurance and attitude of the whole society towards the necessity of the conceptual changes in physics, as existed 100 years ago in respect of the crisis of the 19th-century physics, are certainly absent now. The public attitude, which existed 100 years ago, may be described by a quotation from Lebon's book [42]: "The world is full of things, which we don't see.... Every step forward is possible only in case of sufficient decay of the previous ideas. Specifying of errors and their consequences is equally important as a new discovery. Acknowledgement of the uncertain knowledge as the undeniable truth may be the most dangerous for development of the reason of man.... We should not hesitate and refuse from the earlier investigation of the main scientific principles only because these principles are revised and seems as unchangeable. The great merit of Descartes was in that he doubted in the validity of what was considered before him as undeniable truth. Cult figures of the modern science have no more grounds to believe that they are untouchable as compared with the past cult figures". Here one must point out that that the translation of Lebnon's book into Russian was fulfilled from the 10$^{th}$ (!) French edition, which is the evidence of the open-mindedness of his contemporaries in respect of natural science problems. Nowadays, the society is different. That is why the author does not cherish illusions in respect of quick perception of the suggested approaches concerning the revision of the principal concept of physical reality.

Since the theoretical science is not yet able to give answers to abovementioned questions about the nature of gravity, dark energy, dark matter and so on, a starting position in the search of new ideas for the "revealing the reality exactly how it shows itself *before* science turns to it with its questions" ([43], p.8) can be phenomenology ("phainomenon" is the manifesting; "logos" is a concept, doctrine). Phenomenology can be considered as a general methodology of cognizing the essence of a phenomenon, if one follows the basic maxim of Husserl's realist phenomenology "Back to the things themselves!" without "any attempts of premature systematization", to "studying 'the things themselves', not allowing ... any abuse of the givenness" [44]. Moreover, the analysis of the experimentally obtained information should be based on purely philosophical tradition with the a priori use of transcendental images, new transcendent entities which are outside the direct experimental tests, but reflect the main essence of the phenomenon under study, such as the "thing in itself" (*das Ding an sich*) and "now", just as in the classical mechanics of Newton there were introduced the images of "point-like mass" and "inertial reference system". As the experience of developing the methodology of retrieving information from complex signals shows [45], such a view on phenomenology, while following the indicated philosophical traditions, allows not only focusing on achieving a purely pragmatic goals in solving technical problems and believing that in the framework of phenomenology "the cognition of the laws of nature is to a great extent even excessive" [46], but also setting the goals of penetration into the physical essence of the phenomenon under investigation by introducing new transcendent entities (see also [3]).

In the author's opinion, for the essential understanding of all of the aforementioned cosmological problems and determining of the nature of low energy nuclear reactions, in that number, it is necessary to return to physics of the Universe absolute base medium, a kind of "ether", the reference system with global time $t$ selection, being common for all points of the expanding Universe and measured beginning from the time of $t = 0$, corresponding to the Big Bang. From this point of view, the author completely agrees with the opinion of Prof. Robert B. Laughlin (Nobel Prize winner, 1998) [47]: "Space is more like a piece of window glass than ideal Newtonian emptiness. It is filled with "stuff" that is normally transparent but can be made visible by hitting it sufficiently hard to knock out a part. The modern concept of the vacuum of space, confirmed every day by experiment, is a relativistic ether. But we do not call it this because it is taboo."

II. ELECTROMAGNETIC COMPONENT OF PHYSICAL VACUUM – EM VACUUM AS THE BASIC MEDIUM AND THE ABSOLUTE REFERENCE SYSTEM FOR THE UNIVERSE

2.1. "Casimir polarization" of the EM vacuum in the vicinity of atomic nuclei and electrons. The "Casimir" polaron. Physical essence of mass.

As such a medium ("the relativistic ether"), the author considers the electromagnetic component of the physical vacuum – EM vacuum, "tied" to the expanding Universe. Let us define this reference system as "Mach system", since Mach was the first who introduced the base system connected with "the center of all Universe masses" [5, p. 44]. If space is free from charges, then it may be believed that the electrostatic potential $\varphi = 0$. Then the electromagnetic field may be considered as an ensemble of independent harmonic oscillators with all possible values of the wave vector $\vec{k}$, and the energy of the electromagnetic field will be equal to $\varepsilon = \frac{1}{8\pi}\int (E^2 + H^2)d\vec{r} = \sum_{\vec{k}}\left(n_{\vec{k}} + \frac{1}{2}\right)\hbar\omega_{\vec{k}}$, where $\vec{E}$ and $\vec{H}$ – vectors of intensities of the electric and magnetic fields, respectively; $\omega_{\vec{k}}$ – frequency, $n_{\vec{k}}$ – number of the excited state of the oscillator described by the wave vector $\vec{k}$. In the ground state, all $n_{\vec{k}} = 0$ (quanta are absent) and $\varepsilon_V^e = \frac{1}{2}\sum_{\vec{k}}\hbar\omega_{\vec{k}}$. At that, $\vec{E}$ and $\vec{H}$ do not have any definite value, but their average values are equal to zero.

However, the EM vacuum manifests itself via the fluctuating mean values of the squared intensity of the electric and magnetic fields [48]. This means that a noise electrodynamic component acts from EM vacuum upon the system of concentrated charges and local currents of any material body. The power spectrum of this component, as it should be assumed, corresponds to the "white noise". Such stochastic effects are known (see, for example, [49], p.394) to initiate smearing-out of the point electrons, which leads to the appearance of the natural width of the excited levels of atoms, determines the Lamb shift and a number of other quantum electrodynamic or radiation corrections.

The electromagnetic vacuum, as well as electromagnetic radiation, fully manifest themselves also in the electronic and nuclear subsystems. It is exactly the interaction of the electronic subsystem of the excited atom with the zero-point oscillations of the electromagnetic field in vacuum, the frequency of which is equal to the frequency of the emitted quantum, that is the initiating factor, the cause of its spontaneous emission (the states of the excited atom cease to be stationary). The initiating factor of the zero-point oscillations of the electromagnetic field is also manifested in the spontaneous emission of γ-quanta by the excited nuclei. The extraneous impact of an outburst of X-ray radiation on a set of excited iron-57 nuclei trapped in the structure, which is an X-ray waveguide (the atoms are at the crest of the passing excitation), leads to a six-fold decrease in the time of emission of γ-quanta by the iron-57 nuclei [50].

It is obvious that in the interaction of EM vacuum with a system of charges and local currents of each material object, at the interphase boundary between EM vacuum and the surface of the object, there are formed certain boundary conditions for the tangential and normal components of the electric field strength, on which the respective impacts of EM vacuum on the objects depend. In particular, in the vacuum gap of the width $d$ between the perfectly smooth metal plates (assuming that the tangential component of the strength vector $E_\tau$ of the electric field, created by the zero-point oscillations of EM vacuum on the surface of the plates is equal to zero), there arises a force of attraction (the "Casimir force"), the value of which is determined by the existence in the gap $d$ of one "resonant" frequency of EM vacuum, equal to $c/d$, and the suppression of a wide range of other frequencies. This force per unit area of the plates (pressure) is given by [51-53]:

$$F_C(d) = -\frac{\pi^2}{240}\cdot\frac{\hbar c}{d^4}. \qquad (3)$$

The validity of the relation (3) is established experimentally for the variation of the gap width $d$ from 50 to 500 nm [52].

Guided by the relationship (1), Casimir expressed the idea (see [54]) that the corresponding forces should stabilize elementary particles as well, in particular, electron, keeping a negative charge in it. In a number of subsequent model calculations of such Casimir forces [54, 55], a particle of the mass $m_0$ was represented by a sphere of radius $a$ with metallic conductivity, so that the tangential component $E_\tau$ of the strength vector of the electric field created by the zero-point oscillations of the EM-vacuum on the surface of the particle, assumed to be zero, $E_\tau = 0$, while the boundary condition for the normal component $E_n$ of the electric field strength had the form $d/dr\left[r^2 E_n(r)\right]_{r=a} = 0$. However, the relevant calculations made in [54, 55] using different methods of renormalization have shown that the vacuum energy $U_a$ of such a sphere of radius *a*, from which a certain constant ($\bar{E}_0$) was subtracted, formally infinite and independent of the sphere radius, is positive and equal to $U_a = 0.04618\,\hbar c/a$. The positive value of the calculated energy means that, inside such a sphere, there acts the repulsive forces and the sphere seeks to expand. Note, the similar result could be obtained by using the method of calculation of the Casimir force without the procedure of renormalization [53]. In accordance with [53], the vacuum-to vacuum graphs with the inner lines, that define the zero-point energy, did not enter the calculation of the Casimir force, which instead only involves graphs with external lines.

The subsequent model calculations showed [54] that the stability of various spherical objects, determined by the Casimir pressure, depends both on the material of the object and on the boundary conditions. In particular, for a sphere made of a dielectric material, when Van der Waals forces are introduced, one can expect to receive negative energies and the stability of such objects according to calculation. The concepts of stabilization of particles due to the considered polarization - the "Casimir polarization" of the EM vacuum using the appropriate boundary conditions were used in the simulation of confinement of quarks and gluons in hadrons. According to this model of bags, hadrons are bubbles in the Quantum Chromodynamics vacuum enclosing quarks and gluons whose currents through the walls of the bubble are zero (confinement).

It is obvious that all specified model analysis may pretend only to the elucidation of the essential possibilities of realization of the real spherical objects stability (more specifically – atomic nuclei) owing to the Casimir pressure of EM vacuum (cases $U_a < 0$), as well as to specifying the possibility of the nuclei stability violation (at $U_a \geq 0$) in case of changing of the boundary conditions for $E_\tau$ and $E_n$ components of the EM vacuum intensity vector at the surface of these objects. It is quite possible that the condition for the calculation of $U_a$ value should become the 3$^{rd}$ kind condition, such as $d/dr\left[r^2 E_n(r)\right]_{r=a} = \chi E_n(a)$, where $\chi$ – the phenomenological parameter, defining the stationary state of the atomic nucleus in case of the dynamic unity of the system nuclear matter – the electronic subsystem of the atom. Since, pursuant to the aforementioned results, at $\chi = 0$ the value of vacuum energy $U_a > 0$, then it would appear reasonable that at some value of $\chi_0 > 0$ within the model analysis $U_a = 0$ should be reached, so at $\chi \leq \chi_0$, the value $U_a > 0$ and the atomic nucleus is unstable. Then at $\chi > \chi_0$, the design energy $U_a$ of such spherical object should be negative, $U_a < 0$, while the object itself will remain stable. Subsequently (Sections V, VI), within the framework of the performed phenomenological analysis, only the possibility of changing of the atomic nucleus stability depending on changing of the nuclear matter state will be taken into account.

It is necessary to emphasize once more that the value $U_a$ reflects only that part of the full energy of the particle in EM vacuum, which depends on the size *a* of the spherical object, and that is why it defines the Casimir pressure. The full energy of the spherical object in EM vacuum is analyzed by quantum electrodynamics methods [55]; it includes some formally infinite $\bar{E}_0$ constant not depending on size, which usually does not attract interest. Within the frameworks of phenomenological approach we fall beyond the scope of the existing methods of analysis and considering the problem in general based on the concepts pertinent to EM vacuum as the base medium for every Universe object. As we shall see below, this will manifest itself in the definition of the parameter $\bar{E}_0$.

To be specific, we will consider an atomic nucleus as the material object and *a priori* suppose that the total energy of this object determined by the Casimir potential energy is negative. Casimir force arises from the action of EM vacuum to atomic nucleus treated as a set of distributed charges (quarks in nucleons) and local exchange currents. In fact, we believe that EM vacuum in the vicinity of the atomic nucleus is polarized: the oscillation spectrum of EM vacuum is rearranged, and a unified dynamic system is formed composed by the atomic nucleus and vacuum, an "EM vacuum polaron". This raises the question: is there any reason for such a hypothesis? Here are these arguments.

First of all, the introduction of the EM vacuum polaron allows qualitatively understanding the genesis of *a priori* limitation of the speed of moving material objects with nonzero rest mass by the velocity of light $c$ in EM vacuum which is equal to $c = 3 \cdot 10^{10}$ cm/s. This conclusion is due to the limitation of the rate of reorganization of EM vacuum adjacent to all atomic nuclei of the moving object. It is clear that the rate of such reorganization of the physical vacuum areas adjacent to the boundaries of these nuclei is limited by the speed of light $c$, which makes impossible the movements of material bodies with such speeds in the EM vacuum, which acts as a material medium, a kind of "ether" (see also Section 2.2). An additional rationale for introducing the "Casimir polaron" hypothesis is the possibility of understanding on this basis the quantum essence of the quantum-mechanical image of "wave-particle" [3] (see Section 3.1) and also the physical essence of operators in the Dirac equation for a free electron, represented as a "Casimir polaron" (see Section 3.2).

When introducing the ideas about the EM vacuum polaron, we assume that the expression for the corresponding potential energy of the Casimir field in the neighborhood of a point material object (for definiteness, a particle with the mass $m_i$) has the form:

$$U(\vec{r}) = -\gamma_0 \frac{\hbar c}{r}. \qquad (4)$$

Here $\vec{r}$ is the radius-vector (we connect the coordinate system with EM vacuum and assume that the particle at rest is localized at the origin); $\gamma_0$ is a dimensionless parameter, characterizing the intensity of the introduced interaction.

The solution of the Schrödinger equation in a centrally symmetric field with potential energy in this form (4) is well known [56]. The energy levels $\bar{E}(n_r)$ of the discrete spectrum, reflecting the degree of interconnection of the particle of the mass $m_i$ with EM vacuum due to its polarization, and the corresponding expression $a_B$ for the radius $a_{Vi}$ of the domain of the EM vacuum Casimir polarization in the neighborhood of the particle $i$, have the form:

$$\bar{E}(n_r) = -\gamma_0^2 \frac{m_i c^2}{2n_r^2}, \qquad (5)$$

$$a_{Vi} = \frac{2\hbar}{\gamma_0 m_i c}. \qquad (6)$$

Here $n_r$ is the principal quantum number. For $\gamma_0 = \sqrt{2}$, the position of the lower energy level ($n_r = 1$), characterizing the binding energy of the considered particle with EM vacuum, corresponds in the absolute value to the "rest energy of the considered particle" $\bar{E}_0 = m_i c^2$ in the form suggested by Einstein, where $U(\vec{r})\big|_{r=a_{Vi}} = -m_i c^2$. Within the developed phenomenological ideas, the quantity $\bar{E}_0$ is defined as the "binding energy of the particle with EM vacuum", so that the mass defect in the nuclear processes just characterizes the energy released due to the difference of binding energies of the initial and final products with EM vacuum. Then the total energy $E_{tot}$ of the stable particle is $E_{tot} = -m_i c^2 + U_a = -m_i c^2 - |U_a|$.

Taking the radius $a_{Vi}$ (6) as the boundary with EM vacuum of the considered material object, we identify on the both sides of this boundary the "external" and "internal" regions of the introduced vacuum polaron. The EM quanta at the excited levels, localized in the "external" region of polaron, should be referred to as virtual quanta, for which the wave vector $\vec{k}$ and frequency $\omega$ are independent variables, not connected by the dispersion relation $\omega = kc = 2\pi c/\lambda$, which holds for the real photon. To each excited level, there correspond the virtual EM quanta with the frequencies $\overline{\omega}(n_r) = \overline{E}(n_r)/\hbar$. Moreover, there may "condensed" at each level an arbitrary number of virtual EM quanta.

According to (6), the value of the domain of the EM vacuum Casimir polarization in the neighborhood of the proton equals $a_{Vp} = 2.82\ 10^{-14}$ cm, i.e. corresponds to the action scale of the nuclear forces. Thus, the quantity $q_s^2 \equiv \gamma_0 \hbar c$ for $\gamma_0 = \sqrt{2}$ in the expression (4) can be defined as the squared "elementary charge of strong interaction", so that the dimensionless constant $\alpha_s$ of such interaction, by analogy with the fine structure constant $\alpha_e$, can be naturally represented in the form

$$\alpha_s = q_s^2/\hbar c = \sqrt{2}. \qquad (7)$$

The value of the fine structure constant $\alpha_e$ can be written as

$$\alpha_e = \frac{a_{Ve}}{\sqrt{2} a_B} = \frac{1}{137}. \qquad (7a)$$

Here, $a_{Ve} = 5.2\ 10^{-11}$ cm is the radius of the domain of EM vacuum Casimir polarization in the neighborhood of the electron, i.e., the "Casimir" size of the electron; $a_B = \hbar^2/m_e e^2 = 0.52 \cdot 10^{-8}$ cm is the characteristic size of the simplest hydrogen atom, its "Bohr radius", i.e., the average distance of an electron in a hydrogen atom to its atomic nucleus (proton). This means that the fine structure constant $\alpha_e$ characterizes the degree of overlap ("interaction") of the Casimir polarization domain of electron shell in the hydrogen atom with the nucleus.

If a particle possesses a structure (hadron), then the dependence (4) inside such particle can be treated as a "seed" potential energy of the nuclear forces, which are characterized by the "nuclear charge" $q_s$ and the strong coupling constant $\alpha_s$. It is obvious that, as a result of dynamic mobility of the nuclear matter "inside" such particle, the seed potential of the nuclear forces is shielded, and the effective potentials are formed of the "short-range" nuclear forces, exponentially declining with the distance, of the type of the Yukawa potential

$$U(\vec{r}) = -\frac{q_s^2}{r} \exp(-\kappa_s r), \qquad (4a)$$

where $\kappa_s = m_\pi c/\hbar$ is a screening factor and $m_\pi$ is the mass of π-meson. This corresponds to the standard idea about the dynamic nature of the nuclear forces, which are usually related to the π-meson exchanges by nucleons.

The unexpectedness of the last result is in the effectively revealed at the phenomenological level physical unity of the electromagnetic and strong interactions. Since the nature of the Casimir effect is related to the local spatial charges of the electromagnetic component of the physical vacuum in the vicinity of and inside a material object (with the resonant amplification of the frequencies, characteristic for the structure of the object, and the exclusion of other frequencies, contained in the spectrum of the physical vacuum), then the nuclear forces in such a phenomenological model are a response of the nuclear matter on the spatial-temporal nuclear scale to the action of the electromagnetic component of the physical vacuum. Obviously, such a response is extremely specific and depends both on the specific character of configuration and dynamics of each individual nucleus and its excitation.

The obtained result (5) can be regarded as a kind of "justification" of the use of (4) up to the distances corresponding to the sizes of elementary particles, since it leads to physically meaningful results. First of all, it turns out that all objects of our world are connected with the EM vacuum, considered as a physically distinguished basic environment. It is also becomes clear the physical cause of the appearance in the expression for the "particle rest energy" $\overline{E}_0$ of a characteristic of this basic environment, the speed of light in vacuum as a parameter, included in the definition of the potential energy (4), characterizing, in the product with the Planck constant, the Casimir polarization of EM vacuum and the corresponding reorganization rate of EM vacuum in the vicinity of the particle as it moves in this basic medium. In accordance with the abovementioned idea of Casimir, it is this polarization of the particle in EM vacuum acting as a factor of stabilization of elementary particles and stable isotopes, that may keep from breakup the electron, as well as others elementary particles having a charge. It is possible that the nature of quark confinement inside hadrons [52] should also be linked to the Casimir force. Of interest also are the excited states of the particle localized in the vacuum. In particular, these levels can be manifested as "resonances", the short-lived excited states of hadrons with the characteristic lifetimes of $10^{-22} - 10^{-24}$ s, which are formed in the interaction of π-mesons with nucleons [57].

2.2. Features of the EM polaron movement in EM vacuum. Inertia. Relativistic effects

Let a particle move freely with the velocity $u$ relative to the introduced basic reference system, and let it have the mass $m_0$ for $u = 0$. Let us find out with what changes in the polarization region of the EM vacuum in the vicinity of this material particle there can be linked the appearance of the factor $\eta_u$ in the expression for the total energy $E_u$ of the particle [58]:

$$E_u = \eta_u m_0 c^2, \quad \eta_u = \left(1 - \frac{u^2}{c^2}\right)^{-1/2}. \tag{8}$$

For the first time this phenomenological factor was introduced by Oliver Heaviside in 1889 [59], in the model calculations of the aether drag (a basic medium of the science of the XIX century) by the moving charged spherical particle of the radius $a$ and the mass $m_0$. The relation (8) is postulated as a base relation for the total energy of a freely moving particle in the Special Theory of Relativity [60].

In accordance with the concepts being developed, the EM vacuum, being the basic medium and a physically distinguished system for all the objects in our Universe, is a modern analogue of the aether of the science of the XIX century. Thus, it is natural to try to relate the effect of the energy increasing of a relativistic particle with the changes in the vacuum polarization region in the vicinity of such particle in the direction of its motion, following the general ideas of O. Heaviside and J.J. Thomson. It should be understood that the postulated EM vacuum polarization in the vicinity of any material object, in reality, implies openness in the dynamic sense of every atomic nucleus of this object for EM vacuum in accordance with the boundary conditions introduced at the interface of every nucleus with EM vacuum. In other words, the properties of any elementary particle are formed in the interaction of its inner essence with the EM vacuum.

We will assume that during movement of the considering atomic nucleus relative to the basic reference system, a definite "equilibrium" exchange of virtual photons, which are localized in the EM vacuum polarization domain in the vicinity of the nucleus, and of the virtual photons of the EM vacuum as the basic medium, is realized. It means, the polarization region is characterized by a certain level of "solvation" or condensation of virtual photons on the system of excitation levels of the nucleus, for every velocity $u$ relative to the basic reference system. It is natural to assume that the exchange rate of virtual photons uniquely depends on the velocity $u$, and in the absence of third-party effects on the particle (free movement in the EM vacuum) when the particle is nonrelativistic, there are no factors that could change the considered exchange rate, and hence the speed $u$. It is precisely in such a motion of a particle with a constant magnitude of velocity in the absence of external influences on the particle, that the nature of inertia consists. At the same time, at relativistic particle velocities, when the manifestation of the dynamic Casimir effect [61, 62] with the direct conversion of the fluctuations of virtual photons into real photons in the region of the boundaries of the objects moving with relativistic velocities, the phenomenon of inertia can be violated.

The exchange of virtual photons, realized on the boundary "atomic nucleus – EM vacuum" is disturbed under external influences on the nucleus (see Sections V and VI) or under its relativistic movements. To characterize the considered exchange of virtual photons, we consider, in accordance with [63-65], a boundary condition of the third kind with the introduction of a boundary ad-state or a state I, from which effectively, with the rate constant $k_1$, there take place transitions of the localized virtual photons into the state of the EM vacuum in the vicinity of the nucleus. We also introduce the boundary state II, from which effectively, with the rate constant $k_2$, there occur the inverse transitions of virtual photons: from the state of the EM vacuum into the localized states in the region of polarization of the EM vacuum by the nucleus. Let us also introduce a dynamic variable $\xi(u)$, which characterizes the level of population of the state I by the localized virtual photons, which determines the level of the notional "lubrication" needed for the movement of the particle through the vacuum with the velocity $u$.

We will assume that, with the increasing of $u$ to the relativistic values, the localized virtual photons will be "blown out" from the frontal with respect to the movement of the particle in the polarization region of the EM vacuum, which can be naturally associated with the increase of the rate constant $k_1 = k_1(u)$ as $u \to c$. Furthermore, the value of the "lubrication" level $\xi(u)$ should decrease (the particle is partially "get stripped"), whereas for $\xi(u) \to 0$, when the virtual photons in the frontal polarization region are absent, the particle in the EM vacuum cannot move. In fact, the EM vacuum acts as "reins" on the particle, which tries to break out of the shell polarizing it, so that, as the velocity $u$ grows, the potential energy of such system increases.

Taking into account what is said, the corresponding balance equation for the variable $\xi(u)$ in the stationary case of moving a particle in the EM vacuum with the velocity $u$ can be represented as follows:

$$\frac{d\xi(u)}{dt} = -k_1\xi(u) + k_2(1-\xi) = 0, \qquad (9)$$

so that

$$\xi(u) = \frac{k_2}{k_1 + k_2}. \qquad (10)$$

We assume, in accordance with the idea of J.J. Thompson, that the frontal region of polarization of the EM vacuum in the vicinity of a particle moving with the velocity $u$ is transformed for $u \to c$ from a spherical one to spheroid, the surface of an ellipsoid of rotation, whose minor semi-axis $b$ is oriented in the direction of the particle velocity, whereas the major semi-axis of this ellipse remains equal to the radius $a$ of the spherical region of polarization in the case of a particle at rest or moving with nonrelativistic velocities. The deformation of the polarization region of the EM vacuum as $u \to c$ can be naturally characterized by the ratio $b/a$ of the diminishing small semi-axis $b$ to the large semi-axis $a$, which is equal to $b/a = \sqrt{1-e^2}$, where $e$ is the eccentricity of the ellipse, defined as the ratio of the distance from its center to each focus to half of the major axis. It is the dependence $e = e(u)$ on the velocity $u$ that can be regarded as an indicator of increasing of the level of "nakedness" of the particle and the disappearance of "lubrication" due to the loss of localized photons in the region of EM polarization for $u \to c$. Clearly, $e \to 1$, when $b \to 0$, and the "nakedness" of the particle increases to its maximum. Furthermore, the value

$$\eta = 1/(b/a) = (1-e^2)^{-1/2} \qquad (8a)$$

can be regarded as a factor which characterizes the rate constant of losing by the EM polarization region of localized photons which provide "lubrication" for moving of the particle in the EM vacuum.

With the introduction of the phenomenological relation $e(u) = u/c$ for the eccentricity of the ellipsoid of rotation, whose form is assumed by the polarization region of the EM vacuum in the vicinity of the particle moving with the relativistic velocity $u$ relative to the basic reference system, it follows from comparing (8) and (8a): $\eta = \eta_u$. If one is guided by the relations of the STR and the results of the relevant experimental studies, then one should take $k_1 = k_{10}\eta_u$, where $k_1 = k_{10}\eta_u$ and $\eta_u$ is the Heaviside factor (8). In addition, the rate constant $k_2$ should not depend on $u$. Then expression (10) can be rewritten as:

$$\xi(u) = \frac{k_2 \eta_u^{-1}}{k_{10} + k_2 \eta_u^{-1}} = \frac{k_2 \sqrt{1-u^2/c^2}}{k_{10} + k_2 \sqrt{1-u^2/c^2}}, \qquad (10a)$$

so that

$$\xi(0) = \frac{k_2}{k_{10} + k_2}, \quad \xi(u) \xrightarrow[u \to c]{} \frac{k_2}{k_{10}} \sqrt{1-u^2/c^2}. \qquad (9b)$$

It is with the decrease of the value $\eta_u^{-1}$ with the increasing of the particle velocity and the increasing of the potential energy of the system under the disappearance of "lubrication" needed to move the body in the basic medium that is natural to connect the nature of the relativistic increase in the inertial mass and the inability to move the object in the medium with the speed of light, according to the understanding of the relation (8) by Feynman [66]. This conclusion is fully consistent with the idea of J.J. Thomson that the arising in the movement of the charged particle kinetic perturbations of the surrounding medium turn out to be equivalent to the potential (not kinetic!) energy of the moving particle, causing an increase of exactly this component of energy [59].

Since the dynamic variable $\xi(u)$ characterizes the level of population of the state I by the localized virtual photons for the particle moving with the velocity $u$ relative to the basic reference system, just it is the impossibility of changing this ratio during free movement in the EM vacuum, when the particle is nonrelativistic, can be considered (perceived) as the law of inertia.

Obviously, the introduction of the absolute base medium, a kind of "ether", as a reference system with choosing the global time t, being common for all points of the expanding Universe, requires explanations in connection with the known results of special theory of relativity (STR).

At first, let us dwell upon one more relativistic effect, namely the increase of the lifetime of decaying relativistic particles. The corresponding experiments were carried out at CERN's second Muon Storage Ring (MSR) which stores relativistic muons in a ring in a uniform magnetic field. We will cite the well studied decay processes $\mu^- \to e^- + \tilde{v}_e + v_\mu$ and $\mu^+ \to e^+ + v_e + \tilde{v}_\mu$ [67]. The probabilities of this processes are described with a high degree of accuracy by the following relation of the STR:

$$w = w_0 \eta_u^{-1} = w_0 \sqrt{1 - \frac{u^2}{c^2}}, \qquad (11)$$

where $w_0$ is the decay probability of the particle at rest. In the frame of considered phenomenological approach [3], it is natural for one to attribute the lowering of the decay probability, because of the phenomenological parameter $\eta_u$ decreasing with increasing velocity $u$ of the particle, to the kinetic difficulties involved in the cardinal alteration of the polarization region of the EM vacuum surrounding the particle prior to its decay. Indeed, such decay becomes possible, provided that a quite definite EM vacuum polarization zones are produced in the neighborhood of the electron and electron antineutrino, as well as the muon neutrino, being formed, the particle scattering directions being governed by the laws of conservation of energy and momentum. And the relatively low probability of the formation of the indicated polarization zones necessary for the decay is associated for relativistic muons with a small phenomenological parameter $\eta_u^{-1}$.

Another well-known result of the SRT is associated with the Clock Hypothesis. Pursuant to STR, it is possible to embed an arbitrary number of inertial reference systems, which are steadily and linearly moving with different velocities in relation to the fixed reference system, and for each of the systems, its "own time" is accepted (the Clock Hypothesis). Thus, if the interval of time $\Delta t \equiv t_2 - t_1$ is fixed using immovable clock (in the laboratory system), then the intervale of time, fixed by the clock moving at $u$ speed, will be equal to $t_2' - t_1' = \Delta t \cdot \eta_u^{-1}$ [60]. Since under STR postulates, the laws of nature are the same in all inertial reference systems, i.e. the inertial systems are equivalent in the context of the real processes description, then introducing of the "own times", as well as of the "own (decreased) length" (of the rod) in the moving systems looks like relative (contingent) procedures, since in that reference system, in which the rod is immovable, its length is not distorted and the time is not retarded.

In the frame of considered phenomenological approach, there is one distinguished absolute reference system associated with the EM vacuum of an expanding Universe, and no inertial reference systems with respect to this system can be introduced. Nevertheless, the Einstein time dilation formula was tested in several experiments. The first, during October 1971, four cesium beam atomic clocks were flown on regularly scheduled commercial jet flights around the world twice, once eastward and once westward, to test Einstein's theory of relativity with macroscopic clocks [68, 69]. From the actual flight paths of each trip, the theory predicts that the flying clocks, compared with reference clocks at the U.S. Naval Observatory, should have lost 40 ± 23 nanoseconds during the eastward trip, and should have gained 275 ± 21 nanoseconds during the westward trip, where the errors are the corresponding standard deviations. In accordance with [68, 69], the recorded dependent time differences were in good agreement with predictions.

In other words, in the described experiments, the real but not virtual phenomenon has been detected: the investigated systems, which are moving "steadily" and "linearly" with nonrelativistic speeds (it is believed that such approximation is quite feasible) in relation to the considered base, actually are non-inertial ones. In the same time, this conclusion fully corresponds to the developed concepts pertinent to the existence of the absolute base medium, a reference system with global time $t$, being universal for all points of the expanding Universe. Objectively, the fixed differences in reading of the cesium atomic clock at different velocities of Cs137 nuclei movement may be connected not with the relativistic change of the timescale, but with a definite physical reason, i.e. with the aforementioned physical process of reorganization of the field of Casimir polarization of Cs137 nuclei in the conditions of high velocities of these nuclei in relation to the base system – EM vacuum. At high velocities of these nuclei, partial loss of the localized photons takes place, while these photons provide "lubrication" for the particle movement in EM vacuum. In this case, the process of radioactive decay, at which the appearing decay products should polarize the EM vacuum, while evolving in the form of polarons, must be made difficult. Of course, such reorganizations in the experimental conditions [68, 69] are insignificant. This is evidenced by the extremely small changes of time (tens of nanoseconds) as compared with the long-term jetliner flights (in excess of hundred hours). However, notwithstanding the macroscopic size of the atomic cesium clock, which was used in the performed experiments [68, 69], just reorganizations of the Casimir polarization field of each cesium-137 nucleus have defined the discovered effect. The author believes that if the mechanical clock with the required measuring precision will be used, the results of the works [68, 69] hardly could be reproduced.

Quantitative measurements of relativistic time dilation were carried out in the 1960s, after the discovery of the Mössbauer effect in 1958, and the interest in such measurements lasts to this day. Many trials have been made to measure the transverse second order Doppler shift of the 14.4-keV Mössbauer absorption line of Fe57 in a rotating system as a function of the angular velocity ω. A Fe57 absorber was placed at a radius of 9.3 cm from the axis of the rotor. A Co57 source was mounted on a piezoelectric transducer at the center of the rotor. By applying a triangularly varying voltage to the transducer, the source could be moved relative to the absorber. This arrangement makes possible the observation of the entire resonance line at various values of ω. Early experiments showed [70] that the measured transverse Doppler shift agreed with the predictions of the theory of relativity. However, it was later indicated that the deviation of the shift in the experiment may be due to a longitudinal Doppler shift caused by the acceleration of the absorber. It is necessary to note that time dilation Mössbauer effect experiments have an advantage in identifying relatively small shift due to acceleration, since in these experiments the shift due to the velocity is of second order while the shift due the acceleration is of first order [71]. Apparently, such experiments will be continued. In the author's opinion, reorganizations of the Casimir polarization field of EM vacuum in the vicinity of each Fe57 nucleus in a rotating system at relativistic speeds may be the factor exerting an effect on the Doppler shift of the 14.4-keV Mössbauer absorption line, especially as just the EM vacuum effect initiates the emission of $\gamma$-quanta by the excited nuclei of Fe57 [50].

Thus, we see that the results provided by the Mössbauer spectroscopy experiment of Kündig [70] and the muon experiment of Bailey et al [67] can naturally be related to changes in the Casimir polarization of the EM vacuum in the vicinity of the Mössbauer nuclei Fe57 and the decaying muons, correspondently, at high speeds relative to the reference frame, rather than as confirmation of the Clock Hypothesis. We see also that the rates of nuclear processes (nuclear decay, Mössbauer effect) can change (decrease) due to the effects of a dynamic nature - due to a change in the region of the Casimir polarization of nuclei with respect to the movement of the nuclei relative to the base reference frame. As will be shown in Sections V and VI, external factors of a different nature – the initiating effects of electrons of high (chemical scale) energies can lead to an increase in the decay rate of nuclear processes – including radioactive nuclear decays.

In connection with the traditional binding of the studied objects to the inertial and non-inertial reference systems, it should be noted that this practice has proven its practicality since the times of G. Berkeley ([72]), who proposed to connect a choice of inertial reference systems with the orientation of the coordinate axes to the "motionless stars". The current practice shows that it is possible to introduce virtually "an absolute inertial frame of reference" in which the relative acceleration of the sufficiently distant from each other bodies is no greater than $10^{-10}$ m/s² [73]. Such a system, in particular, is the International Celestial Reference System, for which the origin is taken to be the position of the barycenter, the center of mass of the solar system, whereas its axes are fixed in space with respect to quasars, the most distant objects of the observable Universe. However, the introduction of the EM vacuum as an absolute reference frame with the choice of a global time *t*, uniform for all points of the expanding Universe, means that the binding of the studied objects to the inertial and non-inertial reference systems can be justified only for nonrelativistic speeds.

The introduction of EM vacuum as the base Universe medium, as well as of the Casimir polarization of EM vacuum in the vicinity of nuclei and electronic subsystems makes it possible to reconsider some of the discovered key problems of contemporary fundamental physics, as it was specified in the introductory section. Of course, the phenomenological analysis of such problems, which has been presented in this and the next Sections, should be considered only as the stage preceding the development of adequate theoretical interpretations of these problems; however, it is the necessary stage. Just the phenomenological approaches applied for revealing the base transcendental hypotheses, which do not follow directly from the experiments, should create foundations of the future theories.

The author hopes that the performed analysis of the well-known, however till now not solved problems, will help the reader to understand adequately the new group of phenomena, which are defined as LENR and considered in sections V-VI, the understanding of which is also based on introducing of the concepts pertinent to Casimir polarization of EM vacuum as the base environment in the vicinity of nuclei and electronic subsystems.

III. EM VACUUM AS A BASIC MEDIUM FOR QUANTUM MECHANICAL PHENOMENA

We note that the introduction of the EM vacuum as an active basic medium allows, above all, approaching the understanding of the physical nature of the basic postulates of quantum mechanics. We should note this referring to the remark by R. Feynman made in 1965: "I can safely say that nobody understands quantum mechanics" ([4], p. 129). Can this claim be contested today, after more than 50 years? Such an attempt is made below.

It is known that the main difficulties in the understanding of quantum mechanics have associated with the physical essence of the "wave-particle" image as well as with the introduction of operators of observable quantities instead of these quantities themselves into the mathematical apparatus of quantum mechanics. We will discuss these problems first of all.

3.1. The essence of the "wave-particle" image

In accordance with the introduced concepts of the EM vacuum as the base medium and the Casimir polarization of the EM vacuum in the vicinity of particles, the movement of the particles with the mass $m$ in EM vacuum with the velocity $u$, in fact, means the displacement of the local heterogeneity of EM vacuum with the momentum $p = mu$ and the energy $E = \sqrt{p^2 c^2 + m_0^2 c^4} = mc^2 = \hbar\omega$, where $m_0$ is the mass of the particle for $u = 0$ and $\omega$ is the cyclic frequency of the moving perturbation. The phase velocity $u_{ph}$ of the displacement of such perturbation of EM vacuum with the wave number $k = p/\hbar = 2\pi/\lambda_{dB}$, which we connect with the de Broglie wave ($\lambda_{dB}$ is the corresponding wavelength), is equal to $u_{ph} = \omega/k = E/p = c^2/u$, whereas the group velocity of the displacement equals $u_g = dE/dp = u$. In this case, the domain of the EM vacuum Casimir polarization is "attached" to the moving particle, and the phase velocity $u_{ph}$ of the moving heterogeneity region is uniquely related to the particle velocity $u$. Therefore, the spreading of the domain of the moving perturbation of the EM vacuum will not occur when the perturbation propagates (there is no time dispersion). These wave properties are inherent in each of the moving particle.

Such view of the Broglie wave genesis differs from the accepted one in the canonical quantum mechanics, in which the Broglie waves are simply postulated, while the spatial dispersion is forcedly attributed to these waves. By virtue of the latter fact, the Broglie wave because of the dispersion should decompose at microscopic distances. "Just this approach kept the founders of quantum mechanics from connecting of the Broglie wave with some real property of the stable electron. However, any serial transmission electron microscope operates in spite of the canonical interpretation of the microcosm physics. In electron microscopy practice, it is necessary to accept that the Broglie wave in the microscope actually accompanies electron and passes the substantial distance from the cathode to the detector without decay. Electrons in the beam within the microscope demonstrate corpuscular and wave properties SIMULTANEOUSLY, but not "either – or", as it is commonly claimed" [74].

In the work [75], it was experimentally confirmed that the wave properties are inherent not only to the flow of electrons but also to each electron separately. It is by this fact that the known phenomena are conditioned of the formation of the diffraction patterns for extremely low fluxes of particles, striking on the diffraction grating, when the interaction between the particles is impossible. It was shown that even in case of a weak electron beam, when each electron passes through the instrument independently from others, the diffraction pattern, which appears at long exposure, does not differ from the diffraction patterns, which are obtained at short exposures for the electron flows, the intensities of which are million times as high. The electron (the kinetic energy of which was equal to $E_e$ = 72 keV) passed the instrument during 8,5·10⁻⁹ sec, after which during the period exceeding the mentioned time interval by 30,000 times (!), on average, the instrument stayed empty, and only after elapsing of this period, the new electron passed the instrument. It is evident that at such a big time interval between subsequent electron passes, the probability of a simultaneous pass of at least two electrons is negligible.

3.2. On the reasons for introducing the operators of the observed values into the quantum mechanics mathematical apparatus. Physical meaning of operators in the Dirac equation

Let us assume that any particle in any moment possesses well-defined physical characteristics (space position, momentum, potential energy, etc.) irrespective of the availability of the corresponding measuring devices. Of course, these values cannot be measured simultaneously, if for no other reason than two instruments cannot be placed in one space point. The reason of introducing operators into the quantum mechanics mathematical apparatus lies exclusively in the fact that these particles are not the points because of the Casimir polarization of the EM vacuum. That is exactly why quantum mechanics as an adequate microcosm science has applied the mathematical apparatus with the introduction of linear operators corresponding to the observed characteristics and acting on the wave functions $\psi(\vec{r},t)$. In this case, the *correspondence principle* is introduced, according to which each physical characteristic should have a corresponding operator, and vice versa, each operator in quantum mechanics should have a corresponding physical characteristic, while the transformation formula for the operator of a definite characteristic should be identical to the transformation formula for this characteristic..

The important role of the latter fact was discussed in the work [76] in connection with the problem existing in quantum electrodynamics, which is practically not discussed in the scientific publications: the absence of a quantum-mechanical foundation of the well-known Dirac equation [77] for the unbound particle, which was postulated by Dirac in 1928. In the Schrödinger representation this equation has the form:

$$i\hbar \frac{\partial \psi(\vec{x},t)}{\partial t} = c\left[\sum_{j=1}^{3} \hat{\alpha}_j \hat{p}_j + m_0 c \hat{\alpha}_0 \right] \psi(\vec{x},t). \qquad (12)$$

where: $m_0$ – the rest mass of the electron; $\vec{x} = (x_1, x_2, x_3)$ and $t$ – spatial coordinates and time, respectively; $\hat{p}_j = -i\hbar \partial/\partial x_i$ – three operators of the momentum components (along $x_1, x_2, x_3$); $\psi(\vec{x},t)$ – four-component complex wave function (bispinor); $\hat{\alpha}_0$, $\hat{\alpha}_1$, $\hat{\alpha}_2$, $\hat{\alpha}_3$ – linear operators over the bispinor space, which exert an effect on the wave function. These operators are selected in such a manner that each pair of operators anticommutes, while the square of each operator is equal to one:

$\hat{\alpha}_i \hat{\alpha}_j = -\hat{\alpha}_j \hat{\alpha}_i$, where $i \neq j$, and the indices $i$ and $j$ are varied from 0 to 3;

$\hat{\alpha}_i^2 = 1$ for $i$ from 0 to 3.

In the considered approach, these operators are presented by matrix arrays of 4×4, which are called Dirac alpha-matrixes. It should be emphasized that these operators have been introduced just as mathematical objects, without reference to the correspondence principle:

$$\alpha_0 = \begin{pmatrix} I & 0 \\ 0 & -I \end{pmatrix}, \quad \hat{\vec{\alpha}} = \begin{pmatrix} 0 & \hat{\vec{\sigma}} \\ \hat{\vec{\sigma}} & 0 \end{pmatrix}; \quad \sigma_x = \begin{pmatrix} 0 & 1 \\ 1 & 0 \end{pmatrix}, \quad \sigma_y = \begin{pmatrix} 0 & -i \\ i & 0 \end{pmatrix}, \quad \sigma_z = \begin{pmatrix} 1 & 0 \\ 0 & -1 \end{pmatrix}. \qquad (13)$$

where: 0 and $I$ – 2×2 zero and unit matrixes, respectively; $\sigma_j$ ($j$ = x, y, z) – Pauli matrixes. These $\hat{\vec{\sigma}}$ matrixes were used by Pauli while introducing the spin vector operator $\hat{\vec{s}} = 1/2 \cdot \hbar \hat{\vec{\sigma}}$, in deriving the equation for the 2-component wave function (spinor) $\varphi(\vec{x},t)$ of the nonrelativistic particle with 1/2 spin in the external electromagnetic field:

$$i\hbar \frac{\partial \varphi}{\partial t} = \frac{\left(\vec{p} - \frac{e}{c}\vec{A}\right)^2}{2m_0} \varphi - \frac{e\hbar}{2m_0 c} \vec{\sigma}\vec{H}\varphi + e\Phi\varphi. \qquad (14)$$

where: $\Phi(\vec{x},t)$ and $\vec{A}(\vec{x},t)$ – the scalar and vector potentials of the electromagnetic field, respectively; $\vec{H} = rot\vec{A}$ – magnetic field intensity. In order to solve the problem of introducing the corresponding physical characteristics for the alpha-operators, Fernando Wilf has introduced two dimensional operators instead of $\hat{\alpha}_0$ and $\hat{\vec{\alpha}}$:

$$\hat{\tau}_0 = \frac{\hbar}{m_0 c^2}\hat{\alpha}_0 \equiv \tau_0 \begin{pmatrix} I & 0 \\ 0 & -I \end{pmatrix} \text{ и } \hat{R}_j = \frac{\hbar}{m_0 c}\hat{\alpha}_0 \hat{\alpha}_j \equiv R_0 \begin{pmatrix} 0 & \vec{\sigma} \\ -\vec{\sigma} & 0 \end{pmatrix}, \qquad (15)$$

which were associated with some time interval $\tau_0$ (period) and some radius-vector $\vec{R}$ for some point on some sphere, the center of which corresponds to the center of inertia of the point particle, the movement of which is characterized by momentum $\vec{p} = m_0 \vec{u}$, where: $m_0$ and $\vec{u}$ – the mass and velocity vector of the electron in the base reference system connected with EM vacuum. Using these operators, the Dirac equation may be presented as follows:

$$\left( i\hbar \hat{\tau}_0 \frac{\partial}{\partial t} - \sum_{j=1}^{3} \hat{R}_j \hat{p}_j \right)\psi(\vec{x},t) = \hbar\psi(\vec{x},t). \qquad (16)$$

For the stationary state, characterized by the energy $\varepsilon$, the following equation will be valid:

$$\left[ \varepsilon \hat{\tau}_0 - \hat{\vec{R}}\hat{\vec{p}} \right]\psi(\vec{x},t) = \hbar\psi(\vec{x},t). \qquad (17)$$

Since for the states with definite momentum values, the momentum operator coincides with the momentum vector, then, according to Davydov [78], the 4-component bispinor $\psi$ may be presented via 2-component spinors $\varphi$ and $\chi$ by the following set of equations:

$$\varepsilon \tau_0 \hat{I}\varphi - R_0 \hat{\vec{\sigma}}\vec{p}\chi = 0,$$
$$-\varepsilon \tau_0 \hat{I}\chi + R_0 \hat{\vec{\sigma}}\vec{p}\varphi = 0. \qquad (18)$$

For solving this set of equations (18), its determinant should be equal to zero. With accounting of the operator identity:

$$(\vec{\sigma}\vec{A})(\vec{\sigma}\vec{B}) = \vec{A}\vec{B} + i\vec{\sigma}[\vec{A}\vec{B}],$$

for the arbitrary vectors $\vec{A}$ and $\vec{B}$, commutating with $\hat{\vec{\sigma}}$, one can obtain:

$$\varepsilon^2 = \frac{1}{\tau_0^2}\left( \hbar^2 + R_0^2 p^2 \right) = m_0^2 c^4 + p^2 c^2 \equiv m^2 c^4. \qquad (19)$$

Introducing the concept of the electron as a "Casimir polaron" allows associating the introduced operators $\hat{\tau}_0$ and $\hat{\vec{R}}$ with more straightforward physical characteristics of the electron. The vector operator $\hat{\vec{R}}$ is naturally associated with the concept of the electron, in which the latter is not a point elementary particle, but is a "Casimir polaron" with a characteristic size $R_0 = \hbar/m_0 c$ of EM vacuum polarization field. The scalar operator $\hat{\tau}_0$ may be associated with the specific reorganization time, which is pertinent to the Casimir polarization field of EM vacuum in the vicinity of the electron as it moves in EM vacuum, while this polarization is caused by the interchange with virtual photons, which acts as the "lubrication" for the particle movement in EM vacuum. At relativistic velocities of electron movement, the Casimir polarization of EM vacuum in the movement direction is decreased. In the ultra-relativistic case ($u \to c$), when almost the whole energy is connected with the moving mass, and it can be accepted that $m = m_0 \eta_u \equiv m_0 \left(1 - u^2/c^2\right)^{-1/2}$, the value $R_0 \to \hbar/mc = R_0 \left(1 - u^2/c^2\right)^{1/2}$, the "lubrication" share abruptly decreases, the resistance to movement increases, which leads to an increase in the inertial mass (potential energy) of the electron. In this case, the scalar operator $\hat{\tau}_0$ may be associated with the decreasing time $\tau_0 \to \hbar/mc^2 = \tau_0 \left(1 - u^2/c^2\right)^{1/2}$, which reflects the suppression of reorganization of the Casimir polarization field of EM vacuum, which is necessary for the electron movement.

3.3. Stability of electronic subsystems in quantum mechanics

In connection with introducing the concept of the Casimir polarization field of EM vacuum and the concept of the electron as "Casimir polaron", there appeared new problems, which have not been discussed in quantum mechanics earlier, and which are pertinent to the stability of two- and multi-electron associates in the conditions of integral electroneutrality of the considered systems. Here the author means electrons, which participate in the arrangement of the valence links in molecules, Cooper pairs in superconductor systems, electronic subsystems of atoms, etc. The main requirement for the quantum-mechanical consideration of the energy states of such systems is pertinent to the antisymmetrization of the corresponding wave functions. Such requirement obviously reflects the phenomenon of arrangement of the unified electronic subsystems, bond in which is realized in the consequence of overlapping of the Casimir polarization fields of EM vacuum by electrons with definite spin alignment (according to Ohanian [79], the spin of a particle can be treated as a circulating flow of energy in the wave field of the electron, i.e. electron, being an open system, uses the EM vacuum energy). Apparently, the requirement of the wave functions antisymmetrization not only reflects the necessary condition pertinent to the minimization of the electronic subsystems energy but also reflects the necessity for the definite pressure drop outside and inside such subsystem in order to provide their stability.

Such phenomena as the surprising compactness of many-electron shells of the atoms, proximity of sizes of all atoms of all atoms – from the hydrogen atom (~ 0.05 nm) up to the atoms, the quantity of electrons of which exceeds one hundred (~ 0.15 nm) – may be connected just with the Casimir polarization of EM vacuum in the vicinity of electronic systems. All of this is observed in the conditions of the Coulomb repulsion of electrons. From this point of view, different model problems come down to setting to establishing those factors, which exert an effect on the stability of such electronic subsystems, including the distribution of positive charges in the considered system, the dynamics of the ion subsystem, the effect of different external factors on the stability of electronic subsystems, etc., may be of particular interest.

The author also believes that the introduction of the concept of electrons as "Casimir polarons", which are connected with EM vacuum, opens definite possibilities for solving the technical problem pertinent to the regularization of divergent integrals in calculations of the radiation corrections, which is usually understood as "logical imperfection of the existing quantum electrodynamics" [77, paragraph 110]. In this case, one should consider the possibility of constructing Feynman diagrams with the restriction of the ranges of energy integration of virtual particles by intervals $\left[-mc^2, 0\right]$.

IV. EM VACUUM IN THE DYNAMICS OF UNIVERSE

On the basis of the developed ideas about the physical nature of the relativistic increase in the masses of material particles (nuclei of macroscopic bodies, electrons) and limiting the speed of material particles and macroscopic objects as a set of nuclei and electrons, alternative approach to understanding the dynamics of the Universe can be developed on the basis of which the essence of dark energy and dark matter can be understood, as well as can be resolved "the problem of orders" [9, 12, 13]. True, it was necessary to adopt an alternative model for the dynamics of the Universe, *a priori* considering our Universe as an open system.

4.1. The standard model of the Universe dynamics [7-10, 80]

The basis of the standard model of the Universe dynamics is the Hubble ratio:

$$\dot{a} = Ha, \quad (20)$$

which relates the rate of change of the metric $a$ to the so-called constant Hubble $H$, and the Friedmann equations of the GR theory of relativity for the Euclidean Universe:

$$\left(\frac{\dot{a}}{a}\right)^2 = \frac{8\pi G}{3c^2}(\varepsilon_V + \varepsilon_m), \quad (21)$$

$$\frac{\ddot{a}}{a} = -\frac{4\pi G}{3c^2}(\varepsilon_m - 2\varepsilon_V + 3p). \quad (22)$$

Here $G$ and $c$ are, respectively, the gravitational constant and speed of light in a vacuum; $\varepsilon_V$ and $\varepsilon_m$, respectively, the vacuum energy density and the energy density of gravitationally interacting particles; $p$ is the value of the effective pressure averaged over all galaxies and clusters of galaxies. In accordance with the standard model, the vacuum energy density $\varepsilon_V$ must determine, according to Eq.(2), the cosmological term $\Lambda$ and plays the role of "antigravity". As stated above, according to the present-day data, $\varepsilon_V \approx 0.66 \cdot 10^{-8}$ $erg/cm^3$ [11]. Note also, that according to the estimates [11], in the present epoch $H = 73$ km/(s×Mpc) $\approx 2.36 \cdot 10^{-18}$ s$^{-1}$ and the average energy density of the Universe is $\varepsilon_{tot} = \varepsilon_V + \varepsilon_m \approx 0.9 \cdot 10^{-8}$ erg/cm$^3$.

However, the attempts to link the quantity $\varepsilon_V$, determined on the basis of (2) from the experimentally determined value of $\Lambda \approx 1.37 \cdot 10^{-56}$ cm$^{-2}$, with the parameters of the physical vacuum, were unsuccessful. The differences were 120 orders of magnitude (see below, Section 4.6). Since this problem couldn't be solved in the frame of the standard theory, to the energy density $\varepsilon_V$ was given a different physical meaning (!) as the density of hypothetical "dark energy", which is ~ 73% of the total energy density $\varepsilon_{tot}$ of the Universe and which is uniformly "spilled" in the Universe. Then, the value $\varepsilon_m$ of the energy density of gravitationally interacting particles was divided into two terms, $\varepsilon_m = \varepsilon_b + \varepsilon_{dm}$, where the energy density $\varepsilon_b$ (~ 4% $\varepsilon_{tot}$) corresponds to the baryonic component of matter, and the energy density $\varepsilon_{dm}$ (~ 23% $\varepsilon_{tot}$) corresponds to so called "dark matter". The latter, physically hard-to-imagine substance is introduced into the Friedmann equations of the dynamics of the Universe in order to remove contradictions between the apparent masses of gravitationally bound objects, as well as systems of such objects, and their observable parameters, including the structural stability of galaxies and galactic clusters in the expanding Universe. It should be noted, the quantity $\varepsilon_b$ also includes the energy density of non-baryonic components (electrons, neutrinos, electromagnetic radiations).

In addition, a state equation $p_V/\varepsilon_V = -1$ with negative pressure was introduced for the "dark energy". It is because of the action of negative pressure under the expansion of the Universe that energy is radiated in the amount necessary to "swell" space. The physical essence of such a model is very difficult to understand, rather, impossible. Indeed, every day the volume of the Universe is increased by $10^{18}$ cubic light years. Factually, there's a kind of "free lunch" ([72], Chapter 12).

It should be added that the Universe is spatially homogeneous and isotropic on a large scale (hundreds of Mpc), but the Universe is not homogeneous in time translation: the epochs of the Universe development are very different from each other. Indeed, the specific density of the most powerful energy sources that appeared in the early stages of the Universe evolution was significantly (thousands of times!) higher than the observed corresponding values for subsequent epochs [81]. It was at the early stages of the development of the Universe, which corresponded to redshift of z ~ 1 (~ 6 billion light years or more), quasars were formed, gamma-busters were manifested. In accordance with Noether's theorem, there is no law of the energy conservation in the Universe. In other words, the Universe is an open system, and there is a source of energy that constantly feeds it. But the standard model excludes the existence of such a source.

## 4.2. Phenomenological equations of the dynamics of the Universe as an open system [82]

At first, after the basic frame of reference for our expanding Universe, "tied" to EM vacuum, was introduced, it is possible to return to the basic questions that could not be solved within the framework of the standard model of the Universe dynamics:
1. What is the source of energy for the Universe and what is the essence of "dark energy" and "dark matter"? Is the "lunch", mentioned by Davies, "free"?
2. How is "the 120-order problem" of establishing a connection between the dark energy and the cosmological constant solved?

It is believed that the Universe, which is represented in the form of a ball with the Hubble radius $R_H = c/H$, whose volume is equal to $V_H = \frac{4}{3}\pi R_H^3$, is an open system, and the source of energy that "feeds" the Universe is the "Proto-vacuum" that is outside the ball-Universe and is more energy-intensive medium (like "False vacuum" in the Inflation theory) than the EM vacuum of our Universe. We believe that a Proto-vacuum existed in the Boundless Universe before the Big Bang.

It is assumed that the energetic power that constantly feeds our Universe across the boundary "the Proto-vacuum – EM vacuum" is equal to the Planck power $w_{Pl}$. In this case the total energy $E_{tot}$ received by the Universe during the time $t$ of the action of the Planck energy source with power $w_{Pl} = c^5/2G$ is

$$E_{tot} = w_{Pl} t = \frac{c^5}{2G} \cdot \frac{1}{H} = \frac{4}{3}\pi R_H^3 \varepsilon_{tot}, \tag{23}$$

so that

$$\varepsilon_{tot} = \frac{3c^2 H^2}{8\pi G}. \tag{24}$$

Within the framework of the developed phenomenological approach, the time $t$ is defined as the reciprocal of the Hubble $H$ parameter, $t = H^{-1} \approx 13.4$ billion years, and for the estimator of the Hubble radius of the Universe, we have $R_H = c/H \approx 1.27 \cdot 10^{28}$ cm.

Expression (24) for the average energy-mass density $\varepsilon_{tot}$ obtained from the Planck source of the Universe during the time $t$ of its existence – the expansion in accordance with the empirical Hubble relation (20) for the rate $\dot{a}$ of changing the scale factor $a$ is formally represented as the first Friedmann equation [80] for Euclidean space, but for a different sense of the incoming parameters, $\varepsilon_V^e$, $\varepsilon_b^e$ and $\varepsilon_{dm}^e$:

$$\left(\frac{\dot{a}}{a}\right)^2 = \frac{8\pi G}{3c^2}(\varepsilon_V^e + \varepsilon_b^e + \varepsilon_{dm}^e), \tag{25}$$

where $\varepsilon_V^e$ is the EM vacuum energy density; $\varepsilon_b^e$ is the density of the energy binding to the EM vacuum all the mass components of the Universe in their state of rest with respect to the base EM vacuum reference system (this quantity, characterizing the degree of "freezing" of the resting mass components into the space of the EM vacuum, is negative: $\varepsilon_b^e = -|\varepsilon_b^e|$); $\varepsilon_{dm}^e$ is the energy density component, which is also negative ($\varepsilon_{dm}^e = -|\varepsilon_{dm}^e|$) and which characterizes the relativistic increase in the mass and the degree of "freezing-in" of those star clusters and galaxies that move with relativistic velocities relative to the base reference system (the subscript "dm" is used here to denote that the energy density so introduced is related to "dynamical mass"; the hypothetical "dark matter" is a phantom!).

To evaluate the quantity $p_{eff}$, we consider the amount $\Delta A$ of work expended on the formation of fresh space upon enlargement of the radius $R_H$ of the Universe in a time of $\Delta t$. Considering relation (25), we find

$$\Delta A = \frac{\varepsilon_V^e}{\varepsilon_{tot}} w_{Pl} \Delta t = p_{eff} \Delta V_H \bigg|_{R_H} = p_{eff} \frac{\Delta V_H}{\Delta R_H} \cdot \frac{\Delta R_H}{\Delta t} \bigg|_{R_H} \Delta t = p_{eff} \cdot 3 V_H H \Delta t, \qquad (26)$$

so that for $p_{eff}$ we get

$$p_{eff} = \frac{\varepsilon_V^e c^5}{8\pi \varepsilon_{tot} R_H^3 H G} = \frac{1}{3} \varepsilon_V^e. \qquad (27)$$

Expression (26) actually "replaces" the Friedmann second equation (22) in the standard model. (The pressure $p$ considered in the standard model is made negative in magnitude in order to realize "antigravity").

We will take into account that the energy $\varepsilon_b^e$ and $\varepsilon_{dm}^e$ densities introduced are different from the energy densities $\varepsilon_b$ and $\varepsilon_{dm}$, considered in the standard model, and we also take for the energy density $\varepsilon_{tot}$ the value found in [11] within the framework of the standard model, since the actual basis for determining the total energy $\varepsilon_{tot}$ density in the Universe in the standard model and in the open Universe model under consideration is formally the same general energy balance equation - the 1st Friedmann equation (see Eqs. (21) and (25)). Then a simple recalculation shows that the values of $\varepsilon_V^e = 1.14 \times 10^{-8}$ erg/cm³, introduced on the basis of the modified Friedmann equation (25), will correspond to the values of $\varepsilon_V \approx 0.66 \times 10^{-8}$ erg/cm³ determined earlier on the basis of Friedmann equation (21), but with a different sense of the incoming parameters. We find also, $\Lambda \approx 2.36 \times 10^{-56}$ cm⁻². Note, the value obtained $p_{eff} \approx +3.8 \cdot 10^{-9} \, erg/cm^3$ completely agrees with the values in the standard model adopted for the repulsive pressure. We also indicate that within $\Delta t$ = 24 hours the volume of the Universe increases by an amount $\Delta V_H \approx 6.2 \cdot 10^{18} (light \, years)^3$ (in agreement with the Davis's estimate).

4.3. Weinberg's modified phenomenological relation

There is one more problem in cosmology, which so far is practically not discussed - the determination of the reasons for the differences in the dynamics of the Universe in different temporal epochs. The existence of the most powerful energy sources that appeared in the early stages of the Universe evolution [81] was not fixed in the subsequent stages of the evolution of the Universe. If we proceed from the single essence of the Universe, in accordance with the conceptual position of Mach, the features of the Universe dynamics on the macroscale should be associated with understanding the processes on the microscales. In this case, the differences in the features of the dynamics of the Universe at different stages of its evolution can be attributed to the idea of P.A.M. Dirac on the change in time of world constants $\hbar$, $c$ and $G$. But it is necessary to understand the reasons for such changes! What causes them?

Below we show that for a possible answer to this question is sufficient to modify the known result of S. Weinberg, who noticed the approximate equality [83]:

$$\hbar \approx \frac{1}{2\pi} G^{1/2} m_\pi^{3/2} R_H^{1/2}, \qquad (28)$$

where $m_\pi$ is the mass of $\pi$-meson ($m_\pi \approx 140$ MeV/$c^2$), and represent the expression for $\hbar$ in the form of the equality [82]:

$$\hbar = \frac{1}{2\pi} G^{1/2} m_Q^{3/2} R_H^{1/2}. \qquad (29)$$

Here we introduce a new mass parameter $m_Q$, the value of which ($m_Q \approx 3.6 \cdot 10^{-25}$ g) is defined in such a way that there holds the well-known connection between the de Broglie wavelength and the particle momentum. The corresponding energy parameter $E_Q = m_Q c^2 \approx 202.5$ MeV can be considered as a specific energy of reorganization of the physical vacuum, corresponding to the elementary quantum of action. This value turns out to be corresponding to the energy scale of 200 MeV, considered in quantum chromodynamics [84]: at the corresponding intranuclear temperatures there occurs a phase transition in the nuclear matter, quarks are no longer bound in the nucleons, and the quark-gluon plasma is formed.

The expression (28) can be conveniently represented in the form:

$$G = \frac{(2\pi\hbar)^2 H}{m_Q^3 c} = \frac{2\pi^2 cH}{m_Q} a_Q^2 = 2^{3/2} \pi^2 \frac{\hbar c}{m_Q^2} \frac{a_Q}{R_H}. \quad (30)$$

Here $a_Q = 2^{1/2} \hbar / m_Q c \approx 1.31 \ 10^{-13}$ cm is the domain of the EM vacuum Casimir polarization in the neighborhood of the particle of mass $m_Q$ [3]. We will give to the quantity $m_Q = E_Q/c^2 \approx 3.6 \ 10^{-25}$ g the meaning of a fundamental one, characteristic for the strong nuclear interactions of masses, considering it, as well as the Hubble radius $R_H$ of the Universe, as a universal parameter together with $\hbar$ and $c$. We believe that the introduction of the four basic fundamental parameters $\hbar$, $c$, $m_Q$ and $R_H$ suffices for the representation of all basic interactions in the Universe at the phenomenological level.

The usage of the introduced collection of the basic universal constants $\hbar$, $c$, $m_Q$ and $R_H$ allows to present in a more "compact" fashion the quantity $q_g^2 \equiv G m_Q^2$ which is considered as the squared "elementary gravitational charge" as well as the dimensionless constant $\alpha_g$ of the gravitational interaction, defined as:

$$\alpha_g = \frac{G m_Q^2}{\hbar c} = (2\pi)^2 \frac{\hbar H}{m_Q c^2} = 2^{3/2} \pi^2 \frac{a_Q}{R_H} \approx 2.88 \cdot 10^{-40}. \quad (31)$$

The expression (31) allows to understand that the nature of the unique smallness of gravitational interaction is the smallness of the ratio of the EM vacuum Casimir polarization region characteristic size $a_Q$ in the neighborhood of the atomic nucleus to the characteristic size $R_H$ of the Universe. Thus, the name of "the Law of Universal Gravity" for the relation describing the gravitational interaction of two arbitrary masses is "justified" at the conceptual level.

In addition to the dimensionless constant $\alpha_g$ of the gravitational interaction and introduced in section 2.1 the dimensionless constant $\alpha_s$ of strong nuclear interaction and the fine structure constant $\alpha_e$, we will need the corresponding dimensionless constant $\alpha_F$ of weak nuclear interaction. Here we must bear in mind that weak nuclear interactions are not as weak as it is often assumed: the value of the corresponding dimensionless constant $\alpha_F$ is almost an order of magnitude greater than the value $\alpha_e$ of the fine structure constant [3]. Indeed, if we take $\alpha_s = \sqrt{2}$ as the dimensionless constant of strong nuclear interaction, then taking into account the value of the square of the "elementary charge of weak nuclear interaction" $q_F^2 \equiv G_F / a_Z^2$ [3], where $a_Z = 2^{1/2} \hbar / m_Z c \approx 3.3 \ 10^{-16}$ cm is the characteristic radius associated with the mass of the intermediate $Z^0$ vector boson ($m_Z$ = 91.2 GeV/$c^2$ = 1.62·10$^{-22}$ g), and $G_F = 1.17 \cdot 10^{-5} (\hbar c)^3 / (GeV)^2$ is the Fermi constant of the four-fermion interaction, we obtain: $\alpha_F = q_F^2 / \hbar c \approx 4.9 \cdot 10^{-2}$, so that $\alpha_F/\alpha_s \approx 3.45 \cdot 10^{-2}$. Remind that $\alpha_e = 1/137 \approx 0.73 \cdot 10^{-2}$, so that $\alpha_e/\alpha_s \approx 5.2 \cdot 10^{-3}$ and $\alpha_F/\alpha_e = 6.7$. Unfortunately, very often in the literature, in the estimation of the dimensionless constant of weak nuclear interaction, the proton mass is used as the normalization mass which is almost 100 times smaller than the mass of the $Z^0$ vector boson. For this reason, the value of the constant $\alpha_F$ is underestimated by almost 4 orders of magnitude. The real value of this constant, in accordance with the above estimates, is only 35 times, not 5 orders of magnitude less than the dimensionless constant of the strong nuclear interaction.

In accordance with the Dirac's idea, we will assume that under the introduction of the value $R_H$ or the Hubble parameter $H = 1/t$, the dependence of the basic fundamental parameters $\hbar$, $c$ and $G$ on the global time $t$ is postulated. Each $t$ value corresponds to a specific epoch in the evolution of the Universe and the well-defined set of these fundamental parameters in this era. To find such dependencies $\hbar = \hbar(t)$, $c = c(t)$ and $G = G(t)$, it is first necessary to understand the physical essence of the phenomenon of gravitational attraction. This is done in the next section. On the basis of ideas about the Casimir polarization of the EM vacuum in the vicinity of atomic nuclei and the physical essence of mass as a measure of the connection of a material object with the EM vacuum, the concepts of the essence of gravity were developed, and the reason for the smallness of the gravitational interaction was found [3].

4.4. Gravitation as a key factor in the dynamics of Universe

The main objective of this Section is to show how at the phenomenological level, that is, with the introduction of new transcendental images, and this turns out to be necessary, an expression for Newton's law can be obtained and understood the physical content of this law. At first, for solving these questions it is necessary to use the Weinberg's modified phenomenological relation (28) and introduced the basic universal constants $\hbar$, $c$, $m_Q$ and $R_H$ for obtaining the Planck numbers in the needed and more convenient form:

$$a_{Pl} = 2\pi a_Q \left(\frac{a_Q}{R_H}\right)^{1/2}, \quad t_{Pl} = 2\pi \tau_Q \left(\frac{a_Q}{R_H}\right)^{1/2}, \quad m_{Pl} = \frac{1}{2\pi} m_Q \left(\frac{R_H}{a_Q}\right)^{1/2}, \quad w_{Pl} = \frac{m_Q c^2}{4\pi^2 \tau_Q} \cdot \frac{R_H}{a_Q}, \qquad (1a)$$

where $\tau_Q = a_Q/c \approx 0.44 \cdot 10^{-23}$ s is the time scale, corresponding to the space scale $a_Q$. The relations (1a) clarify the cosmological essence of the "smallness" of the Planck parameters $l_{Pl}$ and $t_{Pl}$, as well as the cosmological scale of the quantity– $w_{Pl}$, thus demonstrating the heuristic justification of the representation (1) for the Planck constant. As for the physical nature of the Planck parameters, it follows from section 4.2, that the Planck parameter $w_{Pl}$ is the energetic power that constantly feeds our Universe across the boundary "the Proto-vacuum – EM vacuum". The physical essence of the mass parameter $m_{Pl}$ will be established below, and the meaning of the length $l_{Pl}$ and time $t_{Pl}$ parameters will be clarified in Section 4.6.

In accordance with [3], first of all, point out that due to the infinite range of action of Casimir potential energy, the presence in the medium of various material objects with the masses $m_i$ inevitably leads to overlapping of the potential fields and, as a result, to forming the fields of the particle attraction – gravity fields. Consider for example the corresponding Casimir potential energy $U(\vec{\xi}; \vec{r}_1, \vec{r}_2)$ of two particles with the masses $m_1$ and $m_2$. As before, we associate the coordinate system with EM vacuum. Then we have:

$$U(\vec{\xi}; \vec{r}_1, \vec{r}_2) = -\frac{\sqrt{2}\hbar c}{|\vec{\xi} - \vec{r}_1|} - \frac{\sqrt{2}\hbar c}{|\vec{\xi} - \vec{r}_2|}. \qquad (32)$$

Introduce the radius-vector $\vec{R}$ of the center of mass of the particles $m_1$ and $m_2$, as well as the radius-vector $\vec{\rho}$ of the difference of the radius-vectors $\vec{r}_1$ and $\vec{r}_2$:

$$\vec{R} = m_{12}\left(\frac{\vec{r}_1}{m_2} + \frac{\vec{r}_2}{m_1}\right), \quad \vec{\rho} = \vec{r}_1 - \vec{r}_2, \quad m_{12} = \frac{m_1 m_2}{m_1 + m_2}.$$

Then

$$U(\vec{\xi}; \vec{r}_1, \vec{r}_2) = -\sqrt{2}\frac{\hbar c}{m_{12}}\left[m_1\left|\vec{\rho} + \frac{m_1}{m_{12}}(\vec{R} - \vec{\xi})\right|^{-1} + m_2\left|\vec{\rho} - \frac{m_2}{m_{12}}(\vec{R} - \vec{\xi})\right|^{-1}\right]. \qquad (32a)$$

If we are only interested in the potential energy of attractive interaction between two particles, then for the exclusion from consideration of the dynamics of the system as a whole, there must be chosen $\vec{\xi} = \vec{R}$. We obtain in this case:

$$U(\vec{r}_1, \vec{r}_2)\big|_{\vec{R}=0} = -\frac{\sqrt{2} m_1 m_2}{m_{12}^2} \cdot \frac{\hbar c}{\rho} . \qquad (32b)$$

In accordance with the abovementioned Mach's basic idea [1] (see Section I), each of the masses $m_1$ and $m_2$ is attracted by other masses of the Universe. In order to obtain Newton's expression for the potential energy of attraction of masses $m_1$ and $m_2$, it is necessary to keep the product of masses $m_1$ and $m_2$ in Eq. (32b) and take in account the effect of "pulling apart" $m_1$ and $m_2$ by the other masses of the Universe, which decreases the strength of attraction the masses $m_1$ and $m_2$ to their center of mass. We assume the last effect can be considered by carrying out the averaging procedure of the factor $1/m_{12}^2$ in Eq. (32b) with the distribution function $f(m_{12})$. Since this function is unknown, we believe that the mass distribution in the Universe is isotropic and introduce the definition $\langle 1/m_{12}^2 \rangle \equiv 1/m_M^2$, where $m_M$ is the "Mach mass" [3]. From the comparison in Newton's formula it follows that:

$$U(\vec{r}_1, \vec{r}_2)\big|_{\vec{R}=0} = -\frac{\sqrt{2}\hbar c}{m_M^2} \cdot \frac{m_1 m_2}{\rho} = -\frac{\hbar c}{\sqrt{2} m_{Pl}^2} \cdot \frac{m_1 m_2}{\rho} = -G \frac{m_1 m_2}{\rho}; \quad m_M = \sqrt{2} m_{Pl} . \qquad (32c)$$

The expression (32c) elucidates the physical essence of the $G$ constant and its relatively small value due to the small spatial extent of the Casimir polarization region of EM vacuum in the vicinity of gravitating masses as well as demonstrates the physical essence of the Planck parameter $m_{Pl}$.

Consequently, for the Generalized law of gravitation, taking into account the arbitrary position of the radius-vector $\vec{R}$ of the center of mass of the attracting masses $m_1$ and $m_2$, we obtain [3]:

$$U(\vec{\xi}; \vec{r}_1, \vec{r}_2) = -G m_{12} \left[ \frac{m_1}{\left| \vec{\rho} + \frac{m_1}{m_{12}}(\vec{R} - \vec{\xi}) \right|} + \frac{m_2}{\left| \vec{\rho} - \frac{m_2}{m_{12}}(\vec{R} - \vec{\xi}) \right|} \right]; \quad G = \frac{\hbar c}{\sqrt{2} m_{Pl}^2} = \frac{\sqrt{2}\hbar c}{m_M^2} . \qquad (32d)$$

The difference of this expression from the commonly used law of gravitation (case $\vec{\xi} = \vec{R}$) can be very significant in the problems of celestial mechanics. From this point of view, interest is the calculations of the perihelion precession of the orbit for each of the solar system planet with using of the most general expression (32d) for the potential energy of the paired interaction of all major masses of the solar system. This circumstance should manifest to the greatest extent for Mercury as the nearest planet to the Sun, because the deviation of the center of mass of the Sun from the barycenter, the center of mass of the solar system, can be up to 2% of the distance from the Sun to Mercury in the perihelion.

The final conclusion can be done after the appropriate calculations using the general expression (32d) for the Generalized gravitation law in the frame of reference associated with the center of mass of the solar system. In this case, $\vec{\xi} = \vec{R}_b$, where $\vec{R}_b$ is the radius vector of the barycenter. And we must bear in mind that due to the specificities of the Sun's center of mass motion relative to the barycenter [85], the value of the calculated precession of the perihelion of Mercury's orbit must change from century to century.

It can be assumed that ideas about the isotropic mass distribution in the Universe, used when mass $m_M$ is introduced and the gravitational constant $G$ is determined by this mass, should be violated for objects outside the orbits of the planets of the Solar System, in particular, in the Heliopause, where the solar wind is completely inhibited by galactic wind and other components of the interstellar medium. When considering the gravitational interaction of such objects with the Sun, an obvious asymmetry manifests itself: in addition to the Sun, the bodies of these "inner regions" of the Solar System exert the main gravitational effects on such objects. On the part of the remaining, "outer part" of the Solar system, including the Oort cloud - the habitat of long-period comets, and the subsequent stellar medium, the gravitational effects on the object in question are weaker. This means that the gravitational constant $G^*$, which determines the interaction of an object in remote regions of the solar system with the Sun, is slightly larger than the value $G$, and the corresponding Mach mass $m_M^*$ is less than mass $m_M$.

Apparently, in such representations one can understand the anomalies of the "Pioneers" - the observed deviation in the trajectory of the spacecraft "Pioneer-10" and "Pioneer-11" from the calculated one using the standard model for space bodies. These vehicles were launched in 1972 and 1973, respectively, and their trajectory was recorded until February 1998. For many years, scientists have known that the Pioneer spacecraft have not been exactly where they thought they should be. Each year the spacecraft falls behind where it should be by about 5000 km. An additional, linearly growing with time, violet shift of the received signal was found, which was interpreted as a manifestation of a very weak force, not taken into account in the calculation, which causes a constant acceleration of the apparatus toward the Sun, equal to $(8.74 \pm 1.33) \times 10^{-10}$ m / s² [86]. The most probable cause of such an anomaly is the technical version, according to which the fixed effect has a thermal nature and is explained by the anisotropy of the intensity of thermal radiation of the energy elements of the apparatus [87]. Despite the arguments presented in favor of such explanation for this anomaly, doubts remain, so the nature, direction, and temporal and spatial variation of the Pioneer anomaly remain an open arena of research [88].

The later launched, in 1977, Voyager-1 and Voyager-2, whose flight trajectory is similar to that of the Pioneers, and with which communication is still maintained, did not show a pronounced deviation effect. This is explained by the fact that the "Pioneers" are in free flight, and their orientation has stabilized due to the proper rotation of the vehicles. For Voyagers, the desired orientation is provided by small shunting of the shunting engines, which affects the trajectory.

If future experimental studies confirm the conclusion proposed in this paper, it would be reasonable to believe that the determining contribution to the magnitude of the gravitational constant in an arbitrary planetary star system is provided by the emerging mass distribution in the "inner region" of this system, including the star and planets, so that stellar-planetary systems may have their specific meanings $G$. It also follows from the above that the gravitational constant characterizing the gravitational interaction in the isotropic regions of the interstellar medium, in which there are no stellar-planetary objects, should be characterized by an even larger value $G$ and a smaller Mach mass $m_M$ than the corresponding values in the Solar System. The entire analysis, the derivation of the expression for the generalized Newton's law, the Einstein's commentary mentioned when introducing the gravitational constant $G$ into the energy-mass tensor and the conclusions above clearly indicate that gravity cannot be considered as a fundamental interaction, and therefore the corresponding gauge boson ("graviton") in Nature is not. This could be mean that there are no gravitational waves in nature. Therefore, we can assume that it is the wave propagation of the EM vacuum perturbation was recorded in the recent LIGO observation [89], and this disturbance could arise in the collision of two neutron stars or by some other large-scale events.

For Newton's law, an alternative formula can be proposed, if we consider $m_Q$ as the fundamental mass characterizing the strong nuclear interaction. In this case, the potential energy of interaction between the two masses $m_1$ and $m_2$ at the distance $\rho$ apart ("the Law of Universal Gravity") can be conveniently represented in the form:

$$U_g(\rho) = -G\frac{m_1 m_2}{\rho} = -\frac{q_g^2}{\rho}\mu_1\mu_2 = -\frac{\alpha_s \hbar c}{\eta_g \rho}\mu_1\mu_2, \quad \mu_i \equiv \frac{m_i}{m_Q}. \tag{33}$$

$$\eta_g = \frac{m_M^2}{m_Q^2} = \frac{2m_{Pl}^2}{m_Q^2} = \frac{\alpha_s}{\alpha_g} = \frac{1}{2\pi^2}\frac{R_H}{a_Q} \approx 0.46 \cdot 10^{40}.$$

The value $\eta_g$ can be defined as a "gravitational permeability" of the EM vacuum. The Universe is a Unified System! The anomalously large value of the introduced parameter $\eta_g$ may mean that, in accordance with the idea of Mach, the masses of the Universe that are at "cosmologically distances" make a contribution to the "gravitational interaction" of the two considered masses. Actually, gravitation can be considered as a manifestation of a strong nuclear interaction ($q_s^2 = \alpha_s \hbar c$) outside atomic nuclei and can be defined as an attractive interaction that arises as a result of overlapping the regions of Casimir polarization of every two atomic nuclei of every two material objects.

As follows from the obtained expression (32d), gravitation can be considered as a manifestation of a strong nuclear interaction ($\sim \alpha_s$) outside atomic nuclei and can be defined as an attractive interaction that arises as a result of overlapping the regions of Casimir polarization of every two atomic nuclei of every two material objects. It means that gravitation is not a fundamental interaction, and therefore there is no gauge boson for gravity.

4.5. EM vacuum in different periods of the evolution of the Universe

The discussed differences in the dynamics of the Universe at different stages of its evolution can be associated with the Dirac idea about the change in time the world constants - $\hbar$, $c$ и $G$. But it is necessary to understand the reasons for such changes! What causes them?

When discussing such possibilities, one should first of all direct one's attention toward the result on the practically negligible value ($\sim 0.6 \times 10^{-6}$) of the relative change $\Delta\alpha_e/\alpha_e$ of the fine structure constant $\alpha_e = e^2/\hbar c \approx 1/137$ for those regions of the Universe, which are characterized by red shifts of $z > 0.4$ [90], when the size of the Universe was $1/(1+z) \approx 0.7$ from the modern size. If we also neglect the time variations in power $w_{Pl} = c^5/2G$ and assume that the value of the elementary charge "$e$" remains constant during the evolutionary expansion of the Universe. We also introduce the dimensionless variable $h(t) = H/H_0 = t_0/t$, where the Hubble parameter $H = t^{-1}$, where $t$ is the age of the Universe, $H_0$ is the value of the Hubble parameter in our epoch ($t_0$ is the "our" time).

Dependencies $\hbar(t)$, $c(t)$ and $G(t)$ will be represented in the form:

$$\hbar(t) = \hbar_0 \cdot h^x(t), \quad c(t) = c_0 \cdot h^y(t), \quad G(t) = \sqrt{2}\frac{\hbar c}{m_M^2} = 4\pi^3\frac{\hbar^2(t)H(t)}{m_Q^3 c(t)} \equiv G_0 h^z(t), \tag{34}$$

where $\hbar_0(t)$, $c_0(t)$ и $G_0(t)$ - the values of the corresponding world constants in our era. From the independence of the quantities $\alpha_e$ and $w_{Pl}$ on the time t, as well as from the indicated representation of the constant $G(t)$, it follows: $x + y = 0$, $5y - z = 0$, $z = 2x - y + 1$.
Solving this system, we obtain the required dependences:
$$\hbar(t) = \hbar_0 h^{-1/8}(t), \quad c(t) = c_0 h^{1/8}(t), \quad G(t) = G_0 h^{5/8}(t). \tag{35}$$
Note, the last conclusion corresponds qualitatively with the results of J. Mould et.al [91].
Note also, the red shift is given by
$$z + 1 = \frac{R(t_0)}{R(t)} = \frac{\lambda_0}{\lambda(t)} = \frac{c_0 \omega(t)}{c(t)\omega_0}, \tag{36}$$

where $\lambda(t)$, $\lambda_0$ and $\omega(t)$, $\omega_0$ are the length wave and cyclic frequency of light emitted by the stellar source at the instant $t$ and received by the observer, respectively; $R(t)$ and $R(t_0)$ are the "sizes" of the Universe at the instants $t$ и $t_0$. By taking into account Eq. (35), it is easy to obtain in this case: $\omega(t) = \omega_0 h^{1/8}(t)$.

According to the developed phenomenology, for $R_H(t)$, we get

$$R_H(t) = c(t)/H(t) = c_0 t_0^{1/8} t^{7/8}; \quad \dot{R}_H(t) \sim t^{-1/8} \qquad (37)$$

(we draw attention to the decrease in the rate of increase in $R_H$ as it evolves, in contrast to the conclusion of the standard model). This relation means that the volume $V_U(t)$ of the Universe increases according to: $V_U(t) \sim R_H^3 \sim t^{21/8}$, and since $\varepsilon_V^e \sim t^{-13/8}$, we have $\varepsilon_V^e(t) \cdot V_U(t) \sim t$, which is a consequence of the linear increase in the total EM vacuum energy content of the Universe (Eq. 23)) due to the Planck source: $E_{tot} = w_{Pl} t$.

It remains to consider the reason for the detected variations with time of the universal constants. From the presented relations, it follows that the main reason lies in decreasing with time, $t$, of the EM vacuum energy density, $\varepsilon_V^e \sim t^{-13/8}$, which leads to a decrease in the intensity of all interactions realized within the Universe, the genesis of which pursuant to the developed concepts is determined by the EM vacuum energy density. For this reason, the $c$ value is decreased, while the $\hbar$ value is increased, which makes it difficult to realize quantum transitions and reduces the probabilities of the corresponding processes. A decrease in the gravitational interactions intensity, $G \sim t^{-5/8}$, with time is connected with an increase in the Mach mass, $m_M^2 \sim t^{5/8}$, with time, resulting in the effective reduction in the mutual attraction of two masses caused by a relative increase in the connection of each of these masses with another masses within the Universe.

In conclusion, it ought to be noted that in the conditions of the considered Universe dynamics, which is determined by the continuous energy input from the border regions of two vacuums – the EM vacuum and the proto-vacuum – the severe character of the existing problems is reduced with general understanding of the Universe baryon asymmetry, that is, the preponderance of matter over antimatter. Really, the consequences of mini Big Bangs at the vacuum boundaries should cause the appearance of not only flows of EM emissions in the direction of the expanding Universe but also the resulting flows of the generated particles with an insignificant flow of antiparticles. It is quite natural to believe that such situation began to develop at the earliest stages of the Universe formation, when just by fluctuation the number of particles in the appearing Universe substantially exceeded the number of antiparticles, which were shifted to the other side of the boundary between two vacuums. If in the result of subsequent Bangs, fluctuations appeared to be not so noticeable as the mentioned one, then the process of the particles accumulation in the Universe continued, and beginning from some moment, even large (as the initial one) fluctuations could not change the situation of formation and growth of baryon asymmetry of the Universe.

4.6. Estimate of the energy density of EM vacuum [92]

The principal difference between the model of the Universe dynamics under consideration and the standard model consists in localization of the energy source, which determines the total Universe energy and its expansion by maintaining the pressure of the EM vacuum, at the Proto-vacuum – EM vacuum boundary, but not in the volume of the Universe. In our model, all energy has an initially electromagnetic nature. Due to linearity of the equations of electromagnetic field, the total energy of any electromagnetic field can be represented as the sum of energies of the field oscillators, and for the average energy per the angular frequency range from $\omega$ to $\omega + d\omega$, there holds [93]:

$$du(\omega,\Theta) = \left(\frac{\hbar\omega}{2} + \frac{\hbar\omega}{\exp\Theta - 1}\right)dN, \quad \Theta \equiv \frac{\hbar\omega}{k_B T}. \tag{38}$$

Here $dN = \dfrac{\omega^2 V_\omega}{2\pi^2 c^3} d\omega$ is the number of field oscillators having frequencies within the indicated range; $V_\omega$ is the volume of the configuration in which the field corresponding to the oscillator with frequency $\omega$ is enclosed; $k_B$ is Boltzmann's constant, $T$ is the ambient temperature. The right-hand side of expression (38) includes two qualitatively different terms. Let us first consider the second term, which is associated with Planck's formula for the spectral distribution of the equilibrium blackbody radiation, and which has a specific binding to the local environmental conditions, to the particular temperature of various extended regions of the Universe.

We can assume that such regions are randomly scattered over the Universe and have different spatial extent. Therefore, when calculating the energy density of radiation associated with the second term in equation (38), the result obtained after integration over the entire possible frequency range will be assigned to one of these regions. The configurational volume of this region is represented as a constant value, $V_\omega = V = \text{const}$, assuming that in this volume there is localized the field produced by the oscillators with all possible frequencies. Obviously, in this case, the integration can be formally carried out over the infinite frequency interval. As a result, we obtain the Stefan-Boltzmann law for the corresponding Planck density $\varepsilon_P$ of the radiation energy, given the fact that to each wave vector there correspond two states of polarization [93]:

$$\varepsilon_P = \frac{\pi^2}{15} \cdot \frac{(k_B T)^4}{\hbar^3 c^3} \equiv \sigma T^4, \qquad \sigma = \frac{\pi^2 k_B^4}{15\hbar^3 c^2} \tag{39}$$

It is believed that the process of this assimilation of the energy of Proto-vacuum with its transformation into the energy of the expanding Universe occurs permanently during the appearance of the sources of Planck power within the border regions of two vacuums: the EM vacuum and the Proto-vacuum. Furthermore, all the freed mass-energy of the Planck source must be emitted into the existing Universe, which is helped by high affinity ("freezing-in") of the formed material particles to the EM vacuum, characterized by the energy of their connection with vacuum.

At the same time, starting from the uniform distribution of the incoming energy across the entire space of the Universe, the length of which is $R_H \sim 1.27 \cdot 10^{28}$ см (Section 4.2), the averaging of the total zero-point energy must be carried out over the volume $V_H = \frac{4}{3}\pi R_H^3$. The main contribution into the total energy of zero-point fluctuations of EM field is formed by the highest frequencies – from $\omega_Q = 2\pi c/a_Q \approx 1.44 \cdot 10^{24} s^{-1}$ to the Planck value $\omega_{Pl} = 2\pi c/a_{Pl} = 2\pi/t_{Pl} = cR_H^{1/2}/a_Q^{3/2} \approx 0.71 \cdot 10^{44} s^{-1}$, to which there corresponds the spatial scale $a_{Pl} = 2\pi c/\omega_{Pl} \approx 2.64 \cdot 10^{-33} cm$.

Since $a_Q \gg a_{Pl}$, as the configuration volume in the calculation of the average density of zero-point energy of EM field we choose $V_{\omega Q} = \dfrac{4}{3}\pi a_Q^3$. Then, in the integration of the first term in the right-hand side of expression (38) over the frequency interval $[\omega_Q, \omega_{Pl}]$ with referring of the obtained energy to the volume $V_H$ and taking into account the fact that to each wave vector there correspond two polarization states, we obtain [92]:

$$\varepsilon_V^e = \frac{2^{1/2}}{16\pi^2} \cdot \frac{m_Q c^2}{a_Q^2 R_H} \approx 1.33 \cdot 10^{-8} \, erg/cm^3, \tag{40}$$

so that
$$\Lambda = 8\pi G \varepsilon_V^e / c^4 = 2^{1/2}\pi / R_H^2 \tag{41}$$

The obtained value $\varepsilon_V^e$ (taking into account certain conditionality of the selection of numerical coefficients in the introduction of the parameters used) is close the above value, $\varepsilon_V^e$ = 1.14×10$^{-8}$ erg/cm$^3$, calculated by recalculating the value of $\varepsilon_V$, obtained on the basis of observational data (see Section 4.2). Such a result can be considered as an argument in favor of the scenario of the Big Bang proposed in this article (see also [82]) and solving the "120 orders" problem.

Previously, for the value of the cosmological $\Lambda$ constant on the basis of its general definition of Eq. (2) and using the value $\varepsilon_V^e$ obtained by recalculating the value $\varepsilon_V$, found from the experiment, $\Lambda \approx 2.36 \times 10^{-56}$ cm$^{-2}$ was obtained. If, when calculating the cosmological constant, use Eq. (41), obtained using the model expression (40) for $\varepsilon_V^e$ calculated for the open Universe, then for such a model expression, which we denote by $\Lambda_{\varepsilon_V^e}$, we find $\Lambda_{\varepsilon_V^e} \approx 2.75 \times 10^{-56}$ cm$^{-2}$. The difference in the calculated values of $\Lambda$ and $\Lambda_{\varepsilon_V^e}$ is about 15%.

It is possible to make and alternative recalculation. If we fix the indicated value $\Lambda \approx 2.36 \times 10^{-56}$ cm$^{-2}$, obtained on the basis of the value $\varepsilon_V$ determined from the corresponding experiments with the corresponding recalculation, and having in mind the value $R_H \sim 1.27 \cdot 10^{28}$ см calculated on the basis of experimental data on the Hubble constant, then the recalculation $R_H$ on the basis of (41) gives magnitude $R_H \sim 1.37 \cdot 10^{28}$ см. A somewhat smaller difference of the values $R_H$ (~ 7.5%) in this case is associated with the quadratic dependence $\Lambda_{\varepsilon_V^e} = \Lambda_{\varepsilon_V^e}(R_H)$. Relatively small differences in the calculated values when using the model relation (40) actually indicate the adequacy of the developed ideas about the dynamics of the open Universe. This gives additional confidence that the problem of the "120 orders" is solved.

It should be emphasized here that when obtaining the estimates (40) and (41), the values of the Planck parameters of length $l_{Pl}$ and time $t_{Pl}$ were used. The physical nature of the mass parameter $m_{Pl}$ and the energetic power $w_{Pl}$ was discussed in Sections 4.2 and 4.4, respectively. This gives grounds to consider the combination of the Planck numbers, as well as the modified Weinberg relation (28), not as numerological relations, but as phenomenological representations that can become a guideline in the construction of theoretical models of the Universe dynamic.

Note, when choosing the standard scenario of the dynamics of the Universe [7-10, 80] and calculating the EM vacuum energy density $\varepsilon_V$ on the basis of (38) with the choice of $V_\omega = V_H$, integrating over frequencies from zero to $\omega_{Pl} = cR_H^{1/2}/a_Q^{3/2}$, we obtain:

$$\varepsilon_V = \frac{1}{(2\pi)^6 (\hbar c)^6} \cdot R_H^2 (m_Q c^2)^6, \qquad (40a)$$

so that

$$\frac{\varepsilon_V}{\varepsilon_V^e} = \frac{1}{2^{3/2} (\hbar c)^3} \cdot R_H^3 (m_Q c^2)^3 \approx 1.1 \cdot 10^{123}. \qquad (40b)$$

It becomes clear that the reason for the existence of the problem "120 orders" is the inadequate choice of the model of the dynamics of the Universe, which should be considered as an open system.

For the subsequent estimation it is convenient to introduce the dimensionless ratio $\varphi$ of the energy density $\varepsilon_V^e$ to the density $\varepsilon_Q$ of intra-nuclear excitation $E_Q = m_Q c^2$, localized in the volume $V_{\omega Q} = 4\pi a_Q^3/3$:

$$\varphi \equiv \frac{4\pi a_Q^3 \varepsilon_V^e}{3 m_Q c^2} = \frac{\varepsilon_V^e}{\varepsilon_Q}. \qquad (42)$$

Then it follows from (29), (32d), (40) and (42) that:

$$G = \frac{\hbar c}{\sqrt{2} m_{Pl}^2} = \frac{\hbar c}{m_Q^2} \alpha_g = 2^{9/2} \pi^4 \frac{a_Q^4}{m_Q^2} \varepsilon_V^e = 24\pi^3 \frac{\hbar c}{m_Q^2} \varphi. \qquad (30a)$$

For the dimensionless constant $\alpha_g$ of gravitational interaction, taking into account the numerical values $\varepsilon_Q \approx 3.64 \cdot 10^{34}$ erg/cm$^3$ and $\varphi \approx 4.1 \cdot 10^{-43}$, we obtain:

$$\alpha_g = \frac{m_Q^2}{\sqrt{2} m_{Pl}^2} = 2^{3/2} \pi^2 \frac{a_Q}{R_H} = 24\pi^3 \varphi \approx 2.88 \cdot 10^{-40}. \qquad (31a)$$

We note also that, with the introduction of the parameter $\varphi$, the representation (1$a$) of the Planck numbers is also simplified:

$$a_{Pl} = 2^{7/4} 3^{1/2} \pi^{3/2} a_Q \varphi^{1/2}, \; t_{Pl} = 2^{7/4} 3^{1/2} \pi^{3/2} \tau_Q \varphi^{1/2}, \; m_{Pl} = \frac{1}{2^{7/4} 3^{1/2} \pi^{3/2}} \frac{m_Q}{\varphi^{1/2}}, \; w_{Pl} = \frac{1}{2^{7/2} 3\pi^3} \frac{m_Q c^2}{\tau_Q \varphi}, \quad (1b)$$

pointing, together with (31$a$), to the possible dependence of all the dimensional parameters, introduced in order to build the models for the structure and dynamics of the Universe, on the dimensionless parameter of the energy density of the EM vacuum.

Phenomenological analysis has shown that the introduction of the EM vacuum as the base medium of the expanding Universe and the absolute reference system allows to approach the understanding of the key problems of contemporary physics - the physical essence of quantum mechanics, the nature of gravity, the extremely controversial problems of the dynamics of the Universe and the seemingly insoluble problem of "120 orders". Moreover, a number of the presented results, including the stability of atomic nuclei and electronic subsystems, caused by the Casimir pressure of the EM vacuum, is a necessary basis for understanding new phenomena for theoretical physics - Nuclear-Chemical processes. Consideration of these processes and an analysis of their possible mechanisms are dealt with in the Sections V-VI of the article.

V. NUCLEAR-CHEMICAL PROCESSES AS A NEW "GAP IN KNOWLEDGE" IN CONTEMPORARY PHYSICS

Two key factors are rather surprising and shocking for the perception of low-energy nuclear reactions (LENR) as the reality, namely: the chemical scale energies (up to 3-5 eV of kinetic energy, $E_e$, for the electrons), which are necessary for initiating such processes, and almost complete absence of the dangerous ionizing EM radiations or neutron fluxes, which usually accompany nuclear transformations. Introducing the EM vacuum concept as the base Universe medium and the concept of the Casimir polarization of the EM vacuum in the vicinity of the atomic nuclei and electrons made it possible to understand more fundamentally the genesis of interrelation between the nuclear and electronic subsystems of the atom, and based on different amazing characteristics of LENR to study the new group of nuclear-chemical phenomena [36, 94-101], which previously escaped notice of theoretical physicists. Before considering the specific feature of the nuclear-chemical processes in principle, let us discuss one historical aspect of the LENR problem.

5.1. Is there a dineutron?

Beginning with the work of Fleischmann, Pons and Hawkins [28] on observation of excessive heat generation during the electrolysis of D$_2$O heavy water with a Pd cathode, when the generation of neutrons and tritium was recorded, it became clear that the problem of establishing the mechanism of the occurring nuclear transformations could become a key to solving the complex of problems that arose in connection with the phenomenon of low-energy nuclear reactions. One of the first attempts to solve this problem was associated with the possibility of the existence of a stable dineutron $^2n$ with a binding energy $\varepsilon_{dn}$ less than 3.01 eV (for the deuteron not to be radioactive) [102]. In the experiment [28], the arising a dineutron as a result of the interaction of an activated electron with a deuteron under conditions of electrolysis of heavy water could be considered as the first stage preceding the reaction of formation of tritium during the interaction of the dineutron with the deuteron. However, there remained questions. It was unclear whether it is possible to actually activate the electrons in the conditions of electrolysis of heavy water on a palladium cathode to the necessary (what exactly?) energies, and whether there really is a dineutron. The possibility of the $^2n$ nucleus existence was discussed back in the early 1960s when studying the $T(d, {}^2n)^3He$ process using the $^{27}Al(^2n, \gamma)^{29}Al$ reaction for the detection of $^2n$ particles [103]. However, in [104], which appeared shortly after the publication of [103], the results of a similar experiment were repeated and it was reported that the yield of $^{29}Al$, confirming the reality of the existence of $^2n$ nucleus, was observed only at the background level.

Moreover, it was known, based on general considerations, that the $^2n$ nucleus can not exist in principle. Since there is a bound state in the neutron-proton system with a binding energy $\varepsilon_D = 2.22$ MeV and spin $S = 1$, and also there is a virtual level with energy 70 keV and spin $S = 0$, only the existence of a virtual state with spin $S = 0$, due to the charge independence of the nuclear forces (isotopic invariance) for the system of two neutrons, is possible [105]. But in this state there should be a weak repulsion, and the state with $S = 1$ can not exist due to the Pauli principle. Nevertheless, in this review [105], based on an analysis of the experimental data available at that time (1965), an upper estimate of the cross section for the formation of a nuclear stable dineutron, s < 0.001-0.01 mb, was still given, although this estimate practically excluded the possibility of manifestation of $^2n$ in any experiments. This estimate was confirmed by the result of a much later experiment [106], in which for the cross section of occurring a stable dineutron in the interaction of cold neutrons with deuterons in the reaction $n + d \rightarrow {}^2n + p$, was obtained: $\sigma \leq 1$ mcb.

With time, it became clear that in order to understand the results of [28], as well as many later works on initiating low-energy nuclear reactions and accelerating radioactive $\alpha$ and $\beta$ decays, including under low-temperature plasma conditions [29-32, and when laser ablation of metals in aqueous media [33-37], the idea of a stable dineutron should be abandoned, and it is necessary to involve other hypotheses. At the same time, the results of [107] in which the spectrum of the "lost mass" in the process $^6Li(\pi^-, p)^5H$ (see Fig. 1) with the kinetic energy of $\pi^-$-mesons equal to 125 MeV was investigated, continue to cause surprise. It follows from Fig. 1, a fixed maximum in the miss mass (MM) region from zero to MM = –3 MeV/c$^2$ could well correspond to a hypothetical dineutron with a binding energy of $\varepsilon_{dn} \approx 3$ MeV. However, in this case, the probability of the occurring a dineutron in the reaction involving a $\pi^-$-meson would have an order of magnitude higher than the values indicated above, for unknown reasons.

In subsequent years, in order to understand the results of many papers in which low-energy nuclear transformations and low-temperature plasma accelerated decays of radioactive nuclei were investigated, attempts were repeatedly made to introduce neutral particles with a baryon number of two (or one) that were the weakly coupled or resonant state of the deuteron (or proton) with an electron or a neutron with a neutrino. Such particles could participate in low-energy nuclear reactions, since for such a type of particles there would be no problem of overcoming the "Coulomb barrier" in nuclear interactions. We do not consider such possibilities here, since in accordance with the concepts of nuclear physics, electron and neutrino localization on nuclear scales $\sim 10^{-13}$ cm would have an abnormally high, physically incredible uncertainty in the momentum [108].

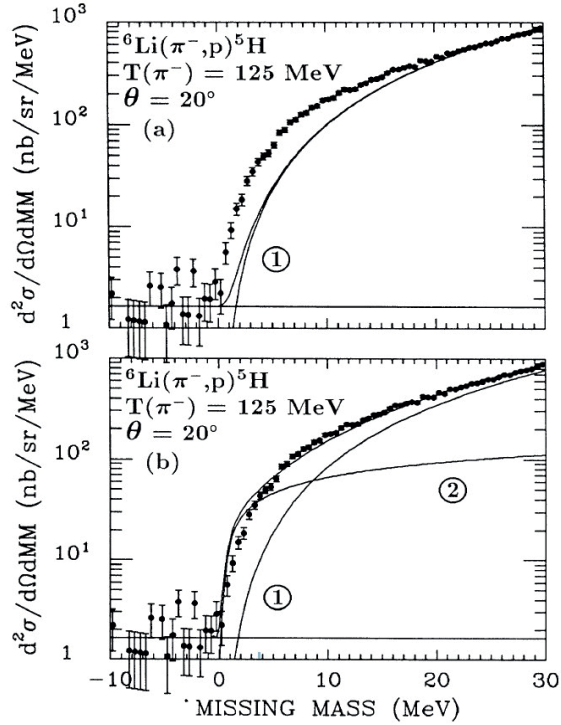

Fig. 1. *The missing-mass spectra for the reaction $^6Li(\pi-, p)^5H$ [107]. The constant line at the bottom corresponds to the background. The phase-space distributions shown are as follows: (a) curve 1, fit for $^5H \to {}^3H + n + n$; (b) best-fit curve as sum of curve 1 for $^5H \to {}^3H + n + n$ and curve 2 for $^5H \to {}^3H + {}^2n$.*

Below we will return (see section 5.3.1) to the problem with the fixed maximum in Fig. 1 in the lost mass in the MM ≈ - 3 MeV/c² region. In the meantime, we discuss the above two factors, which became the basis for understanding the phenomenon of LERN.

5.2. Electronic factor in initiating nuclear processes

According to the phenomenological understanding [95-101], the dynamic interrelation between the electron and nuclear subsystems of atom, realized through the electromagnetic component of the physical vacuum, the EM vacuum, is the key factor in initiating low-energy nuclear reactions [29-34] and the processes of radioactive decay of nuclei [36, 99, 109, 110]. A reflection of this interrelation is, in particular, the experimentally established facts showing that the possibility of radioactive decay of nuclei is determined by the positive difference between the mass of the initial nucleus together with the mass of the electron subsystem of the atom (i.e., the mass of the whole atom, not the nucleus) and the total mass of the decay products [111, 112]. Usually, when considering the mechanisms of these low-energy nuclear processes and the decay of atomic nucleus $^A_Z N$ (Z and A are the atomic number and mass number of the nucleus $N$, respectively), nuclear matter is represented in the form of interacting nucleons. For example, in the *K*-capture, when the electron of the inner shells of the atom interacts with the surface of the nucleus and a new daughter nucleus is formed, the nucleon structure of nuclear matter does not change. At the initial, irreversible stage of this process, the electron emits a neutrino $v$ when interacting with the nucleus surface. The formed virtual vector $W^-$-boson, which is introduced into the nuclear matter, produce a *d*-quark when interacting with the *u*-quark of one of the protons, as a result of which this proton turns into a neutron, and the nucleus $^A_{Z-1}M$ is formed. However, the situation can drastically change when the *K*-capture is energetically forbidden (it is such cases that are considered below), but the electron has a sufficiently large (by chemical standards) kinetic energy $E_e \sim 3-5$ eV, as it can be realized in low-temperature plasma. Under these conditions, when the process of ionization of electron shells by such electrons is not yet realized, then, during scattering of electrons with the indicated kinetic energy and the corresponding de Broglie wavelength $\lambda \approx 0.5$ nm, the oscillation dynamics of the electron subsystems of atoms and ions is initiated on atoms and ions and, thus, the probability of interaction between the electrons of the inner subshells of atoms and ions and the corresponding nuclei increases.

At the first, irreversible stage of such interaction, a neutrino $v$ is emitted and a vector $W^-$-boson is introduced into the nuclear matter of the original nucleus $^A_Z N$ according to the relation:

$$^A_Z N + e^-_{he} \rightarrow ^A_{Z-1} M_{isu} + v. \qquad (43)$$

As a consequence, the nucleon structure of the formed nuclei $^A_{Z-1} M_{isu}$, whose charge is less than the charge of the initial nucleus by one elementary charge, turns out to be locally disrupted. Indeed, the vector $W^-$-boson, when interacting with the $u$-quark of one of the protons of the $^A_Z N$ nucleus, can only yield the formation of a virtual $d$-quark with a subsequent chain of virtual transformations of quarks involving vector $W$-bosons, but a neutron cannot be formed due to the deficit of the total mass of such a nucleus. The emerging state of local anomaly of nuclear matter with a broken nucleon structure is characterized as a metastable state of "inner shake-up" or *isu*-state. In accordance with section 2.1 (see also [94, 99]), the violation of the general stability of nuclear matter in the metastable *isu*-state occurs as a result of a change in the boundary conditions for the components of the electric field intensity vector of the EM vacuum on the surface of the nucleus in whose volume such a violation of the nucleon structure has occurred. The subscript in the notation of the nucleus in the right-hand side of the relation (43) indicates on the *isu*-state of the nucleus being formed. The lower index "$h$" in the notation for the electron in the left-hand side of (43) indicates to the activated nature of this stage of the process. The initiated chain of virtual transformations of quarks with the participation of the vector $W$-bosons should be interrupted in the irreversible decay of the virtual $W^-$-boson with the formation of the initial nucleus, an electron, and an antineutrino $\tilde{v}$:

$$^A_{Z-1} M_{isu} \rightarrow ^A_Z N + e^- + \tilde{v}, \qquad (44)$$

so that the gross process can be represented in the form of inelastic scattering of an electron on the initial nucleus:

$$^A_Z N + e^-_{he} \rightarrow ^A_Z N + e^- + v + \tilde{v}. \qquad (45)$$

The nuclei with the state of nuclear matter in the metastable *isu*-state of the "internal shake" will be called by the "$\beta$-nuclei". The threshold energy of such a process with the production of a $v\tilde{v}$ pair, determined by the rest masses of neutrino-antineutrinos, is about 0.3 eV [113].

As is known, the nucleus is a system of nucleons connected in a single whole by means of exchange interactions by exchanging quarks via pions. Therefore, the formation in the nucleus of three quarks unconnected into a nucleon, which can then be regarded as "markers" of new degrees of freedom, in fact, means that the intensity of nuclear forces is insufficient to provide the traditional, proton-neutron organization of nuclear matter in the system under consideration. The subsequent relaxation dynamics of the locally appeared *isu*-state, which can be transmitted to other nucleons of the nucleus by means of pions, is initiated only by weak nuclear interactions, which are realized through quarks during creation and absorption of the gauge vector neutral $Z^0$- and the charged $W^\pm$-bosons. In the case under consideration, such a relaxation terminates with the decay of the virtual vector $W^-$-boson with the formation of the initial nucleus during the emission of an electron and antineutrino. The lifetime of the formed $\beta$-nuclei in the metastable *isu*-state can be rather considerable (see section 5.3 below), and the nuclei in this state can directly participate in a variety of nuclear processes [97, 99].

It is necessary to take into account here that the relaxation reorganization of nuclear matter with a locally broken nucleon structure in the process of formation of products of such nuclear transformations is carried out, first of all, by the formation of a purely nucleon structure of the nuclei, in accordance with the principle of least action. While in the nuclei with the proton-neutron structure, the relaxation of excited states of the nucleus include the stages of emission of γ-quanta, in the β-nuclei such relaxation is impossible. In this case, the nucleon structure of reaction products is inevitably formed upon relaxation of energy due to the emission of neutrino-antineutrino pairs. Such nuclear processes, in which neutrino is emitted in the process of energy relaxation, were considered by Gamow and Schönberg and were called URCA processes [114].

5.3. Possible mechanisms of Nuclear-Chemical processes

5.3.1. *Mechanism of nuclear fusion*
The simplest $β$-nuclei are the $β$-neutron $^1n_{isu}$ and $β$-dineutron $^2n_{isu}$, which can be formed by the interaction of high-energy (in chemical sense) electrons with protons $p^+$ or deuterons $d^+$, for example, by laser ablation of metals in ordinary or heavy water, and also under conditions of protius-containing or deuteron-containing glow discharge plasma, respectively, according to

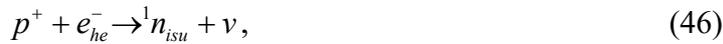
$$p^+ + e^-_{he} \rightarrow {}^1n_{isu} + v, \qquad (46)$$

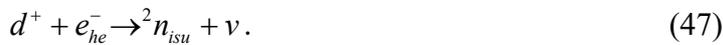
$$d^+ + e^-_{he} \rightarrow {}^2n_{isu} + v. \qquad (47)$$

If the half-lives $T_{1/2}$ of such $β$-nuclei are sufficiently long, the neutral nuclei $^1n_{isu}$ and $^2n_{isu}$, respectively, with baryon numbers equal to one and two, zero lepton charges and rest masses equal to the masses of the hydrogen atom and deuterium atom, can effectively participate in *nuclear fusion processes*, in particular, in the production of tritium. Such possibility was realized in the work [36] devoted to tritium synthesis during the metal laser ablation in heavy water.

In the previous studies of Shafeev *et al*. [33-35], it was shown that in order to initiate the nuclear processes at laser ablation in water solutions, it is sufficient to apply laser pulses of nano- or picosecond duration with a peak intensity of $J_E$ ~ $10^{10}$-$10^{13}$ W/cm$^2$, which is orders of magnitude less than that required for the direct initiation of nuclear processes, $J_E$ ~ $10^{18}$–$10^{19}$ W/cm$^2$ [115, 116]. In the conditions of laser ablation in the metal subsurface region, the liquid is transformed in vapor and frequently is ionized, while in vapor fields non-equilibrium "low-temperature plasma" is generated, electrons in which are characterized by high (on a chemical scale) kinetic energy $E_e$ ~ 3-5 eV [117].

Different targets were used for exposure of bulk metal targets in a heavy water with some tritium content (D$_2$O + ξDOT), where ξ stands for molar fraction of D$_2$O with T. The content of T was measured according to its beta-activity with the use of a Perkin-Elmer beta-spectrometer Tri-Carb 3110TR Liquid Scintillation Analyzer with fluorescent dye ULTIMA GOLDTM cocktail. The accuracy of measurements was 1%. Deuterium content was 98.8 %. The laser source was a Nd:YAG laser DuettoTM with pulse duration of 10 ps and repetition rate of 50 kHz. Three harmonics of this laser, first at 1064 nm, second at 532 nm, and third at 355 nm were used for target exposure. The estimated peak power on the target was $10^{13}$ W/cm$^2$, 7 $10^{11}$ W/cm$^2$, and 2 $10^{11}$ W/cm$^2$ at 1064, 532, and 355 nm, respectively.

Laser beam was focused on the target from bottom through a window made of fused silica to a spot on the target of 50 mcm in diameter. The targets had the shape of rods with diameter about 1 mm. The sides of the targets were electrically isolated from the solution. Typically, 2 ml of heavy water was placed inside a glass cell cooled by flowing water. The inner diameter of the glass cell was 1 cm. In another set of experiments the metallic target was cathodically biased with respect to another electrode made of Platinum wire 0.5 mm in diameter. The applied voltage was 25 V DC. To ensure the conductivity of the solution, 15 mg of metallic Na were added to 5 ml of D$_2$O thus producing conducting solution of NaOD.

The ablation of targets was carried out in 99.9% pure D$_2$O (supplied by Euriso-top) without Tritium content. In this case the content of DOT molecules in D$_2$O was at the background level of the beta-spectrometer, which corresponds to $\xi = 3.54\cdot 10^{-14}$. In one set of experiments the voltage was not applied to the target, in the second set of experiments the ablation was carried out simultaneously with electrolysis. Application of voltage to electrodes is accompanied by reduction of Deuterium (and Tritium) on the target as bubbles. Isolation of the sides of target from the solution confined the emission of reduced gas to ablated area. Typical exposure time of the targets was 1 hour.

It was found that laser exposure of various targets in $D_2O$ leads to formation of Tritium, and its content depends on several experimental parameters. The results are presented in Table 1. Laser ablation of targets without electrolysis leads to the increase of Tritium activity to the level which is about 20 times above the background ($10^{-7}$ Curie/l). The synthesis of Tritium is even more pronounced if the ablated target is cathodically shifted. Ablation of either Ti or Au targets with electrolysis results in synthesis of Tritium to the level, which is $10^3$ times above the background. The largest change of activity is observed at laser wavelength of 1064 nm. The optical density of the solution at the wavelength of the third harmonics of a Nd:YAG laser (355 nm) is high, especially after generation of some amount of nanoparticles, so the intensity of the laser beam on the target is much lower than at 1064 nm. Some amount of gas (both $D_2$ and $T_2$) could be lost from the cell during electrolysis. Estimations show that the mass of gas reduced on the ablated cathode at average current of 25 mA during 1 hour exposure is $2.2\cdot 10^{-3}$ g, which is negligible compared to the total mass of the sample.

Table 1. Laser exposure of targets in pure $D_2O$ with addition of NaOD ($\xi = 3.54\cdot 10^{-14}$).

| № | Material | Wavelength of laser beam, nm | Intensity of laser radiation, $W/cm^2$ and pulse width | Time of irradiation, min | Electrolysis | Activity, Ci/l | Change with respect to initial level |
|---|---|---|---|---|---|---|---|
| 1 | Initial $D_2O$ | - | - | - | - | $8.64\cdot 10^{-8}$ $\xi=3.54\cdot 10^{-14}$ | - |
| 2 | Au | 532 | $7\cdot 10^{11}$ | 60 | + | $4.06\cdot 10^{-4}$ $\xi=1.66\cdot 10^{-10}$ | Increase $4.7\times 10^3$ |
| 3 | Au | 1064 | $10^{13}$ | 32 | - | $1.62\cdot 10^{-6}$ $\xi=6.63\cdot 10^{-13}$ | Increase by factor 18.8 |
| 4 | Ti | 532 | $7\cdot 10^{11}$ | 62 | + | $4.74\cdot 10^{-4}$ $\xi=1.94\cdot 10^{-10}$ | Increase $5.5\times 10^3$ |
| 5 | Ti | 1064 | $10^{13}$ | 90 | - | $1.57\cdot 10^{-6}$ $\xi=6.42\cdot 10^{-13}$ | Increase by factor 18.2 |
| 6 | Ti | 355 | $2\cdot 10^{11}$ | 61 | + | $9.62\cdot 10^{-8}$ $\xi=3.94\cdot 10^{-14}$ | Increase by factor 1.11 |
| 7 | Pd | 355 | $2\cdot 10^{11}$ | 75 | + | $2.46\cdot 10^{-6}$ $\xi=10^{-12}$ | Increase by factor 28.2 |
| 8 | Pd | 1064 | $10^{11}$ | 62 | + | $2.31\cdot 10^{-6}$ $\xi=9.47\cdot 10^{-13}$ | Increase by factor 26.7 |

Presented results on the nuclear fusion of tritium upon the laser excitation of metals in heavy water, which is naturally associated with reaction (47) of formation of β-dineutrons and the subsequent process

$$d^+ + {}^2n_{isu} \rightarrow t^+ + n + Q(3.25 MeV) \qquad (48)$$

can be considered as experimental confirmation of the hypothesized existence of β-dineutrons as long-lived nuclei. Its half-life should be at least comparable to the time of the associated experiment on tritium fusion, and be no less than tens of minutes.
not less than tens of minutes

In connection with the proposed mechanism for the synthesis of tritium, one should also mention data on the features of the nanostructure of the metal surface formed in laser ablation conditions, presented in Ref. [118], when the energy density of the laser beam on the target is of order of several J/cm². The corresponding energy is more than sufficient to melt targets of any material listed in Table 1. The nanostructures are molten at each laser pulse that arrive to the target surface and re-appear after melt solidification. The analysis performed in [118] for the irregularities that emerged in the structure of metal surfaces when using different modes of laser ablation in aqueous media on the basis of atomic force microscopy data demonstrated the interrelation between the efficiency of the nuclear transformations observed in [36] and the structural parameters of the resulting surfaces, especially the "spikiness" parameters as a measure of the most drastic changes in surface profile on the nanometer scale. Indeed, it is precisely in the regions of the sharpest surface relief alterations, in which high electric field strengths (can reach $10^9$ V/m at cathodic displacements) are realized, and mechanical tensile stresses arise that lower the work function of electrons from the metal, high-energy electrons are produced and conditions are created for the initiation of nuclear-chemical processes [33-35]. Perhaps, the well-known fact of nonreproducibility in a number of laboratories [119] of the Fleischmann and Pons experiment results [28] on the electrolysis of heavy water with the synthesis of tritium and the resulting neutrons, is associated with the nature of the roughness of the cathodes used in these experiments.

In the conditions of laser ablation, a number of other *nuclear-chemical fusion processes* in heavy water were also investigated. In particular, the laser action on a suspension of mercury nanodroplets in D₂O was studied in [33], and the transformation (~ 10%) of the stable isotope $^{196}$Hg (a sample enriched to 55.6% with this isotope was used) into $^{197}$Au was observed during laser treatment. The stages of this process could be represented as follows:

$$^{196}_{80}Hg + {}^2n_{isu} \rightarrow {}^{197}_{80}Hg + n + Q(3.78 \text{ МэВ}), \qquad (49)$$

$$^{197}_{80}Hg + e^- \rightarrow {}^{197}_{79}Au + \nu + Q(0.6 \text{ МэВ}). \qquad (50)$$

It should be noted that the formation of HDO molecules in heavy water (there were none in the initial solution) was observed in these experiments by means of Raman spectroscopy. We may assume that protons appear in this medium when deuterons are split by neutrons formed in reaction (49) with the participation of different stable isotopes of Hg, facilitated by higher releases of energy. So, for reaction between $^2n_{isu}$ and isotope $^{199}$Hg in particular, $Q = 5.02$ MeV.

Here it should be noted that the introduction of a hypothetical *β*-dineutron also makes it possible to determine whether under powerful electric discharges in tubes containing mixtures of deuterium-inert gases, neutrons are generated [17, 18]. Fluxes of *β*-dineutrons formed in these experiments can be considered as fluxes of neutrons due to the ($^2n_{isu}, n$) processes on nuclei of $^{107}$Ag and $^{109}$Ag of the detector target. Other processes can occur in the detector target as well:

$$^{108}_{47}Ag + {}^2n_{isu} \rightarrow {}^4_2He + {}^{106}_{46}Pd + e^- + \tilde{\nu} + Q(13.01 MeV), \qquad (51)$$

$$^{108}_{47}Ag + {}^2n_{isu} \rightarrow {}^{12}_6C + {}^{98}_{42}Mo + e^- + \tilde{\nu} + Q(13.64 MeV). \qquad (52)$$

In analyzing the data of [34], where the qualitative difference between the initiated transmutation of $^{232}$Th nuclei in solutions of D₂O and H₂O was established, it should be also considered the ($^2n_{isu}$, n) processes in order to understand the totality of the obtained experimental data:

$$^{232}_{90}Th + ^2n_{isu} \rightarrow ^{233}_{90}Th + n + Q(1.78 \text{ МэВ}), \quad (53)$$

$$^{230}_{90}Th + ^2n_{isu} \rightarrow ^{231}_{90}Th + n + Q(2.11 \text{ МэВ}), \quad (54)$$

The partial exclusion of $^{137}$Cs isotopes from the solution upon laser initiation of this system was understood in a similar manner:

$$^{137}_{55}Cs + ^2n_{isu} \rightarrow ^{138}_{55}Cs + n + Q(1.41 \text{ МэВ}), \quad (55)$$

this allowed us to explain the fixed increase in the initially small concentration of barium in the solution due to the β-decay of $^{138}$Cs ($T_{1/2}$ = 33.4 min):

$$^{138}_{55}Cs \rightarrow ^{138}_{56}Ba + e^- + \tilde{\nu} + Q(5.37 \text{ МэВ}). \quad (56)$$

In accordance with these data, the half-life period of nucleus $^2n_{isu}$ should obviously be at least commensurable with the specified time for the completion of (53)–(55) in an exposure time of 1–4 h, and possibly exceed it.

A question then arises: Can β-dineutrons be observed not indirectly, as in the laser ablation of metals, but directly, in nuclear-physical experiments? The answer apparently lies in selecting processes in which the quark structure of nucleons is manifested in more content. From this point of view, experiments with beams of π⁻-mesons could be of undoubted interest. One such experiment may already have been carried out. In the above-mentioned work [107], the differential missing mass (MM) of spectra for the process $^6$Li(π⁻, p)$^5$H at specified (in the laboratory system) angles θ of proton emission was measured at fixed T energies of π⁻-mesons. We may assume that the relatively low peak in the spectrum at missing mass MM ≈ –3 MeV/c² at θ = 20° and T(π⁻) = 125 MeV, which was pronounced in Fig. 1, was due to the joint formation of β-dineutrons and tritium. It should be stressed that the value of this peak is less than half the maximum value of the spectrum at MM > 0, measured in the same experiment.

### 5.3.2. Mechanism of e⁻-catalysis

Of special interest are the cases when the formation of *isu*-states in nuclear matter is initiated in the initially radioactive nuclei, since the relaxation process with the decay of the vector W⁻-boson can initiate a general radioactive decay of the nucleus in the *isu*-state with the creation of the daughter products of the decay of the initial radioactive nucleus. Let's look at an example the initiating interaction between an electrons and $^{238}$U and $^{235}$U nuclei. Such interaction can initiate the formation of β-protactinium, $^A_{91}Pa_{isu}$, according to the reaction:

$$^{238}_{92}U + e^-_{he} \rightarrow ^{238}_{91}Pa_{isu} + \nu \rightarrow ^{234}_{90}Th + ^4_2He + e^- + \nu + \tilde{\nu} + Q(4.27 \text{ MeV}), \quad (57)$$

$$^{235}_{92}U + e^-_{he} \rightarrow ^{235}_{91}Pa_{isu} + \nu \rightarrow ^{231}_{90}Th + ^4_2He + e^- + \nu + \tilde{\nu} + Q(4.68 \text{ MeV}). \quad (57a)$$

Such processes of radioactive decay are defined as being realized under the *mechanism of e⁻ catalysis*. Indicators of the instability of the $^A_{Z-1}M_{isu}$ nuclei arising during processes (57) and (57a) are the absolute values of the structural energy deficit $\Delta Q$ ($\Delta Q < 0$) of these nuclei β-protactiniums in the metastable *isu*-states, defined as $\Delta Q = (m_{^A_Z N} - m_{^A_{Z-1}M})c^2$. In these cases, the every mass of the $^A_{Z-1}M_{isu}$ nucleus is taken to be $m_{^A_{Z-1}M_{isu}} = m_{^A_Z N} + m_e$, where $m_{^A_Z N}$ is the mass of the $^A_Z N$ nucleus and $m_e$ is the rest mass of the electron. For (46) and (46a), we have $\Delta Q \approx -3.46$ MeV and $\Delta Q \approx -1.41$ MeV, respectively.

As an example, let us consider the process of initiation of nuclear decays of $^{238}$U and $^{235}$U investigated in [97] under conditions of laser ablation of beryllium in aqueous solutions of uranyl chloride UO$_2$Cl$_2$ with concentrations of 20 to 5 mg/ml. During ablation, the cuvette containing the solution was cooled with flowing water. The typical time of laser irradiation of the target was 1 h. A neodymium laser (Nd:YAG laser, 6 ns, $10^{10}$ W/cm$^2$), 1064 nm, with an pulse duration of 10 ps and a pulse repetition frequency of 50 kHz (energy per pulse of 0.2 mJ; $10^{13}$ W/cm$^2$) was used. In the presence of decaying nuclides, the nanoparticles formed in the ablation process as a rule become unstable with respect to sedimentation and precipitate onto the bottom of the cuvette.

Main isotope $^{238}$U decays with the emission of $\alpha$ particles with a half-life longer than $10^9$ years. Isotope $^{235}$U decays via the same mechanism to $^{231}$Th with a half-life longer than $10^8$ years. The effect of laser radiation can be described with good precision by measuring the activities of the most intense peaks of the nuclides present in the solution. These include $^{234}$Th, $^{234m}$Pa, and $^{235}$U. A measure of activity is the number of decays of a given radionuclide per unit of time. Since the spectrum of $\gamma$ quanta emitted upon the decay of a nuclide is characteristic of it specifically, its content in the sample can be measured from the $\gamma$ activity before and after laser radiation. In the families of $^{238}$U and $^{235}$U, these nuclides are arranged as

$$^{238}_{92}U \rightarrow {^{234}_{90}Th} \rightarrow {^{234m}_{91}Pa} \rightarrow {^{234}_{92}U} \rightarrow \dots,$$

$$^{235}_{92}U \rightarrow {^{231}_{90}Th} \rightarrow {^{231}_{91}Pa} \rightarrow \dots$$

$^{235}$U belongs to another radioactive branching, and is not associated with $^{238}$U in any way. Nuclides $^{234}$Th and $^{234m}$Pa decay through $\beta$-decay with half-lives of 27 days and 1.17 min, respectively. The decay of $^{234}$Th is accompanied by the emission of $\gamma$ quanta with energies of 92.5 keV, the number of which is measured before and after the laser irradiation of solutions of uranyl chloride UO$_2$Cl$_2$ containing various metallic targets. $^{235}$U decays to $^{231}$Th with the emission of $\alpha$ particles that in turn emit $\gamma$ quanta with energies of 186 keV. Radionuclide activity was determined from the area under the peaks of nuclide $\gamma$ spectra. The $\gamma$ activity of the samples before and after laser radiation was measured on an Ortec-65195-P semiconductor $\gamma$ spectrometer with an error of ±5%.

The samples of solutions with nanoparticles were placed into polymeric Petri dishes that were then sealed. The laser ablation of beryllium targets in aqueous solutions of uranyl chloride is accompanied by a substantial increase in the activity of $^{234}$Th, with respect to its activity in the initial sample (Fig. 2a). A more than threefold increase of thorium activity with respect to the initial sample is observed at low concentrations of uranium salts. The effect of the elevated activity of $^{234}$Th was more noticeable at a low (5 mg/mL) concentration of uranyl chloride in water solution than at a high one (20 mg/mL), due probably to a reduction in the degree of disassociation of the uranyl chloride molecules in aqueous solution, and thus the relatively low concentration of uranyl ions in low-temperature plasma. Note that the activity of $^{235}$U in the same solution remains unchanged after laser irradiation in the investigated range of uranyl chloride concentrations (Fig. 2b). This difference can naturally be attributed to a larger value of the energy deficit $\Delta Q$ and a greater imbalance in the structure of the nucleus $^{238}_{91}Pa_{isu}$ than the nucleus $^{235}_{91}Pa_{isu}$.

Most fascinating is the continuing increase in the activity of $^{234}$Th after long periods of time after laser irradiation. The dynamics of the changes in the activity of this nuclide over times significantly longer than its half-life is shown in Fig. 3. These experimental values for the relative changes in the periods of $^{234}$Th and $^{235}$U activity were obtained by studying the activity of solid-phase samples that were collections of metal nanoparticles separated by centrifuging them from a solution of uranyl chloride after laser ablation. In these dependences, the ratio of the activities of $^{234}$Th and $^{235}$U at the initial moment of prepared sample activity measurements was taken as unity. The emergence of activity indicated that the radioactive nuclei in the solution were captured in the volume of the resulting nanoparticles.

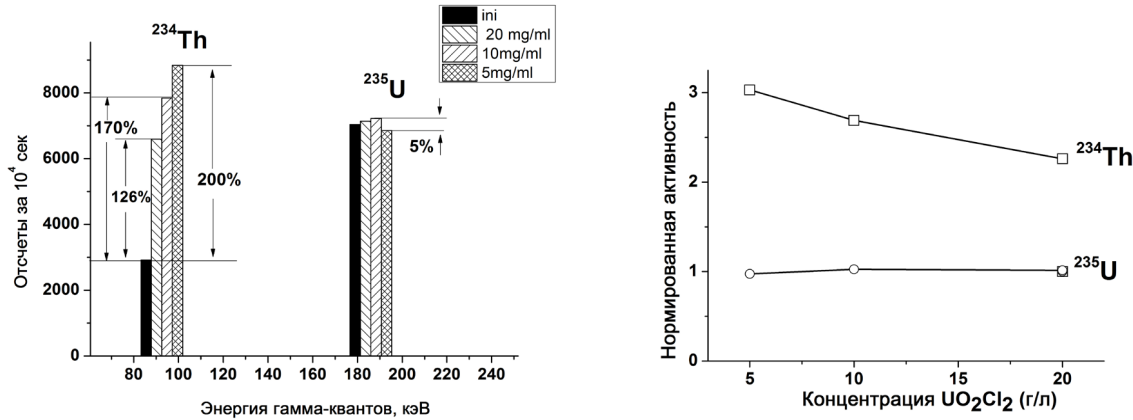

Fig. 2. (a) Change in the activities of $^{234}$Th and $^{235}$U ($N$ denotes counts per $10^4$ s) after the laser irradiation of a beryllium target in an aqueous solution of uranyl chloride. (b) Change in the activities of $^{234}$Th and $^{235}$U, depending on the concentration of uranyl chloride, normalized according to the activities of the respective nuclides in the initial solution. The activity was measured 10 days after laser radiation. *1* shows the activity of the nuclides in an aqueous solution of uranyl chloride before irradiation; *2*, *3*, and *4* show the activities of [UO$_2$Cl$_2$] = 20, 10, and 5 mg/mL, respectively. $E$ denotes the energies of γ-quanta (Nd:YAG laser, 6 ns, $10^{10}$ W/cm$^2$).

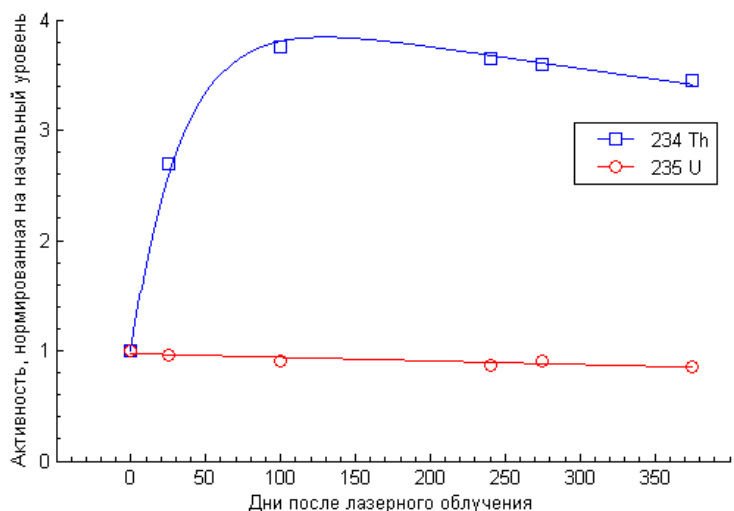

Fig. 3. Changes in the activities of $^{234}$Th (squares) and $^{235}$U (circles) over time (τ denotes the time after laser radiation). Uranyl chloride concentration, 20 mg/mL; duration of laser irradiation, 1 h (the same condition as in Fig. 2); *a* denotes the activities of nuclides, normalized according to the values in the initial measurements of sample activity after irradiation. The solid lines are results of calculations (see text after Eq. (60*a*)).

The dynamics of the emergence (2 h after initiating the process) of $^{238}$U decay products and the formation of isotope $^{214}$Pb as a result of a chain of sequential decays of nuclei, beginning with the decay of $^{238}$U and the violation of the secular fraction of $^{234}$Th, were observed under the experimental conditions in [120]. Among the radioactive isotopes preceding formation of $^{214}$Pb, we should note $^{234}$U and $^{230}$Th, whose half-life periods are $2.45 \times 10^5$ and $7.7 \times 10^4$ years, respectively, under natural conditions. The broad process characteristic of the sequence of transformations from the decay of $^{238}$U to the formation of $^{214}$Pb, including the separation of six α-particles, can be presented schematically as

$$^{238}_{92}U \xrightarrow{e^-_{he}} {}^{214}_{82}Pb + 6 \cdot {}^{4}_{2}He + 2e^- + \nu + 3\tilde{\nu} + Q(31.92\ MeV), \tag{58}$$

abbreviated somewhat by placing $e^-_{he}$ under the arrow. Since the activity of $^{238}$U itself almost did not change in this process, it could be concluded that in the conditions of the performed experiments, the accelerated decay of $^{238}$U with $^{234}$Th formation and subsequent formation of $^{234}$U, $^{230}$Th, and $^{214}$Pb affected only a small share of the initial $^{238}$U atoms. In the conditions of the metal laser ablation in uranyl solutions, the acceleration of decay of any small part of $^{235}$U nuclei has not been detected.

$$^{235}_{92}U \xrightarrow{e^-_{he}} {}^{207}_{82}Pb + 7 \cdot {}^{4}_{2}He + 4e^- + \nu + 5\tilde{\nu} + Q(31.92\ MeV). \qquad (59)$$

In accordance with the data, presented in Fig. 3 (see also [97]), it is naturally to suppose that the observed post-ablation activity can be associated with the nuclei of $^{238}$Pa in the *isu*-state that form upon interaction between $^{238}$U nuclei and electrons in accordance with the process (57) and are encapsulated in metal nanoparticles.

The equation for number $N_T(t)$ of $^{234}$Th nuclei formed over time $t$ upon the decay of encapsulated nuclei of $^{238}_{91}Pa_{isu}$ has the form:

$$\frac{dN_T(t)}{dt} = -k_2 N_T(t) + k_1 P_0 \exp(-k_1 t), \quad N_T(0) = 0, \qquad (60)$$

where $k_1$ and $k_2$ are constants of the decay of $^{238}_{91}Pa_{isu}$ and $^{234}$Th nuclei, respectively; $P_0$ is the number of $^{238}_{91}Pa_{isu}$ nuclei formed upon laser ablation. Since the number of $^{234}$Th nuclei remaining in the nanoparticles after laser ablation and before measuring the activity of the solid-phase samples was not recorded, Eq. (60) considers the kinetics for the newly formed nuclei of $^{234}$Th. Since the duration of the kinetics in this case is one year and more, and the half-life of $^{234}$Th is $T_{1/2}$ = 24.1 days, this analysis is sufficient on the qualitative level. According to the kinetics shown in Fig. 2, $k_2 \gg k_1$. The solution to Eq. (60) has the form

$$N_T(t) = A[\exp(-k_1 t) - \exp(-k_2 t)], \quad A = k_1 P_0 / (k_2 - k_1). \qquad (60a)$$

This dependence is in good agreement with experimental data presented in Fig. 3 from observing the increase and subsequent slow reduction in the activity of $^{234}$Th after the initiation of $^{238}$U decay upon the laser ablation of gold. Constant $k_1$ = 0.000776 days$^{-1}$ corresponds to the half-life of $^{238}_{91}Pa_{isu}$: $T_{1/2} = \ln 2 / k_1 = 893.2$ days $\approx 2.45$ years, only 37 times higher than the corresponding value for $^{234}$Th, $k_2$ = 0.02876 days$^{-1}$. We see that the effective rate constant of such initiated decays of $^{238}$U nucleus increases by 9 orders of magnitude in the realization of the "$e^-$-catalysis".

Based on the dependence for the kinetics of the weaker change in $^{235}$U activity over time shown in Fig. 3, and considering that the decay of $^{235}$U could also be initiated by interactions with electrons, leading to formation of $^{235}_{91}Pa_{isu}$ at the first stage of decay, we obtain a decay rate constant of 0.0003285 days$^{-1}$ by approximating linearly the dependence for $^{235}$U and estimating a half-life of $T_{1/2}$ = 2110 days $\approx 5.58$ years. These values should naturally be considered as rough estimates that can be adjusted using data from longer experiments with better monitoring of the state of the initial system. We should note too that noticeable differences in the initial rate of $^{238}$U and $^{235}$U decay could be due to the above differences in the values of the structural energy deficit in the in-shake-up state: $\Delta Q$ = –3.46 MeV and $\Delta Q$ = – 1.41 MeV for $^{238}$U and $^{235}$U, respectively.

In conclusion of this section, let us pay attention to some unexpectedness of the presented result about the possibility of external influences of electrons on the dynamics of the decay of a radioactive nucleus. At the same time, as experience shows, the external excitation of the radioactive nucleus as an integral system (under the influence of $\gamma$-radiation, in particular) can not affect the rate of radioactive decay, and therefore, of the discussed phenomenon of initiating of the metastable state of the nucleus. In these cases, nuclear matter manifests itself as an integral system of interacting nucleons with inherent individual characteristics. It turns out that although the electrons can not interact with the nucleons of the nucleus as fragments of nuclear matter, they can initiate (through vector W⁻-bosons) local disturbances in the nucleon structure of the nucleus through weak nuclear interaction.

It should be borne in mind here that according to section 4.3, weak nuclear interactions are not as weak as it is often assumed. The value of the corresponding dimensionless constant $\alpha_F$ is almost an order of magnitude greater than the value $\alpha_e$ of the fine structure constant [3], although in the literature it is often stated that $\alpha_F$ is less than the constant $\alpha_e$ by 3 orders of magnitude. There is one more small comment. Since the Fermi constant, $G_F$, characterizes the 4-fermionic interaction with participation of the external (in relation to the nucleus) particles – electron and neutrino, it makes sense to take into account the fine structure (7a) constant $\alpha_e$ that characterizes the extent of overlapping between the fields of Casimir polarization of the electron at the inner shell and nucleus, and to present the non-dimensional Fermi constant $\alpha_F$ in the form: $\alpha_F = \alpha_e \cdot \alpha_{sW}$, where $\alpha_{sW}$ – the parameter, characterizing the interaction of $u$- and $d$-quarks with vector bozons, $W^{\pm}$ and $Z^0$. As was shown in Section 4.3, $\alpha_{sW}$ = 6.7. This value is close by value to the non-dimensional constant $\alpha_{QCD} \approx 10$ pertinent to the strong interaction in quantum chromodynamics at low energies, when the interaction via pion exchange appears to be the strongest one [94] should be considered as the phenomenological parameter, characterizing the interaction of pions as quark-antiquark ensembles with nucleons as the ensembles of 3 quarks, while the value $\alpha_{sW}$ may characterize the direct interaction of vector bozons ($W^{\pm}, Z^0$) with $u$- and $d$-quarks.

### 5.3.3. Harpoon mechanism

The extreme difficulties in understanding the mechanism of low-energy nuclear processes are the reactions between many-electron atoms. Such processes are usually considered in connection with the study of transformation processes in native systems [121-124]. However, it was recently shown [38] that reactions of this type can occur during the initiation of self-propagating high-temperature synthesis (SHS) processes [125]. The composition of condensed combustion products of thermite powder mixtures (Al + Fe$_2$O$_3$) in air was investigated in [38]. The purity of the starting materials was (99.7-99.9)% mass. It was shown that in the process of termite combustion (flame temperature exceeded 2800 K) calcium is formed and stabilized in an amount up to 0.55% mass. In the initial thermite powder systems (Al + Fe$_2$O$_3$), calcium was absent. According to [38], calcium could be formed in nuclear reactions

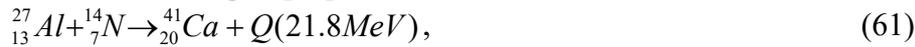
$$^{27}_{13}Al + ^{14}_{7}N \rightarrow ^{41}_{20}Ca + Q(21.8 MeV), \tag{61}$$

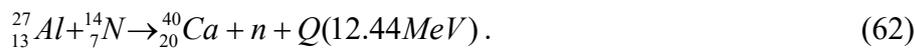
$$^{27}_{13}Al + ^{14}_{7}N \rightarrow ^{40}_{20}Ca + n + Q(12.44 MeV). \tag{62}$$

The phenomenon of calcium formation in experiments [38] may indicate that in the flame of iron-aluminum thermite combustion in air, the electron temperature can be much higher than the flame temperature, fixed by the energy of atoms and ions. It is the latter situation that is typical of a low-temperature glow discharge plasma. In this case, the interaction of high-energy electrons with nuclei $^{27}_{13}Al$ and $^{14}_{7}N$ could lead to the formation of nuclei $^{27}_{13}Mg_{isu}$ and $^{14}_{6}C_{isu}$, accordingly. The largest activity in nuclear interactions from these nuclei is inherent to the nucleus $^{27}_{13}Mg_{isu}$, since the deficit of its energy relative to the nucleus $^{27}_{13}Mg$ is $\Delta Q$ = - 2.61 MeV, whereas for the nucleus $^{14}_{6}C_{isu}$ the corresponding value is much less and equal to -0.16 MeV.

We will assume, following [100], that if the nucleus of an atom (or ion) is in a metastable, pre-decay $isu$-state (we assume that this is $^{27}_{13}Mg_{isu}$), then the lability of the electronic subsystem increases, and the probability of its partial overlap with electronic subsystems of the located near atom (in this case, the nitrogen atom), grows for this atom. It is obvious that large values of the energy releases of the gross processes (61) and (62) should act as an initiating factor to the manifestation of the spin-spin interaction of the electronic subsystems of both atoms and the formation of common "molecular" orbitals under the correcting action of spin electron-nuclear interactions for each of the atoms. The emerging bonds pull both atoms together, and the formation of common orbitals is more intense as the nuclei are brought closer together. As a result, a sort of a "*harpoon mechanism*" is observed when a neighboring atom is captured by the atom with the nucleus in the *isu*-state. The complete integration of the electronic subsystems of both atoms initiates the fusion of the nuclear matter of the nucleus in the *isu*-state (in our case $^{27}_{13}Mg_{isu}$) and the adjacent nucleus (in our case $^{14}_{7}N$). The corresponding gross-process can be presented as:

$$^{27}_{13}Al + {}^{14}_{7}N + e^{-}_{he} \rightarrow {}^{41}_{20}Ca + e^{-} + \nu + \tilde{\nu} + Q(21.8 MeV), \qquad (61a)$$

$$^{27}_{13}Al + {}^{14}_{7}N + e^{-}_{he} \rightarrow {}^{40}_{20}Ca + n + e^{-} + \nu + \tilde{\nu} + Q(12.44 MeV). \qquad (62a)$$

Earlier, the harpoon mechanism was considered in connection with the processes of nuclear transmutations in native systems [100].

Because of the manifestation of weak nuclear interactions in the formation of the nuclear matter of the final nucleus as a set of interacting nucleons, a significant part of the energy release can also be realized by emitting neutrinos and antineutrinos if the final nucleus can be formed in the ground state by virtue of spin and parity conservation laws. Of course, in cases where the final nuclei are formed in an excited state, along with non-ionizing radiation of neutrinos and antineutrinos, X-rays or gamma quanta will be emitted. Under the experimental conditions [38], X-rays were detected.

Our phenomenological analysis shows that in order to understand the essence of the observed nuclear transformations during burning of thermite mixtures, when the harpoon mechanism is realised, it is actual to develop new theoretical approaches not to the calculation of the quantum mechanical probabilities of certain processes, but rather to modeling the dynamics of nuclear processes on the basis of quantum-chemical analysis. By this we mean calculations of the electronic structure of an atom upon the formation of nuclei in the *isu*-state with disturbed nucleon structure; calculations for modeling the spatial instability of the electronic subsystem of an atom that emerges due to the loss of nucleus stability; and calculations for the dynamics of the overlapping of such mobile orbitals with the electron orbitals of neighboring atoms and the formation of molecular orbitals that initiate the convergence and fusion of the respective nuclei. Kramer's activation mechanism in its discrete version (the random walk on the energy levels of the system to reach a set boundary) [126] generally used in physicochemical kinetics could be useful in analyzing the dynamics of nuclear radioactive decay. Here we mean the dynamics of the accumulation of energy by the nucleus in the unstable *isu*-state on the "last" bond, the disruption of which means the decay of the nucleus along a certain channel.

5.3.4. *Nuclear-Chemical processes in native systems*

The possibility of nuclear transformations occurring in biological systems in vivo [109, 110, 121-124] must be considered one of the most intriguing puzzles of modern bio- and nuclear physics. Indeed, the experimental data obtained using modern investigation techniques (e.g., Mössbauer spectroscopy, mass spectrometry, and laser time-of-flight mass spectrometry) clearly demonstrate that the following nuclear transformations occur in the *Saccharomyces cerevisiae* culture in a sugar–saline nutrient medium containing all necessary macro- and microelements:

$$^{55}_{25}Mn + {}^{2}_{1}d \rightarrow {}^{57}_{26}Fe + Q(15.6 MэB), \qquad (63)$$

$$^{23}_{11}Na + {}^{31}_{15}P \rightarrow {}^{54}_{26}Fe + Q(22.3 MэB). \qquad (64)$$

It was also shown [110] that half-life $T_{1/2}$ (the reciprocal value of the decay rate constant) of radioactive elements can change markedly in the concentrated mass of different types of metabolically active microorganisms when the microelements, macroelements, and other components of the nutrient blend needed for the growth of a culture are contained in a medium.

For example, the half-lives of the β-active nuclei of cesium-137 and barium-140 (30.1 years and 12.8 days, respectively) are reduced to approximately 380 and 2.7 days under the indicated conditions. At the same time, the half-life of cobalt-60 (1925 days) remains virtually the same. The effect adding salts (NaCl, KCl, $CaCO_3$) to the cuvette has on the rate of radioactive decay was established during investigations of cesium-137 β-decay. It was found that the presence of $CaCO_3$ caused an even greater acceleration of the decay rate (the $T_{1/2}$ value fell to 310 days), while adding NaCl and KCl to the system inhibited β-decay, which was quite moderate for NaCl ($T_{1/2} \approx 480$ days) and considerable for KCl ($T_{1/2} \approx 10$ years).

Although the conditions for initiating nuclear-chemical transformations in native systems are obviously quite different from those required in low-temperature plasma, we believe that a key role in initiating and governing by such processes plays to the dynamic relationship that exists between the electronic and nuclear subsystems of an atom thanks to the EM vacuum. Is this nuclear-chemical process safe for biological systems with respect to the functioning of cells, and to the participants of an experiment, especially since the experimenters in the growing biological systems usually last tens of days? We attempt to answer these questions below.

In studying nuclear transformations native systems, we must consider aspects of initiating energy-dependent processes *in vivo*, particularly active transmembrane ion transport with "soft" use of the energy of the hydrolysis of adenosine triphosphate (ATP), the universal accumulator of energy in any cell. The change in free energy during ATP hydrolysis is $\Delta G_0 = -0.32 эB$ [127]. According to [128], the processes initiated by ATP hydrolysis in various transport enzymes, and those triggered by photoexcitation in native systems, occur through the energy-dependent redistribution of charge density and the creation of intramembrane local electric fields with directions opposite to the vector of the average electric field intensity in membranes. This possibility was illustrated in [128] using simple models of the partial transfer of electron density governed by the absorbed energy, accompanied by the formation of additional pairs of separated (by ~1 nm) charges $D^{\delta+}$–$A^{\delta-}$ in the vicinity of channels in which ion transfer is mediated by transport enzymes (chromophore–protein complexes of bacteriorhodopsin, $H^+$ ATPase, $Na+$ ATPase, $K^+$ ATPase, and $Ca^{2+}$ ATPase). It was shown that the Coulomb field formed in this manner in the vicinity of a pair of charges in an ion channel could significantly alter the profile of the potential energy of ions transported through the channel and facilitate their migratory transfer in the direction opposite the one prescribed by the average electric field of the membrane. For the transmembrane active transport of ions, this situation need only be produced in isolated key stages of ion transport along a specific channel. We should also mention that studies conducted in recent years on the distribution of electron density and the spatial structure of transport ATPases [129-131] with high resolution (up to ~0.5 nm) open up the possibility of refining common model conclusions on mechanisms for harnessing the energy of ATP hydrolysis and using it for active transmembrane transport of ions [132].

In connection with the data discussed in [128] it must be noted that *Saccharomyces cerevisiae* yeasts are capable of the intense synthesis of endo-inulinase [133, 134], an enzyme that cleaves fructose-containing oligosaccharides and polymers (fructans) into virtually pure fructose and fructooligosaccharides with different degrees of polymerization. We may therefore assume that in systems containing *Saccharomyces cerevisiae* culture, the key role of initiating nuclear processes *in vivo* should be attributed to the yeast plasma membrane $H^+$-ATPase, which is the main protein of the plasma membrane not only functionally but also structurally. As for the data on the acceleration of radioactive decay in an association of different types of microorganisms, here too the role of $H^+$-ATPase as the main ion transport enzyme could be key.

We may assume that the changes in the magnitudes and directions of local electric field intensities $F_{loc}$ during the indicated shifts of electron density, and the high absolute values of average field intensities in biological membranes ($\bar{F} \sim 10^5 V/cm$) that occur in characteristic times of $10^{-10}$ s [128] could have an impact effect on nearby groups of atoms and initiate processes of interaction between the inner-shell electrons of these atoms and their nuclei, leading to the formation of *β* -nuclei. It is these *β* -nuclei atoms with their structure in the metastable radiation-active state [99] that could effectively interact with surrounding atoms, making possible processes of types (63) and (64).

As for the simpler reaction (63) between deuterons and multinucleon nuclei, the neutral *β*-nucleus $^2n_{isu}$ (*β*-dineutron) that forms during interaction between the high-energy electrons and deuterons $d^+$ could play the role of a *β*-nucleus in the *isu*-state.

It is because of its long lifetime that the $^2n_{isu}$ nucleus can interact directly with surrounding atoms, facilitating the process:

$$^{55}_{25}Mn + {}^2n_{isu} \rightarrow {}^{57}_{26}Fe + e^- + \widetilde{\nu} ,\qquad (63a)$$

so the corresponding gross process can be presented as

$$^{55}_{25}Mn + {}^2_1d + e^-_{he} \rightarrow {}^{57}_{26}Fe + e^- + \nu + \widetilde{\nu} + Q(15.6 MэB) .\qquad (64b)$$

Hence, the initiation of reaction (63) is a result of sequential processes: the initiating action of the local electric field of the pair of charges $D^{δ+}$–$A^{δ-}$, which originates during hydrolysis of the ATP molecule in the vicinity of one of the cell's ion channels, on the localized charge of the deuterium atom $A^{δ-}$; the formation of the fairly mobile neutral nucleus $^2n_{isu}$ formed as a result of this interaction; and the subsequent interaction between this nucleus and a manganese atom (ion) also located in the region of the considered channel.

It is important to note here that weak nuclear interactions occur due to the quark structure of the local excitation of the *isu*-state of *β*-nucleus $^2n_{isu}$ during the formation of the nuclear matter of final nuclei as a set of interacting nucleons. In this case, as indicated in section 4.2, the relaxation mechanism of the formed products inevitably includes energy loss through the emission of neutrino–antineutrino pairs or the URCA process [114], rather than through the emission of *γ* -quanta by an excited nuclei composed of protons and neutrons, and the nuclear process (63) is safe for cells.

In the case of reaction (64), after this initiating action of the changing electric field of the $D^{δ+}$–$A^{δ-}$ pair of charges to one of the atoms initial for of this reaction (for the sake of clarity, we assume that this is a sodium atom, and the distance between the considered atoms is approximately 0.5 nm), one of the inner shell electrons of this atom interacts with the nucleus according to the equation

$$^{23}_{11}Na + e^- \rightarrow {}^{23}_{10}Ne_{isu} + \nu .\qquad (65)$$

The *β*-nucleus of neon formed in the *isu*-state in process (65) is unstable with respect to the subsequent *β*-decay. We may assume that if the nucleus is in the unstable *isu*-state, this atom displays increased lability of the electronic subsystem, and the probability of its partial overlap with electronic subsystems localized in the vicinity of the atoms (in this case, phosphorous atoms) grows for this atom. Most of the energy released in gross process (64) - the fusion of two atoms with formation of a new element, obviously serves as an initiating factor for spin–spin interaction in the electronic subsystems of both atoms, and for the formation of common molecular orbitals during the correcting action of electron–nuclear spin interactions in each of the atoms. The emerging bonds pull both atoms together, and the formation of common orbitals is more intense as the nuclei are brought closer together. As a result, the *harpoon mechanism*, which was realized under calcium formation in experiments [38] (see section 5.3.3), is observed when a neighboring atom is captured by the atom with the nucleus in the *isu*-state. The complete integration of the electronic subsystems of both atoms initiates the merging of the nuclear matter of the nucleus in the *isu*-state and the regular nucleus. The corresponding gross process can be presented as

$$^{23}_{11}Na + {}^{31}_{15}P + e^-_{he} \rightarrow {}^{54}_{26}Fe + e^- + \nu + \widetilde{\nu} + Q(22.3 MэB) .\qquad (64a)$$

Due to the weak nuclear interactions that occur during the formation of the nuclear matter of the final nucleus as a set of interacting nucleons, the release of energy here also occurs through the emission of a neutrino and an antineutrino if the final nucleus, by virtue of the laws of spin and parity conservation, can form in the ground state. When the final nuclei are formed in the excited state, gamma quanta will obviously be released along with the non-ionizing radiation of neutrino and antineutrino. It is the author's opinion that only these assumptions allow us to understand the existing data on processes of types (63) and (64) occurring *in vivo*.

In analyzing the kinetics of the radioactive decay of cobalt-60, cesium-137, and barium-140 in the concentrated masses of different types of metabolically active microorganisms investigated in [110], we shall also assume that the considerable acceleration of the β-decay of cesium-137 and barium-140 nuclei initiated in vivo is due to the impact of the inner-shell electrons of these atoms on their nuclei during ATP hydrolysis, or to photoexcitation if the indicated atoms are adsorbed on the respective fragments of H$^+$-ATPase in the cytoplasmic membranes of microorganisms. The cobalt-60 nuclei are obviously also subjected to this action of electrons. By analogy with initiated processes of $\alpha$-decay (57) and (58), those of the *β*-decay of cobalt-60, cesium-137, and barium-140 in native systems can be presented in the form:

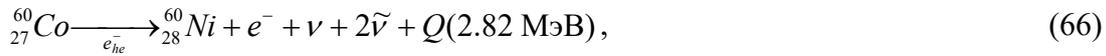
$$^{60}_{27}Co \xrightarrow{e^-_{he}} {}^{60}_{28}Ni + e^- + \nu + 2\tilde{\nu} + Q(2.82 \text{ МэВ}), \tag{66}$$

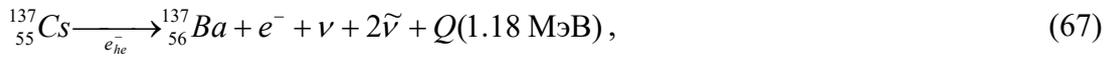
$$^{137}_{55}Cs \xrightarrow{e^-_{he}} {}^{137}_{56}Ba + e^- + \nu + 2\tilde{\nu} + Q(1.18 \text{ МэВ}), \tag{67}$$

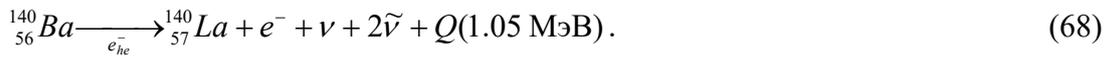
$$^{140}_{56}Ba \xrightarrow{e^-_{he}} {}^{140}_{57}La + e^- + \nu + 2\tilde{\nu} + Q(1.05 \text{ МэВ}). \tag{68}$$

(Here we use more compact presentations of *β*-decay processes emphasizing their the $e^-$-catalytic nature).

It is assumed that processes (66), (67), and (68) occur through the formation of intermediate nuclei $^{60}_{27}Co$, $^{137}_{55}Cs$ and $^{140}_{56}Ba$, respectively. The magnitudes of structural energy deficit $\Delta Q$ required for the formation of the $^{60}_{26}Fe$, $^{137}_{54}Xe$ and $^{140}_{55}Cs$ nuclei are −0.237, −4.17, and −6.22 MeV, respectively. It can be expected that the initiating influence of electrons on the β$^-$-decay of nuclei will be manifested to the greatest extent in the cases when the "mismatch" in the absolute magnitude of the deficit $\Delta Q$ of the structural energy for the forming nuclei in the *isu*-state is the greatest. Therefore, in the cases under consideration, the effect of accelerating radioactive decay should have been manifested for $^{137}_{55}Cs$ и $^{140}_{56}Ba$ nuclei, while for $^{60}_{27}Co$ nuclei, it should have been minimal. The available experimental data [110] on the initiated decays of $^{137}_{55}Cs$, $^{140}_{56}Ba$ and $^{60}_{27}Co$ agree with this conclusion: the half-life of *β*$^-$-active cesium-137 and barium-140 nuclei, equal to 30.1 years and 12.8 days, respectively, decreased to about 380 days and 2.7 days, while the half-life of cobalt-60, equal to 1925 days, remained practically unchanged.

As for the effect of salts additionally introduced into the investigated system (NaCl, KCl, and CaCO$_3$) on the decay rate, the revealed in [110] promoting effects of Ca$^{2+}$ ions and inhibiting effect of Na$^+$ (moderate) and K$^+$ (significant) could be due to the sorption of the indicated ions close to the mouth of ion channels. However, the available experimental data on these processes is insufficient to establish the mechanisms of these effects.

It is possible to believe, the formation of atoms with nuclei in metastable *isu*-states as a result of the initiating action of inner shell electrons on predecessor nuclei in biological systems *in vivo* during ATP hydrolysis could be of functional necessity. Due to the high lability of the electronic subsystem of *β*-atoms, the kinetic difficulties in initiating such energetically advantageous processes as the formation of different complex bonds could be eliminated more effectively.

## 5.4. Specific nuclear-chemical processes
### 5.4.1. *The Cherdyntsev–Chalov phenomenon*

In 1954, Professor V.V. Cherdyntsev and his graduate student P.I. Chalov observed for the first time the separation of even isotopes of uranium upon their transition from the solid to the liquid phase (the Cherdyntsev–Chalov effect) [21, 22]. It was found later that in almost all of the Earth's hydrosphere, the activity of isotope $^{234}$U exceeds that of $^{238}$U, due to the interaction between natural waters and mountain rocks, assuming that both isotopes are involved in a single chain of radioactive transformations (a uranium–radium series) [23, 24]. The activity $\eta_i$, introduced as the decay rate of the *i*-th uranium isotope, is defined as $\eta_i = k_i N_i$, where $k_i$ and $N_i$ are the decay rate constant and the number of *i*-th isotope nuclei to be decayed, respectively. Note that the fraction $\theta_i$ of the uranium-234 isotope in natural uranium ores is as low as $\theta_{234} \approx$ 0.0055%, with a half-life $T_{1/2}(^{234}U) \approx 2.45 \cdot 10^5$ years. At the same time, the fraction of the uranium-238 isotope is $\theta_{238} \approx 99.3\%$, yielding $\chi \equiv \theta_{234}/\theta_{238} \approx 5.54 \cdot 10^{-5}$, with a much longer half-life $T_{1/2}(^{238}U) \approx 4.47 \cdot 10^9$ years. The corresponding decay rate constants are related by the formula $k_{234} = k_{238} \cdot \chi^{-1}$.

It means that in undisturbed minerals older than several million years, the abundances of $^{238}U$ and its intermediate $\alpha$-decay product, $^{234}U$, reach a state of secular equilibrium. Under these conditions, the activity ratio (AR), $\eta_{234}/\eta_{238} = \dfrac{T_{1/2}(^{238}U)\theta_{234}}{T_{1/2}(^{234}U)\theta_{238}} \equiv {}^{234}U/{}^{238}U\text{ AR}$, will equal unity. However, natural waters, especially in seismically active regions, typically are enriched in $^{234}U$ with $^{234}U/^{238}U$ AR between 1 and 10 [25, 135]. The uranium concentrations and $^{234}U/^{238}U$ AR ratios in saturated-zone and perched ground waters were used to study the hydrologic flow in the vicinity of Yucca Mountain [25]. The *U* data were obtained by thermal ionization mass spectrometry for more than 280 samples from the Death Valley regional flow system. Wide variations in both U concentrations (commonly 0.6–10 μg l$^{-1}$) and $^{234}U/^{238}U$ AR (commonly 1.5–6) were observed on both local and regional scales. The ground water beneath the central part of Yucca Mountain had intermediate U concentrations but a distinctive $^{234}U/^{238}U$ AR of about 7–8. It is necessary to add that about 600 seismic events have occurred near the site in the last 20 years alone, with a 5.6-magnitude earthquake that happened as recently as 1992. There is also an evidence of relatively recent volcanic activity in the area.

Similar results were reported elsewhere [135], where the measurements of the $^{234}U/^{238}U$ AR in groundwater samples were used for monitoring the current deformations in the active faults at the Kultuk polygon, West Shore of Lake Baikal, for earthquake prediction. It was observed that the $^{234}U/^{238}U$ AR fluctuated in time, with the duration of cycles and amplitudes of $^{234}U/^{238}U$ AR fluctuations were variable in the range of 1.5-3.3, and the cycles of $^{234}U/^{238}U$ AR in water were synchronized in the lines of the monitoring stations in the sublatitudinal and submeridional direction at the time intervals when seismic shocks occurred at the Kultuk polygon. The U concentrations in the ground-water samples of the Kultuk polygon ranged from 0.0087 to 5 mcg/l. The basic scenario of $^{234}U/^{238}U$ AR variations in groundwater, recorded in the Kultuk polygon during the monitoring session, was examined in connection with the seismogenic activation of the western end of the Obruchev fault.

It is commonly believed that $^{234}U$ enters solutions preferentially as a result of several mechanisms related to its origin by radioactive decay of $^{238}U$ [25]. These mechanisms include damage of crystal-lattice sites containing $^{234}$U and the preferential release of $^{238}U$ not bound to the crystal lattice from the defects of minerals, as well as direct ejection of the recoil nucleus into the water near the boundaries of mineral grains.

At the same time, the results of [25, 135] suggest that the mechanochemical processes in relatively small volumes of uranium ore in ore deposits located in the geologically active, including seismically and volcanically active, zones of the Earth's crust are the important factor that can account for the significant changes of $^{234}U/^{238}U$ AR under study. These zones can be characterized by the emergence of high mechanical stresses, initiated shifts in the ore, and the formation of cracks and fissures. These processes in the U ore at high local mechanical pressures can not only change the structure of groundwater flows in the zone, but also give rise to high local electric fields and initiate the decomposition of water molecules and the formation of high-energy (on chemical scales) electrons. In this case, the concept developed in this paper allows us to expect that the formation of cracks and fissures in a uranium ore can initiate the radioactive decay of uranium-238 nuclei by the $e^-$-catalytic mechanism, producing *isu*-state *β*-protactinium nuclei. Note that it is the phenomenon of mechanically activated nuclear processes discovered in the works of Deryagin et al. [26, 27] that can be regarded as the starting point in the new stage of studying LENRs, which is usually attributed to the work of Fleischmann and Pons [28]. For instance, it was experimentally recorded that the destruction of targets made of a heavy ($D_2O$) ice by a metal striker with an initial velocity of 100–200 m/s produces neutrons, and their number is several times higher than the background level [27]. In contrast, no new neutrons were recorded when the same action was applied to the target made of an ordinary ($H_2O$) ice.

Assume that when a fissure is formed, a fraction of uranium atoms leaves the fissure surface layer of the uranium ore and is dissolved in the aqueous phase, with each isotope dissolved according to its abundance in the ore. Additionally, assume that a very small fraction $\xi$ ($\xi \ll 1$) of $N_{238}$ nuclei of the main uranium-238 isotope that pass to the aqueous medium is activated in the fissure formation by the $e^-$-catalytic mechanism and converted to *isu*-state *β*-protactinium nuclei. Without this activation, the activity level of $N_{238}$ nuclei of the atoms of uranium-238 isotope in the aqueous medium, $\eta_{238} = k_{238}N_{238}$, was equal to the activity level $\eta_{234} = k_{234}N_{234}$ for the $N_{234}$ nuclei of uranium-234 isotope that passed to the aqueous medium. Section 5.3 implies that in the initiated radioactive decay, the effective decay rate constant of $^{238}U$ nuclei, $\tilde{k}_{238}$, for a relatively small number $\xi$ of $N_{238}$ nuclei in the aqueous medium can dramatically change. It is wise to use the above in considering the simplified decay of uranium-238 and uranium-234 isotopes. In this case, the decay of "intermediate" thorium-234 and protactinium-234m isotopes with short lifetimes, which are also involved in the radioactive uranium/radium series, is taken out of consideration. The balance equations for the numbers of $N_{238}$ and $N_{234}$ nuclei at a steady-state concentration of uranium-234 isotope in the aqueous medium can be written as

$$\frac{dN_{238}}{dt} = -k_{238}(1-\xi)N_{238} - \xi\tilde{k}_{238}N_{238} = -k_{238}^{eff}N_{238}, \qquad (69)$$

$$\frac{dN_{234}}{dt} = -k_{234}N_{234} + k_{238}(1-\xi)N_{238} + \xi\tilde{k}_{238}N_{238} = 0. \qquad (70)$$

Here,

$$k_{238}^{eff} = k_{238}\left[1 + \xi\left(\frac{\tilde{k}_{238}}{k_{238}} - 1\right)\right] \qquad (71)$$

is the effective rate constant for the decay of the uranium-238 isotope when the radioactive decay of the fraction $\xi$ of uranium-238 nuclei is initiated by external factors and characterized by the decay rate constant $\tilde{k}_{238}$. Equations (69) and (70) yield the desired formula for the ratio of activity levels of uranium-234 and uranium-238 isotopes in open systems in which the initiated accelerated decay of uranium-238 is effectuated:

$$\eta_{234}/\eta_{238} = \frac{k_{234} N_{234}}{k_{238} N_{238}} = {}^{234}U/{}^{238}U \text{ AR} = 1 + \xi\left(\frac{\tilde{k}_{238}}{k_{238}} - 1\right). \quad (72)$$

In this case, the apparent higher activity level of the uranium-234 isotope cannot be attributed to the fact that the groundwater is directly enriched with $^{234}U$ nuclides because its release to the aquatic medium is easier due to the decay of the main $^{238}U$ isotope, as is usually assumed [25, 135]. The increased content of $^{234}U$ nuclei in the aqueous medium is a result of the decay of $^{238}U$ nuclei initiated by the formation of cracks and fissures, which produces $\beta$-protactinium nuclei by the $e^-$-catalytic mechanism; their release to the aqueous medium, and their subsequent decay along the chain of the radioactive uranium/radium series. The reference value of $\tilde{k}_{238}/k_{238} \sim 10^9$ estimated in [97], showing a possible increase by 9 orders of magnitude of the decay rate constant for the $^{238}U$ nuclei in the laser ablation, implies that for the ratios $^{234}U/{}^{238}U \text{ AR} \sim 5\text{-}10$, characteristic for the system studied in [25], to take place the fraction $\xi$ of activated $^{238}_{91}Pa_{isu}$ nuclei relative to $^{238}U$ nuclei in the aqueous media must be $\sim (0.5\text{-}1)\cdot 10^{-8}$.

The acceleration of the radioactive decay of uranium-238 during the mechanochemical activation of uranium-containing ore allows us to consider the possibility of developing new sources of nuclear power based on uranium ores that differ from their conventional application, which requires enrichment with uranium-235 in order to initiate the chain process. We could create thermal power using the existing wastes of uranium production that contain uranium-238 in order to activate it under conditions of the mechanochemical processing of these wastes in aqueous media to yield $^{238}_{91}Pa_{isu}$, the half-life of which is several years. In 2008, on the territory of Kyrgyzstan in particular, where three of the former Soviet Union's eleven large enterprises for uranium processing are located (including the active Kara-Baltinskii ore mining enterprise), there were tailing dumps and piles with more than 5 10$^7$ m$^3$ of waste from uranium production that had a total radioactivity of nearly 10$^5$ Ci [136]. Wastes of this type could become a basic and virtually inexhaustible raw material for the considered new, environmentally friendly, source of nuclear power. Further experimental studies of the mechanochemical activation of uranium-238 might show how the development of this type of technology with the controlled delivery of fuel to a reactor in the form of an aqueous system containing $^{238}_{91}Pa_{isu}$ - $^{238}_{92}U$ nuclei could be desirable.

### 5.4.2. *Radioactivity initiated in a glow discharge* (*Irina Savvatimova phenomenon*)

In connection with the abovementioned nearly 100-year-old dispute between Wendt and Rutherford (see Introduction) on the possible formation of helium upon an electrical explosion of tungsten wire in a vacuum bulb, let us discuss the results of [99], devoted to an analysis of experimental data of Irina Savvatimova with co-workers [31, 32] on the initiated decay of W-nuclei in the near-surface layers of a tungsten cathode (foil) upon glow discharge. Note first of all that all of the five stable tungsten isotopes, $^{180}_{74}W$, $^{182}_{74}W$, $^{183}_{74}W$, $^{184}_{74}W$, and $^{186}_{74}W$, potentially $\alpha$-radioactive isotopes capable of $\alpha$-decay:

$$^{A}_{74}W \rightarrow {}^{A-4}_{72}Hf + {}^{4}_{2}He + Q_A, \quad (73)$$

where heat release values upon the radioactive $\alpha$-decay of tungsten nuclei with mass numbers $A$ equal to 180, 182, 183, 184, and 186, are 2.52, 1.77, 1.68, 1.66, and 1.12 MeV, respectively. Based entirely on energy considerations, it was possible to anticipate $\alpha$-decays with the formation of several α particles for the indicated stable isotopes of tungsten, including decay with the formation of nine α particles for the tungsten-180 isotope.

In investigating the processes that occur under the conditions of glow discharge in a deuterium-containing gaseous medium in the near-surface layer of a tungsten cathode, the formation of new elements (particularly erbium, ytterbium, and hafnium isotopes not contained in the initial cathode) was detected after plasma treatment lasting 4 to 7 h [31, 32], and the formation of these elements was observed more than 5 months after initiating the tungsten cathode in plasma.

Following the example of [99], let us explain how we can understand the experimental results obtained in [31, 32] using our concepts of initiating nuclear transformations under the conditions of low-temperature plasma by introducing a hypothetical β-dineutron. Two types of processes that initiate α-decay are possible in such cases. First of all, the fairly long-lived $^2n_{isu}$ particles can diffuse along grain boundaries deep into a cathode and interact with tungsten nuclei in its near-surface layers. Excited $^{A+2}_{74}W^*$ nuclei can thus appear in the initial stage upon interaction between $^2n_{isu}$ nuclei and $^A_{74}W$ isotopes. In addition to the overall excitation energy (denoted by "*") of 10 MeV with respect to the main state of the specified nuclei in the considered cases, the nuclear matter of such nuclei can partially form by fusing with $^2n_{isu}$ in the unbalanced *isu*-state with a loss of stability in the nuclear bulk; this leads to subsequent transformation with the emission of α particles and daughter isotopes. In contrast to the nuclear reactions that occur upon the collision of reagents in the gas phase, it should be emphasized that the energy factor alone is sufficient for the considered nuclear transformations in the area of grain boundaries of a solid metal phase, due to the possible effects of the medium (without spin and parity matching conditions for the colliding and final nuclei). The formation of the stable erbium, ytterbium, lutetium, and hafnium isotopes (i.e., the products of the initiated decay of different tungsten isotopes observed in [31, 32]) can be associated directly with this type of process:

$$^{182}_{74}W + {}^2n_{isu} \rightarrow {}^{168}_{68}Er + 4\,{}^4_2He + 4e^- + 4\tilde{\nu} + Q\,(18.19\ MeV)$$

$$^{180}_{74}W + {}^2n_{isu} \rightarrow {}^{170}_{68}Er + 3\,{}^4_2He + Q\,(16.34\ MeV)$$

$$^{184}_{74}W + {}^2n_{isu} \rightarrow {}^{170}_{68}Er + 4\,{}^4_2He + 2e^- + 2\tilde{\nu} + Q\,(17.85\ MeV)$$

$$^{182}_{74}W + {}^2n_{isu} \rightarrow {}^{171}_{70}Yb + {}^9_4Be + {}^4_2He + 2e^- + 2\tilde{\nu} + Q\,(10.43\ MeV)$$

$$^{182}_{74}W + {}^2n_{isu} \rightarrow {}^{172}_{70}Yb + 3\,{}^4_2He + 2e^- + 2\tilde{\nu} + Q\,(16.88\ MeV)$$

$$^{184}_{74}W + {}^2n_{isu} \rightarrow {}^{173}_{70}Yb + {}^9_4Be + {}^4_2He + 2e^- + 2\tilde{\nu} + Q\,(11.22\ MeV)$$

$$^{184}_{74}W + {}^2n_{isu} \rightarrow {}^{174}_{70}Yb + 3\,{}^4_2He + 2e^- + 2\tilde{\nu} + Q\,(17.11\ MeV)$$

$$^{182}_{74}W + {}^2n_{isu} \rightarrow {}^{175}_{71}Lu + {}^9_4Be + e^- + \tilde{\nu} + Q\,(8.72\ MeV)$$

$$^{182}_{74}W + {}^2n_{isu} \rightarrow {}^{176}_{70}Yb + 2\,{}^4_2He + Q\,(13.53\ MeV)$$

$$^{184}_{74}W + {}^2n_{isu} \rightarrow {}^{177}_{72}Hf + {}^9_4Be + 2e^- + 2\tilde{\nu} + Q\,(8.97\ MeV)$$

$$^{186}_{74}W + {}^2n_{isu} \rightarrow {}^{180}_{72}Hf + 2\,{}^4_2He + 2e^- + 2\tilde{\nu} + Q\,(15.56\ MeV)$$

The formation of radioactive ytterbium and hafnium isotopes under the conditions of plasma initiation in a tungsten cathode, observed in [31, 32] immediately after the end of plasma treatment of tungsten cathodes, can also be associated with this type of process:

$$^{180}_{74}W + {}^2n_{isu} \rightarrow {}^{169}_{70}Yb + {}^9_4Be + {}^4_2He + 2e^- + 2\tilde{\nu} + Q\,(10.09\ MeV)$$

$$^{184}_{74}W + {}^2n_{isu} \rightarrow {}^{169}_{70}Yb + {}^9_4Be + 2\,{}^4_2He + 4e^- + 4\tilde{\nu} + Q\,(11.6\ MeV)$$

$$^{182}_{74}W + {}^2n_{isu} \rightarrow {}^{171m}_{70}Yb + {}^9_4Be + {}^4_2He + 2e^- + 2\tilde{\nu} + Q\,(10.34\ MeV)$$

$$^{182}_{74}W + {}^2n_{isu} \rightarrow {}^{172}_{72}Hf + 3\,{}^4_2He + 4e^- + 4\tilde{\nu} + Q\,(14.01\ MeV)$$

$${}^{184}_{74}W + {}^2n_{isu} \to {}^{178}_{70}Yb + 2{}^4_2He + Q\,(12.28\ MeV)$$

$${}^{182}_{74}W + {}^2n_{isu} \to {}^{180}_{70}Yb + {}^4_2He + 2e^- + 2\tilde{\nu} + Q\,(6.87\ MeV)$$

$${}^{182}_{74}W + {}^2n_{isu} \to {}^{180m}_{72}Hf + {}^4_2He + Q\,(10.26\ MeV)$$

The emergence and subsequent increase in the peak with a mass of 9, registered in the mass spectra of products in addition to major peaks with masses ranging from 168 to 180, is considered in the schematic illustration of these processes. We should also explain that the missing product with a basic mass of 4 corresponding to helium nuclei in the mass spectra obtained in [31, 32] was due to the extremely low solubility of helium in tungsten [137] and its high diffusion coefficients in the area of the foil's intergranular boundaries. It is obvious that in order for such processes to proceed, the lifetime of the ${}^2n_{isu}$ nuclei must be long enough for the diffusive permeation of these neutral nuclei into the near-surface areas of the foil along the grain boundaries. This agrees with the conclusion in [36] that the time must be no less than several tens of minutes when synthesizing tritium under conditions of the laser ablation of metals in heavy water.

Another way of initiating the α-decay of tungsten isotopes under the conditions of glow discharge is also possible during $e^-$-catalysis (see Section 5.3.2), when electrons with kinetic energy $Ee \sim 3$-$5$ eV interact directly with stable isotopes of tungsten. Possible examples of such processes are given below:

$${}^{183}_{74}W + e^- \to {}^{183}_{73}Ta_{isu} + \nu \to {}^{171m}_{70}Yb + 3{}^4_2He + 3e^- + 3\tilde{\nu} + \nu + Q\,(5.58\ \text{МэВ})$$

$${}^{184}_{74}W + e^- \to {}^{184}_{73}Ta_{isu} + \nu \to {}^{172}_{72}Hf + 3{}^4_2He + 5e^- + 5\tilde{\nu} + \nu + Q\,(3.41\ \text{МэВ})$$

$${}^{184}_{74}W + e^- \to {}^{184}_{73}Ta_{isu} + \nu \to {}^{168}_{68}Er + 4{}^4_2He + 3e^- + 3\tilde{\nu} + \nu + Q\,(7.59\ \text{МэВ})$$

$${}^{186}_{74}W + e^- \to {}^{186}_{73}Ta_{isu} + \nu \to {}^{174}_{70}Yb + 3{}^4_2He + 3e^- + 3\tilde{\nu} + \nu + Q\,(7.17\ \text{МэВ})$$

$${}^{183}_{74}W + e^- \to {}^{183}_{73}Ta_{isu} + \nu \to {}^{175}_{71}Lu + 2{}^4_2He + 2e^- + 2\tilde{\nu} + \nu + Q\,(3.96\ \text{МэВ})$$

As was shown above, the formation of products from the nuclear decay of tungsten isotopes was observed more than 5 months after their initiation in plasma over 4 to 7 h. We may assume that the above processes with participation of ${}^2n_{isu}$ nuclei proceeds faster, while major long-term changes in the isotope composition of the products from the initiated decay of tungsten isotopes are determined by $e^-$-catalysis, i.e., by the decay of the long-lived ${}^A_{73}Ta_{isu}$ compound nuclei with the unstable state of nuclear matter. As follows from the experimental data, the half-life period of such nuclei (i.e., the time necessary for the rearrangement of nuclear matter in accordance with possible decay through different channels) can be a month or more.

Based on the above data, we may conclude that Wendt [16] was probably right in his dispute with Rutherford. Helium could in fact appear upon the electric explosion of a tungsten wire. Wendt and Irion's experiment [14] should be repeated to register not only helium but also the heavy elements found in the experiments [31, 32]. As concerns Rutherford's experiment [15], in which the formation of helium was not observed after prolonged irradiation of tungsten targets by electrons with a kinetic energy of 100 keV, the considered mechanism of the initiated decay of tungsten could itself be the reason for such a result. Almost all losses of the kinetic energy of electrons in Rutherford's experiment were of an electromagnetic nature (e.g., ionization and electromagnetic radiation), and could also have been partially due to the formation of defects. The channel of losses through weak interaction was basically excluded under these conditions.

The data allow us to conclude that the nuclear decay of initially nonradioactive tungsten isotopes with the formation of lighter elements (erbium, lutetium, ytterbium, and hafnium), initiated under the conditions of low-temperature plasma (glow discharge), can be considered as a new type of artificial radioactivity that differs from the induced radioactivity initiated by nuclear reactions (e.g., with alpha particles or neutrons that lead to the formation of radioisotopes). It should be remembered that stable isotopes of many nuclei, from neodymium to bismuth and including the tantalum-181 isotope in particular, for which initiated decay processes similar to the ones described were also observed in [31, 32], are potentially α-radioactive in the same sense as tungsten isotopes.

### 5.4.3. Nuclear-Chemical processes in Andrea Rossi's E-Cat reactor

The above concept of initiating low-energy nuclear-chemical reactions by the mechanisms of nuclei fusion and $e^-$-catalysis can be used to physically interpret the results of testing A. Rossi's energy E-Cat reactors as well [39, 94]. Let us briefly discuss the results of testing the E-Cat working chamber of the Rossi's reactor, presented by a group of international experts [40]. The working chamber was a hollow ceramic cylinder 2 cm in diameter and 20 cm long, into which the researchers loaded a fuel: about 0.9 g of finely dispersed nickel powder with all stable isotopes present ($^{58}_{28}Ni$, $^{60}_{28}Ni$, $^{61}_{28}Ni$, $^{62}_{28}Ni$, and $^{64}_{28}Ni$ of 67, 26.3, 1.9, 3.9 and 1 %, respectively), and 0.1 g of LiAlH$_4$ powder ($^6_3Li$ and $^7_3Li$ isotopes of 8.6 and 91.4%, respectively). The cylinder was sealed and then heated. The tests were carried out for 32 days at chamber heating temperatures up to 1260ºC (first half of the time) and 1400ºC (second half of the time). The energy released in the tests was measured using the value of the heat flux produced by the chamber. In the tests, the overall excess energy of 1.5 MWh was produced, corresponding to the chamber efficiency higher than 3.5. The researchers recorded changes in the isotopic composition of the main fuel components (nickel, lithium), for which the initial composition of stable elements was close to the tabulated natural composition. After the tests, the isotopic composition of the recorded components was dramatically changed: almost all nickel powder, more than 98%, was a nickel-62 isotope (about 4% initially); the fraction of lithium-7 dropped to about 8% and lithium-6 jumped to about 92%. The isotope abundances of the initial fuel and final "ash" in the tests are listed in Table 1 [40].

Table 1. The isotope abundances of the initial "fuel" and "ash" after the tests [40].

| Ion | Fuel Counts in peak | Measured abundance [%] | Ash Counts in peak | Measured abundance [%] | Natural abundance [%] |
|---|---|---|---|---|---|
| $^6$Li$^+$ | 15804 | 8.6 | 569302 | 92.1 | 7.5 |
| $^7$Li$^+$ | 168919 | 91.4 | 48687 | 7.9 | 92.5 |
| $^{58}$Ni$^+$ | 93392 | 67 | 1128 | 0.8 | 68.1 |
| $^{60}$Ni$^+$ | 36690 | 26.3 | 635 | 0.5 | 26.2 |
| $^{61}$Ni$^+$ | 2606 | 1.9 | ~0 | 0 | 1.8 |
| $^{62}$Ni$^+$ | 5379 | 3.9 | 133272 | 98.7 | 3.6 |
| $^{64}$Ni$^+$ | 1331 | 1 | ~0 | 0 | 0.9 |

According to the above concept, the recorded change in the isotopic composition of main fuel components, nickel and lithium, in the presence of the hydrogen given off in the decomposition of LiAlH$_4$ at the above temperatures may be caused by the formation of a protium-containing plasma in the reaction volume and the occurrence of neutral metastable nuclei $^1n_{isu}$. Like neutrons, these neutral nuclei can interact with the nuclei of elements constituting the fuel, accounting for the changes occurring in its elemental and isotopic composition, which is accompanied by the corresponding energy release:

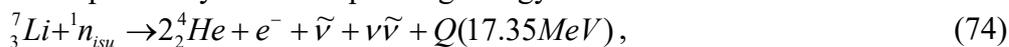
$$^7_3Li + {}^1n_{isu} \rightarrow 2\,{}^4_2He + e^- + \tilde{\nu} + \nu\tilde{\nu} + Q(17.35 MeV), \tag{74}$$

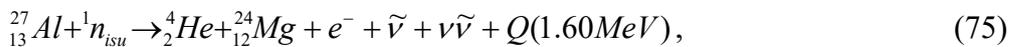
$$^{27}_{13}Al + {}^1n_{isu} \rightarrow {}^4_2He + {}^{24}_{12}Mg + e^- + \tilde{\nu} + \nu\tilde{\nu} + Q(1.60 MeV), \tag{75}$$

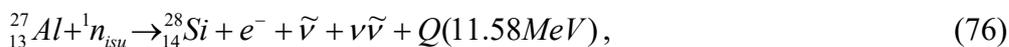
$$^{27}_{13}Al + {}^1n_{isu} \rightarrow {}^{28}_{14}Si + e^- + \tilde{\nu} + \nu\tilde{\nu} + Q(11.58 MeV), \tag{76}$$

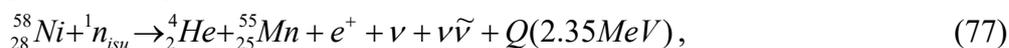
$$^{58}_{28}Ni + {}^1n_{isu} \rightarrow {}^4_2He + {}^{55}_{25}Mn + e^+ + \nu + \nu\tilde{\nu} + Q(2.35 MeV), \tag{77}$$

$$^{60}_{28}Ni + ^{1}n_{isu} \rightarrow ^{4}_{2}He + ^{57}_{26}Fe + \nu\widetilde{\nu} + Q(0.47 MeV), \tag{78}$$

$$^{61}_{28}Ni + ^{1}n_{isu} \rightarrow ^{4}_{2}He + ^{58}_{26}Fe + \nu\widetilde{\nu} + Q(2.80 MeV), \tag{79}$$

$$^{62}_{28}Ni + ^{1}n_{isu} \rightarrow ^{4}_{2}He + ^{59}_{27}Co + e^{-} + \widetilde{\nu} + \nu\widetilde{\nu} + Q(0.34 MeV), \tag{80}$$

$$^{64}_{28}Ni + ^{1}n_{isu} \rightarrow ^{65}_{29}Cu + e^{-} + \widetilde{\nu} + \nu\widetilde{\nu} + Q(7.45 MeV), \tag{81}$$

The above list of reactions implies that the specific (per component unit mass) energy release is the highest for the lithium-7 nuclei. At the same time, when the mass fraction of the lithium-7 isotope in the system is low, the total contribution to the heat release of the nuclear reactions of $^{1}n_{isu}$ nuclei with all other fuel elements, such as aluminum and nickel isotopes, can become dominating. The almost complete disappearance of isotopes $^{7}_{3}Li$ and $^{58}_{28}Ni$ in the ashes, which was recorded after the chamber was tested for more than a month, implies that the values of rate constants are rather high not only for the processes (74) and (77), but also for the other nuclear processes in which the new chemical elements are formed.

To understand the specific mechanisms accounting for the major changes in the fuel composition during the E-Cat operation, including the almost complete exhaustion of the lithium-7 isotope and the dominant growth of the nickel-62 isotope in the ash, it is necessary to consider the other nuclear reactions, which can also change the isotopic composition of the initial nickel. In these reactions, the energy carried away by the formed neutrinos and antineutrinos can noticeably reduce their heat releases, as compared to the above reactions:

$$^{58}_{28}Ni + ^{1}n_{isu} \rightarrow ^{59}_{28}Ni + \nu\widetilde{\nu} + Q(8.22 MeV), \quad T_{1/2}(^{59}_{28}Ni) = 7.6 \cdot 10^{4} yr, \tag{82}$$

$$^{60}_{28}Ni + ^{1}n_{isu} \rightarrow ^{61}_{28}Ni + \nu\widetilde{\nu} + Q(7.04 MeV),, \tag{83}$$

$$^{61}_{28}Ni + ^{1}n_{isu} \rightarrow ^{62}_{28}Ni + \nu\widetilde{\nu} + Q(9.81 MeV),, \tag{84}$$

$$^{62}_{28}Ni + ^{1}n_{isu} \rightarrow ^{63}_{28}Ni + \nu\widetilde{\nu} + Q(6.05 MeV), \quad T_{1/2}(^{63}_{28}Ni) = 100.1 yr, \tag{85}$$

$$^{64}_{28}Ni + ^{1}n_{isu} \rightarrow ^{65}_{28}Ni + \nu\widetilde{\nu} + Q(5.32 MeV), \quad T_{1/2}(^{65}_{28}Ni) = 2.52 h, \tag{86}$$

The long half-life of the $^{59}_{28}Ni$ isotope practically excludes the process in which the other nickel isotopes decayed in the tests are "replenished" with the $^{58}_{28}Ni$ isotope, whose fraction is twice the fractions of the other nickel isotopes. Therefore, the almost complete absence of the $^{60}_{28}Ni$ isotope in the ash should be attributed to the processes (78) and (83). It can also be assumed that the processes (79) and (84) account for the disappearance of the $^{61}_{28}Ni$ isotope in the ash; meanwhile, the process (84) brings the isotope $^{62}_{28}Ni$ to the ash, providing its prevailing abundance among the other nickel isotopes in the ash. The additional contribution to this prevailing abundance is made by the "low" value of rate constant for the decline of the $^{62}_{28}Ni$ isotope in the reaction (80), which describes the formation of cobalt with a low energy release in the process. It is also important to note that the long half-life of the $^{63}_{28}Ni$ isotope practically prevents from increasing the abundance of the isotope $^{64}_{28}Ni$ in the ash, and the process (81) isotope in the initial nickel to the copper-65 isotope.

Admittedly, the above arguments can only qualitatively explain the ash composition recorded in the test. In this case, of high interest could be a comparative study of the elemental and isotopic composition of the ash and initial fuel by inductively coupled plasma mass spectrometry. Here, it is important to study the changes in isotope ratios for various elements in the ash and initial fuel, primarily for the base element (nickel), as well as for the element formed in the processes (75) - (81), such as magnesium, silicon, manganese, iron, cobalt, and copper.

VI. RADIOACTIVE DECAY AS A FORCED NUCLEAR CHEMICAL PROCESS

## 6.1. The basic hypothesis

According to the main dynamic law of radioactive decay, the number $N(t)$ of statistically independent unstable nuclei decaying in an investigated macroscopic object at moment in time $t$ is proportional to the number of available unstable nuclei:

$$dN(t) = -kN(t)dt, \quad N(t) = N_0 \exp(-kt); \quad T_{1/2} = \ln 2 / k. \qquad (87)$$

where $N_0$ is the number of radioactive nuclei at an arbitrary moment in time ($t = 0$), $k$ is the decay kinetic constant, and $T_{1/2}$ is the half-life period. We therefore assume that the radioactive decay of each nucleus, which occurs due to its initial instability, proceeds spontaneously at a moment in time determined by unidentified internuclear reasons. Soddy noted as far back as 1949 that this simple law of decay, which is the same for all radioactive elements and is of a probabilistic nature, is basically inexplicable. "It can be imagined in the form of the spirit of destruction, who randomly disintegrates a certain amount of existing atoms at each moment of time, without taking care of the selection of those of them which are close to disintegration" [138]. A question is being asked here: Is there any material reason for the initiation of possible changes in the state of a nucleus? If there is, what is the difference between a nucleus ready to decay and a nucleus in which the processes leading to its disintegration have not yet begun? Such questions are normally not discussed. In theoretical studies of the radioactive decay processes, attention is focused on quantum-mechanical calculations of the probabilities of the α- and β-decays per time unit. At the same time, it is considered that the dynamic state of nuclear matter in all nuclei of a specific radioactive element is the same, and the process of radioactive transmutation with changes in the composition of the initial nucleus occurs suddenly at a moment in time determined by hidden features of nuclear matter dynamics.

Nevertheless, some attempts to understand whether radioactive decay precede any changes in the state of nuclear matter, and whether external influences can influence the rate of radioactive decay, have been made. Here we should note Yefimov's work, in which the radioactive decay was interpreted as the decay of the quasi-stationary or unstable state of a nucleus. In the model conditions considered in [139], such a state forms upon the stationary resonant scattering of certain particles on a nucleus. A nucleus prepared in this manner begins to decay at a certain moment of time ($t = 0$) when the external source of particles is switched off, and the particles that accumulate in the nucleus start to fly away. A nonstationary Schrödinger equation was used to describe this kind of process. The obtained relationships (Eqs. (6.65) and (6.66) in [139]) describing the temporal changes in the probabilities of detecting decaying particles at some distance from the source, reflect with high accuracy radioactive decay law (87), in which $k = \Gamma/2\hbar$, where $\Gamma$ and $\hbar$ are the resonance width that characterizes a quasi-stationary state and the Planck constant, respectively. If we try to correlate the proposed model with the formation and decay of real radioactive nuclei, a number of questions arise about the nature of the generated quasi-stationary state and the methods of its pumping under such model consideration. However, the very idea of nuclei decaying due to the formation of metastable decaying states inside them at certain moments in time would seem to be constructive. Below, we will propose a variant of developing this idea on the basis of the results presented in Sections 5.2 and 5.3, and general conclusions will be made on the radioactive decay of nuclei as a forced nuclear-chemical process.

First of all, the mere fact of realizing the possibility of the influence of low-energy effects on the rate of radioactive decay of nuclei gives the reason to doubt that the dynamic state of the nuclear matter of all the nuclei of a given radioactive element is the same on the eve of the decay, and the process of radioactive transformation takes place suddenly. In addition, the above data set leads us to believe that the resulting interaction between such nuclei and one of the $e_i^-$ electrons of an atom's internal electron shells precedes the spontaneous α- and β-decays of $_Z^A N$ nuclei (where $Z$ and $A$ are the order number and mass number of the $N$ nucleus, respectively).

As a result of such interaction, a neutrino $\nu$ is emitted and a nucleus $_{Z-1}^{A}N_{isu}$ in the metastable isu-state of inner shake up is formed:

$$_{Z}^{A}N + e_i^- \rightarrow\, _{Z-1}^{A}N_{isu} + \nu. \quad (43a)$$

The characteristic time of the occurrence of the corresponding fluctuation (0.3 eV and more) can be quite long: there is no extraneous initiation in this case, since it takes place under the conditions of low temperature plasma. Furthermore, the characteristic lifetime of the *β*-nucleus $_{Z-1}^{A}N_{isu}$, which is determined by weak interaction processes, can be long as well.

If the initial nucleus $_{Z}^{A}N$ is α-radioactive, it is the nucleus $_{Z-1}^{A}N_{isu}$ that undergoes α-decay:

$$_{Z}^{A}N + e_i^- \rightarrow\, _{Z-1}^{A}N_{isu} + \nu \rightarrow\, _{2}^{4}He +\, _{Z-2}^{A-4}N + e_i^- + \nu + \widetilde{\nu}, \quad (88)$$

and the total α-decay process after the emission of the $e_i^-$ electron in the atom's inner electron shell takes the schematic form

$$_{Z}^{A}N \rightarrow\, _{2}^{4}He +\, _{Z-2}^{A-4}N + \nu + \widetilde{\nu}. \quad (89)$$

It is based on such concepts of $e^-$-catalysis, we understood (see Section 5.3.2) the dynamics of the protracted formation (over more than a year) of that is observed when we monitor the activity of a specially-prepared sample containing nanoparticles formed upon the laser ablation of beryllium in an aqueous solution of $UO_2Cl_2$ for 1 h and then isolated by centrifugation. By analogy, the production of the erbium, ytterbium, and hafnium isotopes as decay products of isotopes $_{74}^{A}W$, which continues 5 months after the initiation of a tungsten cathode in plasma (see section 5.2 and [31, 32]), is associated with the decay of tantalum $_{73}^{A}Ta_{isu}$ isotopes in the unbalanced *isu*-state.

If the initial nucleus $_{Z}^{A}N$ is prone to β-decay, the decay of the $_{Z-1}^{A}N_{isu}$ nucleus is described as

$$_{Z-1}^{A}N_{isu} \rightarrow\, _{Z+1}^{A}N + e_i^- + e^- + 2\widetilde{\nu}, \quad (90)$$

and the overall *β*-decay process can be described by the scheme:

$$_{Z}^{A}N \rightarrow\, _{Z+1}^{A}N + e^- + \nu + 2\widetilde{\nu}. \quad (91)$$

The conclusion that the *α*- and *β*-decays are associated with destabilization of the nuclear matter inside a $_{Z-1}^{A}N_{isu}$ nucleus formed as a result of the interaction between the electronic subsystem of the initial radioactive atom and the nuclear matter via the weak interaction
channel is perhaps more general. The cluster radioactivity, a phenomenon of the spontaneous emission of nuclear fragments (clusters) heavier than an *α*-particle, could also be due to the initiating role of the electronic subsystem of a nucleus. This was observed for the first time in 1984 [140]. In studying the α-radioactivity of $_{88}^{223}Ra$ nuclei [140], it was found that they sometimes emit $_{6}^{14}C$ nuclei instead of *α*-particles (with a probability lower by ten orders of magnitude), and cluster decay is observed:

$$_{88}^{223}Ra \rightarrow\, _{6}^{14}C +\, _{82}^{209}Pb + Q(31.84 MeV). \quad (92)$$

At present, it was experimentally identified more than 20 nuclei ranging from $_{56}^{114}Ba$ to $_{96}^{242}Cm$ that sometimes emit nuclei heavier than an *α*-particle during radioactive decay. Among these emitted particles were found nuclei: $_{6}^{14}C$, $_{8}^{20}O$, $_{10}^{24}Ne$, $_{10}^{26}Ne$, $_{12}^{28}Mg$, $_{12}^{30}Mg$, $_{14}^{32}Si$, and $_{14}^{34}Si$ [141]. However, the probability of such processes occurring is extremely low: in fact, 10 to 17 orders of magnitude lower than the probability of the emission of α-particles by the same nuclei.

It is normal to associate cluster decay with the initiation of nuclear matter dynamics upon the formation of nuclei in the *isu*-state, which leads to the conformational rearrangement of their structure on different internuclear scales, and to the formation of clusters of mononucleon and multiquark natures, the concepts of which were introduced in considering models of cluster decay [142]. The interaction between nuclear matter and an atom's electronic subsystem via the weak interaction channel that is discussed in this work should probably be considered as initiating the rearrangement needed for the formation of $_{Z-1}^{A}N_{isu}$ nuclei in the metastable *isu*-state of the inner shakeup.

In this work, we did not intend to go beyond the scope of the common phenomenological approach to establishing the role of the electronic factor in initiating radioactive decay processes. Nevertheless, we did find a number of problems whose solving might allow us to consider the α- and β-decays of radioactive nuclei and their cluster decay as phenomena controllable by external factors. First of all, model calculations of the dynamics of the considered stages of radioactive decay (i.e., the formation of the metastable $_{Z-1}^{A}N_{isu}$ nucleus in its reactive isu-state and its decay dynamics) were of interest. Note in connection with the latter that the concept of a tunnel mechanism of cluster decay with the emission of nuclei heavier than α-particles can lead to doubt. It therefore makes sense to analyze Kramers' activation mechanism in its discrete version (wandering over the energy levels until a certain limit is reached) [126] as a possible alternative for calculating the probability of the decay of an *isu*-state with the accumulated energy of a decaying bond, which is determined by the type of an emitted particle.

6.2. Radioactive decay of nuclei and the formation of elements in the Universe

Radioactive decay plays an important role in the formation of elements in the Universe, especially in the formation of elements heavier than iron. According to the concepts of the standard model of nucleosynthesis [80, 143], the first stage of the synthesis of elements in the Universe is connected with the first three minutes after the Big Bang, when the first group of light elements was formed: H, D, $^3$He, $^4$He, and $^7$Li. Variations with the additional synthesis of $^9$Be and $^{11}$B are sometimes debated. The second stage of the synthesis of elements, which began with the appearance of the first stars and continues today, is the nuclear reactions of elements in stars that lead to the formation of elements of the periodic system up to the iron group elements [143, 144]. The synthesis of heavier nuclei after that becomes impossible, since the nuclei of the iron group have the highest bond energy per one nucleon (8 MeV). The third stage of the synthesis of elements is connected with the explosions of supernova stars. The elements from Si to Fe are generated upon the explosions of Type I supernovas, where the fraction of elements from O to Mg changes only slightly, and the elements whose weights exceed that of the iron nucleus are formed only in the explosions of heavier Type II supernovae, and in the stages preceding such explosions. The key role in such processes is usually attributed to (*n*, γ)-reactions and *β*-decay. Meanwhile, neutrons form in the $_{10}^{22}$Ne(α,n)$_{12}^{25}$Mg and $_{6}^{13}$C(α,n)$_{8}^{16}$O reactions. If the initial nucleus after the capture of a neutron and an increase in the mass number by unity basically undergoes β-decay (at low neutron flux densities or the relatively high β-decay rates of the nuclei that form upon the capture of a neutron), the process associated with the increase in the mass of the nuclei is defined as an *s*-process (a slow mass increase). At high neutron flux densities or relatively low β-decay rates, the decay of a nucleus formed upon the capture of a neutron, when the nucleus has enough time to capture several neutrons prior to its β-decay, is defined as an *r*-process (a rapid mass increase). The *β*-decay process thus acts as a factor separating the nucleosynthesis processes. There is also a less studied third process: the so-called *p*-process [80], during which neutron-depleted nuclei form. These nuclei, whose abundance is approximately two orders of magnitude lower than that of similar nuclei produced by the neutron capture route, are referred to as by-passed or *p*-nuclei. It is believed [145] that the (γ, *n*) and (*p*, *n*) reactions play a crucial role in *r*-processes; however, radioactive decay processes can contribute appreciably to the synthesis of *p*-nuclei. However, we should remember that atoms participating in the *s*-, *r*-, and *p*-processes become multiply ionized, since these processes take place under the conditions of ultrahigh temperatures. This can substantially affect not only the probabilities of the above *β*-decay processes but the bound-state *β⁻*-decay channels as well [111, 112].

Nevertheless, all of the existing nucleosynthesis models are generally based on the concept of the synthesis of elements during nuclear collisions, which requires the kinetic energy that can be obtained under the conditions of high-temperature plasma in the interiors of stars or in the explosions of supernovas. However, such concepts do not allow us to understand fully a number of anomalies revealed in analyzing the abundance of some light elements in our Galaxy and in intergalactic space. Some of problems connected with the synthesis of heavier nuclei, including the formation of heavy elements during supernova star explosions also remain controversial. As is shown below, some of the problems related to the synthesis of light elements and the synthesis of heavy nuclei can be solved using the concept of nuclear transformations initiated by interaction between an atom's electronic and nuclear subsystems, including radioactive decay processes.

6.2.1. Problems Related to Nucleosynthesis of Light Nuclei

Let us first consider the current problems related to the nucleosynthesis of light nuclei. According to [143], there is currently no definitive estimate of the $D/H$ concentration ratios of deuterium and hydrogen, respectively, registered in different parts of the Universe. It is especially difficult to understand the discrepancies between the values of this ratio measured in the local interstellar medium and determined from the absorption lines of different quasars, which can be as great as an order of magnitude.

It is no less difficult to interpret the data on the dynamics of temporary changes in the relative fraction of other light elements, and on the dynamics of changes in the isotope composition of such elements in the stellar atmospheres accessible for analysis, especially the solar atmosphere. One of the Sun's mysteries [146] in particular is associated with the beryllium-7 radioactive isotope, whose abundance in the solar atmosphere exceeds the content of the lithium-7 stable isotope; the latter is also formed from beryllium-7 upon *K*-capture with a half-life period of 53 days. Content ratio $\eta = {}^{7}_{4}Be/{}^{7}_{3}Li$ of the specified isotopes changes, growing with an increase in solar activity and falling somewhat at the solar activity minimum while remaining higher than unity [146]; this can be explained if beryllium-7 is generated continuously in the solar atmosphere. In addition, the isotope formed in the solar atmosphere contributes appreciably to the radioactivity of the air near the Earth's surface, as was shown in [147, 148]. Variations in the content in the atmosphere are thus connected to solar activity, and display a characteristic seasonal trend and latitudinal dependence.

It is usually assumed that these isotopes are formed in the areas of powerful solar flares as a result of nuclear processes that occur between the protons accelerated in these areas and carbon, nitrogen, oxygen, and iron nuclei, and of the fusion of helium-4 and helium-3 nuclei, since the concentration of helium-3 in the area of powerful flares can be comparable to that of helium-4 [149]. A number of questions remain that are connected with the possibility of the quantitative support of such conclusions upon the analysis of the dynamics of nuclear processes in the solar atmosphere. The rather low abundance of the above elements (carbon, nitrogen, oxygen, and iron) in the solar atmosphere, along with the low probability of the formation of beryllium-7 in the abovementioned nuclear reactions (which is determined not only by the corresponding Coulomb barriers but also by the threshold energies of these processes, ranging from 10 to 25 *MeV*) should also be borne in mind. The only above-threshold process among these is beryllium-7 synthesis during the fusion of helium-4 and helium-3 nuclei, but this reaction can be completely suppressed by a higher Coulomb barrier.

The introduction of a new class of nuclear-chemical processes that occur under the conditions of low-temperature plasma does not fit the current concept [80, 143, 144] of the synthesis of elements in the Universe. When protium, deuterium, and tritium nuclei are present in a plasma and ${}^{b}n_{isu}$ nuclei form according to processes (46), (47), and

$$t^+ + e^-_{he} \to {}^3n_{isu} + v \to {}^3_2He + 2e^- + v + 2\widetilde{v} + Q(0.019 MeV),, \qquad (93)$$

where $b$ is a baryon number equal to 1, 2, and 3 for *β*-neutrons, *β*-dineutrons, and *β*-trineutrons, respectively, the following processes become possible in low-temperature plasma:

$$^A_Z N + {}^b n_{isu} \to {}^{A-A_1+k}_{Z-Z_1} N + {}^{A_1}_{Z_1+1} N + (b-k)n + e^- + \widetilde{v} + Q_1, \qquad (94)$$

$$^A_Z N + {}^b n_{isu} \to {}^{A-A_1+k}_{Z-Z_1} N + {}^{A_1}_{Z_1} N + (b-k)n + Q_2, \qquad (94a)$$

on condition $k \leq b$. As was shown above (see also [36]), there are indications that the $T_{1/2}$ periods for the $n_{isu}$ and ${}^3n_{isu}$ nuclei, and the corresponding period for ${}^2n_{isu}$, are quite long (no shorter than several tens of minutes), so the introduced $n_{isu}$, ${}^2n_{isu}$, and ${}^3n_{isu}$ β-nuclei can participate in the nuclear fusion processes that occur in stellar atmospheres. Examples of such processes related to the synthesis of light nuclei under the conditions of low-temperature plasma in the solar atmosphere, and the formation of nuclei heavier than iron nuclei (excluded in principle in the standard model of nucleosynthesis) were presented in [98].

The above problems of the formation of beryllium-7 radioisotopes in the solar atmosphere within the scope of our concept could due to the presence of $n_{isu}$ and ${}^2n_{isu}$ β-nuclei, and of ${}^6_3Li$ nuclei, which could be synthesized upon interaction between ${}^4_2He$ and ${}^2n_{isu}$:

$$^4_2He + {}^2n_{isu} \to {}^6_3Li + e^- + \widetilde{v} + Q\ (1.47\ MeV) \qquad (95)$$

$$^6_3Li + n_{isu} \to {}^7_4Be + e^- + \widetilde{v} + Q\ (5.61\ MeV), \qquad (96)$$

$$^6_3Li + {}^2n_{isu} \to {}^7_4Be + n + e^- + \widetilde{v} + Q\ (3.38\ MeV). \qquad (97)$$

As for the abovementioned problem related to the considerable increase in the concentration of ${}^3_2He$ in the area of solar flares [149], it is natural to associate this phenomenon with the initiation of the formation of helium-3 according to (93) and

$$d^+ + {}^2n_{isu} \to {}^3_2He + n + e^- + v + \widetilde{v} + Q(3.27 MeV). \qquad (98)$$

Since the concentration of ${}^3n_{isu}$ nuclei can grow in the area of solar flares, the possibility of ${}^7_4Be$ forming due to the following process cannot be excluded:

$$^4_2He + {}^3n_{isu} \to {}^7_4Be + 2e^- + 2\widetilde{v} + Q\ (1.6\ MeV). \qquad (99)$$

The continuously maintained relatively high concentration of beryllium-7 radioisotope in the solar atmosphere at rather low concentrations of lithium-6 (10 times lower than the concentration of lithium-7) can be also understood on this basis [146]. It is natural to associate the more intense burnup of lithium-7 in the area of solar flares with the interaction between lithium-7 and ${}^2n_{isu}$ β-nuclei, and the formation of beryllium-9:

$$^7_3Li + {}^2n_{isu} \to {}^9_4Be + e^- + \widetilde{v} + Q\ (16.7\ MeV). \qquad (100)$$

Thus, at the qualitative level, both observed phenomena are clear: an increase in the ratio η of the content of the isotopes of beryllium-7 and lithium 7 with an increase in solar activity and a certain decrease (with the value of *η* > 1 remaining), with a minimum of solar activity [146], and variations the content of lithium-7 in the solar atmosphere from the minimum values (in the solar cycle maxima) to the maximum (in solar activity minimums [150]).

### 6.2.2. Synthesis of the ${}^{12}_6C$ nuclei

It may be believed that the nucleosynthesis processes under the activating effect of electrons occur not only in stellar atmosphere and in subsurface regions of stars, but in the interiors of massive stars as well. As a first example, let us consider the mechanism of nuclear synthesis of the ${}^{12}_6C$ nucleus, a key element for the emergence of life. According to [151, 152], such a synthesis occurs in "helium flares" in the core of red giants, where at densities of ~ $10^5$ g/cm$^3$ and temperatures of ~ (1-2)×$10^8$ K, a triple helium reaction is realized. One would think that the fusion of two $_2^4He$ nuclei with creation of $_4^8Be$ nucleus characterized by $T_{1/2} \sim 10^{-16} s$ half-life period could be considered as the first stage of the $_6^{12}C$ nucleus synthesis. However at the specified temperatures, to which the mean thermal energy of 17 keV corresponds, the energy deficiency for such fusion reaction amounts 92 keV, that is more than 5 $kT_{rg}$, and thus probability of such $_4^8Be$ nucleus creation is small. The probable mechanism of $_6^{12}C$ nuclei synthesis in the considered conditions of the red giant nucleus was suggested as early as in 1950-th years by Fred Hoyle. Hoyle suggested the hypothesis, pursuant to which in "helium flares" conditions three $_2^4He$ nuclei in the core of red giants may create the excited state of $_6^{12}C^*$ nucleus at subsequent collisions, while the energy of this state is corresponding to the system, which is created from $_4^8Be$ and $\alpha$ – particle and which transforms into $_6^{12}C$ nucleus after $\gamma$ – quantum release:

$$_4^8Be + _2^4He \rightarrow _6^{12}C + \gamma, \quad Q = 7.37 MeV. \tag{101}$$

The corresponding excited state $_6^{12}C^*$ (7.65 MeV), energy of which only slightly (by 0.28 MeV) exceeds those necessary for the indicated in Eq. (101) mass defect reaction, subsequently was discovered by Fowler and his colleagues for $\beta$ – decay of $_5^{12}B$ nucleus. The existence of such excited level of $_6^{12}C^*$, with which the carbon synthesis is associated, is usually considered as the contingency, which has determined the emergence of life in our Universe (refer, for instance, to [4, 153]).

Within the frameworks of the developed perceptions on possibility of initiating of the local disturbances of the nuclear matter's nucleonic structure in case of electrons interactions with nuclei, the mechanism of $_6^{12}C$ nuclei creation in the red giants may be presented as more universal process. First of all let us mention just the new qualitative state of the high-temperature plasma in the interiors of massive stars, which should be revealed in case of considering of $\beta$ – nuclei with accounting of *isu*-state characterized by the locally disturbed nucleonic structure. Since such states of the nuclear matter are initiated in case of electrons collisions with fully ionized nuclei and ions in plasma, it may be supposed that in such plasma definite "equilibrium" parts of $\beta$ – nuclei of all elements present in plasma should be created. In case of high-temperature plasma of the red giants such $\beta$ – nuclei are presented by $_1^4H_{isu}$ nuclei, which are created in plasma of the "helium flares" at electrons interaction with $_2^4He$ nuclei:

$$_2^4He + e^- \rightarrow _1^4H_{isu} + \nu.$$

Let us suppose that the high-temperature plasma of helium flares is strongly non-equilibrium and the local temperature of electronic component of such plasma by order of magnitude and even more exceeds the temperature of ionic component, as it was supposed in the above considered examples of the low-temperature plasma (at glow discharge, at laser ablation of metals in aqueous solutions). In such conditions creation of $_4^8Be$ nuclei as the initial element for the $_6^{12}C$ nuclei synthesis may occur via interaction of $_1^4H_{isu}$ nuclei with $_2^4He$ nuclei under the initiating effect of high-energy (ultrarelativistic) electrons:

$$_1^4H_{isu} + _2^4He + e_{he}^- \rightarrow _4^8Be + 2e^- + \widetilde{\nu} + \nu\widetilde{\nu} + Q(-0.603 MeV). \tag{102}$$

Such process is possible if the total energy of electrons in the helium flares exceeds the threshold energy, $E_{th.e} = 0.603$ MeV. First of all, we note that the excitation of $_2^4He$ nucleus seems to be improbable in the conditions under consideration, since the excitation energy of its first level is equal to 20.21 MeV [154]. It may be supposed, the mechanism of such initiation by electrons involves the energy transfer to the $_1^4H_{isu}$ nucleus through inelastic scattering of electrons by quarks un-bound into neutron, when the interaction is effected through the intermediate $Z^0$ vector boson. Such energy transfer while keeping the $_1^4H_{isu}$ nucleus in the *isu*-state is completely acceptable due to an anomalously large deficit of its energy relative to the $_1^4H$ nucleus, which is $\Delta Q = -23.5$ *MeV*. It is necessary to notice, that for the above-mentioned characteristics of the high-temperature plasma in the "helium flares" of red giants, the Coulomb barrier for the (102) process happens to be insignificant [152]. As will be shown below, the mechanism of such energy transfer from ultra-relativistic electrons to quarks un-bound into neutron in the $_1^4H_{isu}$ nucleus by means of the intermediate $Z^0$ vector boson can be realized also in the synthesis of heavy and super heavy nuclei in the explosion of supernovae.

We believe that $_6^{12}C$ nuclei synthesis occurs in case of interaction of beryllium-8 nuclei following the (102) process (their characteristic life period is equal to $\sim 10^{-16}$ s) with the $_1^4H_{isu}$ nuclei:

$$_4^8Be + {_1^4H_{isu}} \to {_6^{12}C} + e^- + \tilde{\nu} + \nu\tilde{\nu} + Q(7.37 MeV) \qquad (103)$$

The subsequent synthesis of $_8^{16}O$, $_{10}^{20}Ne$, $_{12}^{24}Mg$ nuclei may occur with participation of $_6^{12}C$ nuclei synthesized in the red giant "helium flares". Particularly:

$$_6^{12}C + {_1^4H_{isu}} \to {_8^{16}O} + e^- + \tilde{\nu} + \nu\tilde{\nu} + Q(7.16 MeV). \qquad (103a)$$

Hereinafter we shall show, that participation of $_1^4H_{isu}^*$ and $_1^4H_{isu}$ nuclei in nucleonic synthesis processes may be considered as rather common phenomena, if refer to processes, which occur not only in the high-temperature plasma of the helium flares, but in the interiors of the more massive stars as well, which are characterized by more high (by order of magnitude and even more) temperature, as well as during supernova explosion. This circumstance somehow lowers the above-mentioned mysteriousness aureole in the problem of carbon-12 synthesis, which is substantiated by the reputable references [4, 153].

6.2.3. Possible role of the $_1^4H_{isu}$ and other *isu*-nuclei in the nucleosynthesis of elements heavier than iron

It is well known, that the nucleosynthesis of heavy elements, down to iron and nickel, occurs in the interiors of massive stars during the late stages of their evolution, when burning of *He*-4 with *C*-12 (as well as with *O*-16, *Ne*-20, *Mg*-24) creation is finished already and the interiors are heated to nuclear temperatures $\sim (2-3) \times 10^9$ K, so that the *O*-16 and *Si*-28 nuclei participate in the nucleosynthesis processes ("burnt"). Elements heavier than iron are synthesized at the explosions of supernova stars in subsequent *s*- and *r*-processes, when under the shock wave effect the explosion-type burning is initiated in conditions of even more high-power and more non-steady energy flows. The principal aim, which is pursued in this section, is to reveal the importance of the $_1^4H_{isu}$ nuclei and other *isu*-nuclei in the nucleosynthesis of the elements heavier than iron, with accounting of *s*- and *r*-processes, especially keeping in mind the above mentioned phenomena, following which definite "equilibrium" parts of $\beta$ – nuclei of all elements present in plasma are created in high-temperature plasma of the considered stars.

The existing knowledge pertinent to composition of the high-temperature plasma in the interiors of massive stars, with accounting of *isu*-nucleus existence in plasma, allows to present another point of view on "burning" stages of the above-mentioned elements within the interiors of stars. First of all let us notice, that above presented data on possibility of initiating of $\alpha$ – decay of large number of elements, from neodymium to bismuth, suggest that $_2^4He$ nuclei will be present in plasma at all stages of the star evolution, so full burning of helium does not occur. Now this means, that the nucleosynthesis of heavy nuclei with $_1^4H_{isu}$ $\beta$ – nuclei participation should be also considered at different stages of the massive stars evolution.

Let us cite several possible reactions involving $_1^4H_{isu}$ nuclei in which nuclei with masses heavier than iron mass nuclei are formed:

$$_{25}^{55}Mn + {_1^4H_{isu}} \to {_{27}^{59}Co} + e^- + \tilde{\nu} + \nu\tilde{\nu} + Q(6.94 MeV),$$

$$^{59}_{27}Co + {}^{4}_{1}H_{isu} \rightarrow {}^{63}_{29}Cu + e^- + \tilde{\nu} + \nu\tilde{\nu} + Q(6.78 MeV),$$

$$^{63}_{29}Cu + {}^{4}_{1}H_{isu} \rightarrow {}^{67}_{31}Ga + e^- + \tilde{\nu} + \nu\tilde{\nu} + Q(3.73 MeV),$$

$$^{63}_{29}Cu + {}^{4}_{1}H_{isu} \rightarrow {}^{67}_{30}Zn + \nu\tilde{\nu} + Q(3.73 MeV),$$

$$^{56}_{26}Mn + {}^{4}_{1}H_{isu} \rightarrow {}^{60}_{28}Ni + e^- + \tilde{\nu} + \nu\tilde{\nu} + Q(6.29 MeV),$$

$$^{60}_{28}Ni + {}^{4}_{1}H_{isu} \rightarrow {}^{64}_{30}Zn + e^- + \tilde{\nu} + \nu\tilde{\nu} + Q(3.96 MeV),$$

The ${}^{4}_{1}H_{isu}$ nuclei can directly participate in nucleosynthesis processes up to obtaining nuclei with $A \approx 115-120$:

$$^{114}_{48}Cd + {}^{4}_{1}H_{isu} \rightarrow {}^{118}_{50}Sn + e^- + \tilde{\nu} + \nu\tilde{\nu} + Q(4.06 MeV),$$

If the electronic subsystem in the high-temperature plasma is characterized by the abnormal excitation, as it takes place in case of supernova star explosion, which causes existence of rather high concentration of the ultrarelativistic electrons in plasma, than in the nucleosynthesis processes with ${}^{4}_{1}H_{isu}$ participation heavy and super heavy nuclei may be created. Let us give some examples indicating the threshold energy of electrons for initiating these processes:

$$^{218}_{86}Rn + {}^{4}_{1}H_{isu} + e^-_{he} \rightarrow {}^{222}_{88}Ra + 2e^- + \tilde{\nu} + \nu\tilde{\nu}, \quad E_{th.e} = 7.19 MeV;$$

$$^{230}_{90}Th + {}^{4}_{1}H_{isu} + e^-_{he} \rightarrow {}^{234}_{92}U + 2e^- + \tilde{\nu} + \nu\tilde{\nu}, \quad E_{th.e} = 5.37 MeV;$$

$$^{231}_{91}Pa + {}^{4}_{1}H_{isu} + e^-_{he} \rightarrow {}^{235}_{93}Np + 2e^- + \tilde{\nu} + \nu\tilde{\nu}, \quad E_{th.e} = 6.70 MeV.$$

$$^{235}_{92}U + {}^{4}_{1}H_{isu} + e^-_{he} \rightarrow {}^{239}_{94}Pu + 2e^- + \tilde{\nu} + \nu\tilde{\nu}, \quad E_{th.e} = 5.76 MeV.$$

It is important to emphasize that all the given threshold energy values $E_{th.e}$ are noticeably smaller than the absolute value $|\Delta Q| = 23.5 MeV$ of the ${}^{4}_{1}H_{isu}$ nucleus energy deficit relative to the ${}^{4}_{1}H$ nucleus. This means that in all the synthesis processes under consideration, even the heaviest elements, the ${}^{4}_{1}H_{isu}$ nuclei remain $\beta$–nuclei, that is, they can not decay beyond the characteristic for ${}^{4}_{1}H$ nuclear decay time ($\sim 10^{-22} s$) even at excitation energies of $\sim 10$ MeV. This is a very important point, since preserving the nucleon structure of nuclei at the stages of nuclear matter rearrangement may require much greater internal excitations in the formation of daughter nuclei. Here we mention only a couple of examples demonstrating the statement made and pointing to a kind of catalytic role of the *isu*-states of nuclei in nuclear fusion. Thus, in the restructuring of the nucleon structure of a nucleus with $A = 8$, associated with a change in the charge state of only one nucleon (in the $\beta$–decay and subsequent emission of a $\gamma$–quantum: ${}^{8}_{3}Li(g.s., 2^+) \rightarrow {}^{8}_{4}Be^*(3.03 MeV, 2^+) + e^- + \tilde{\nu} + Q(12.98 MeV)$), the energy of the system changes by 16.01 MeV [155]. With the aforementioned analogous $\beta$–decay process with $\gamma$–quantum emission for a system with $A = 12$, when ${}^{12}_{5}B(g.s., 1^+) \rightarrow {}^{12}_{6}C(g.s., 0^+)$, the energy of the system changes by 13.37 MeV [156].

It can be assumed that non-bound quarks in nuclei catalyze the structural rearrangement of nuclear matter in the formation of the nucleon structure of nuclear transformation products. One of the arguments in favor of such a conclusion is the phenomenon of an increase in the order of magnitude of the rate constants of radioactive decays in the activative actions of electrons on radioactive atoms.

Herewith we exclude from consideration the common hypothesis on possibility of realization of the nuclear reactions between multiply charged nuclei, such as ${}^{12}_{6}C + {}^{12}_{6}C$, ${}^{16}_{8}O + {}^{16}_{8}O$, up to ${}^{28}_{14}Si + {}^{28}_{14}Si$ [152], during the supernova explosions. To what extent these perceptions are adequate and to what extent they may supplement (or change) the well-known views pertinent to *s*- and *r*-processes, which occur during the heavy elements synthesis in massive stars or during the supernova explosions, may be elucidated only in the subsequent experimental investigations.

Nevertheless, the number of isotopes sequentially formed as result of neutron captures may be changed significantly because of the *r*-process rate constant changes under the radioactive decay activation by high energy electrons. If takes in mind the change in nucleosynthesis dynamics as a result of taking part of the $^4_1H_{isu}$ nucleus in these processes, the general conclusion on the synthesis of heave and very heavy nuclei may be changed significantly.

The only experimental results, which may serve as the reference point for estimation of adequacy of the suggested, as well as of the existing mechanisms of the heavy elements synthesis during the supernova explosion, are the data, which have been obtained during explosion of SN 1987A supernova on February 23, 1987, in the Large Magellanic Cloud, at the distance of 150000 light years from the Solar system [80]. Simultaneously with the optical flare, three neutrino stations have detected the neutrino flow during 13 seconds – 24 neutrino and antineutrino, which substantially exceeded the background. The corresponding calculation has shown that during SN 1987A explosion $10^{58}$ neutrino was emitted with total energy of $10^{46}$ Joules. Neutrino and antineutrino flows have caused the process of quick and effective cooling of the star depths while removing the energy from the whole volume of the star nucleus being transparent for the low-energy neutrino. As it was noticed in section 5.2, mechanism of the neutrino cooling of stars was suggested by Gamow and Schönberg and was called URCA process [114].

The standard URCA processes are the simplest neutrino-emitting processes and are thought to be central in the loss of energy in the explosion of supernova as well as in the cooling of neutron stars. These processes are usually presented as:

$$^A_ZN \rightarrow\, ^A_{Z+1}N_2 + e^- + \widetilde{\nu}, \qquad ^A_{Z+1}N + e^- \rightarrow\, ^A_ZN + \nu.$$

The corresponding analysis of possible URCA processes during SN 1987A explosion has shown that the $\nu\widetilde{\nu}$ flows should remove up to 99% of the whole energy, which have released during the explosion. However, in connection with the latter result, as well in connection with the common perceptions pertinent to role of the URCA processes during the explosions of stars some questions have arisen. First of all, it still remains disputable why the URCA processes, which are characterized by weak nuclear interactions, are overriding the excited state relaxation, which is realized in the nuclear processes, by $\gamma$–quanta emission, that is "photon cooling". Which mechanisms of such relaxation processes suppression are realized in this case? Why the highly-effective URCA processes of energy takeoff from the reaction zone are not realized in the normal nuclear processes? It would seem that less effective energy takeoff via "surface" photon cooling should merely suppress the explosive evolution of supernova.

We believe that URCA processes occurrence in heavy elements synthesis during the supernova explosions becomes possible only because of existence in the high-temperature plasma in the explosion zone of the atomic $\beta$-nuclei characterized by the local "non-nucleon" state of nuclear matter. It should be noticed, that part of the $\beta$-nuclei before its relaxation $\beta$–decay [94] with creation of the initial nuclei "has time" to participate in nuclear reactions. When such nuclei participate in the synthesis processes, the state of the created nuclei at the initial stages also is characterized by the local *isu*-disturbances of the nuclear matter structure, and their relaxation during creation of the final nucleus-products is inevitably associated with energy loss by the emission of neutrino-antineutrino pairs. As far as part of $\beta$-nuclei is concerned, which takes part in the processes of inelastic scattering of electrons by nuclei with the emission of neutrino-antineutrino pairs $\nu\widetilde{\nu}$ (see (43)-(45)), just likely via such mechanism, the major part of the energy neutrino emission during the explosion of supernova is realized. The author believes that accounting of the above-mentioned phenomena may bring to changing of the reference estimates pertinent to the part of energy released during flare, which is taken off by neutrino.

The "neutrino cooling" problem, that is effective energy takeoff in the form of $\nu\widetilde{\nu}$–pairs, also appears in study of the young neutron stars, which are created after the supernova stars explosion [157]. In connection with the latter problem, it is possible to suppose, the formation of a stable nuclear matter of neutron stars [105] is realized through appearance of neutral *isu*-nuclei $^1n_{isu}$, $^2n_{isu}$, $^3n_{isu}$, $^4n_{isu}$,…, as the possible components of the neutron star, which effectively participate in the neutrino cooling processes.

6.3. *Nature of the anomaly recorded in the Beryllium-8 decay* [94]

As one else example, which reveals the initiated nature of radioactive decays, we consider the anomalies in the angular correlations between the positrons and electrons emitted in the radioactive decays of excited $^8_4Be^*$ nuclei [158]. We assume that the decay of a $^8_4Be^*$ nucleus is preceded by its interaction with one of the electrons in the inner electron shells of the atom, which emits a neutrino $v$ and produces an excited metastable *isu*-state $^8_3Li^*_{isu}$ nucleus:

$$^7_3Li + p \to {^8_4Be^*}, \quad {^8_4Be^*} + e^-_{he} \to {^8_3Li^*_{isu}} + v \to 2\,{^4_2He} + 2e^- + e^+ + v + \tilde{v}. \qquad (104)$$

Based on the above decay diagram for the excited $^8_4Be^*$ nucleus, we can suggest a new formation mechanism for a correlated $e^+e^-$ pair in the reaction $^7Li(p,\gamma)^8Be$ when two states of the $^8_4Be^*$ nucleus, 17.64 MeV and 18.15 MeV, are excited, as in the experiment [144], which is alternative to the one proposed in [159, 160].

As noted above, the nature of the excitation of $^8_4Be$ nuclei initiated by the collisions with nucleons, when the entire nucleonic subsystem of the nucleus is excited, is substantially different from the nature of the local metastable *isu*-excitation caused by a shake-up in the nucleonic structure of $^8_3Li_{isu}$ nucleus and the loss of the overall stability of these $\beta$-nuclei. The latter depends on the absolute value of structural energy deficit $\Delta Q = (m_{^8_4Be} - m_{^8_3Li})c^2$ and the difference in the masses of $^8_4Be$ and $^8_3Li$ nuclei in the ground state. Therefore, we can assume that the excitation energy of the nucleonic subsystem of the $^8_4Be^*$ nucleus can be almost completely kept in the nucleonic subsystem of the $^8_3Li^*_{isu}$ nucleus if the latter has the corresponding excited state. In this case, it becomes obvious that the efficiency of the decay of the excited $^8_3Li^*_{isu}$ nucleus that emits two alpha particles and a correlated $e^+e^-$ pair will depend on how close one of the excited states of the $^8_3Li$ nucleus approximates the excited state of the $^8_4Be$ nucleus [155].

Here we must take into account that the probability of emitting $\gamma$ quanta by excited nuclei, which depends on the width of the corresponding excited state, and the probability of emitting photons in the transition of a single atom from the excited state to the ground state are initiated by the zero-point oscillations of the EM vacuum [161]. Virtually, the main factor is the average of squared fluctuating values of the electric field intensity for the EM vacuum. As noted above, when a metastable *isu*-state with a local shake-up in the nucleonic structure is initiated in the nuclear matter, the irreversible loss of the nucleus stability is likewise accounted for by the EM vacuum as a result of changes in the boundary conditions at the nucleus surface [3, 99]. However, these two emissions are independent of each other.

As the ground-state energy for the $^8_3Li$ nucleus is 16.005 MeV higher than that for the $^8_4Be$ nucleus [155], the excited states of 1.635 and 2.145 MeV for the $^8_3Li$ nucleus could formally correspond to the excited states of 17.64 and 18.15 MeV for the $^8_4Be$ nucleus. For the $^8_3Li$ nucleus, the excited states closest to the ground one are 0.891 MeV, which is not high enough for producing an $e^-e^+$ pair, and 2.255 MeV, which is 0.11 MeV higher than the above value of 2.145 MeV. If the anomaly in angular correlations between positrons and electrons recorded in [158] is effectuated by the above excited states of $^8_4Be$ and $^8_3Li_{isu}$ nuclei, it implies that in this case the width of the 2.255 MeV state for the $^8_3Li_{isu}$ nucleus is larger than 0.11 MeV, and we can speak about a direct correspondence between the 18.15 MeV excited state for the $^8_4Be$ nucleus and the 2.255 MeV excited state for the $^8_3Li_{isu}$ nucleus. Obviously, this correspondence needed for the anomaly recorded in [158] to take place may be achieved by adjusting the kinetic energy $E_p$ of the bombarding protons, though not always.

Assume that the anomaly recorded in [158], which is the formation of correlated $e^-e^+$ pairs at their opening angles $\Theta \sim 130\text{-}140^0$, is mostly due to the exchange of $d$- and $u$-quarks localized in the region of non-nucleonic metastability of the $isu$-state nucleus, which can migrate over the nucleus, and $d$- and $u$-quarks of the superpositions $d\tilde{d}$ and $u\tilde{u}$ among the quark-antiquarks pairs produced in the decay of vector $Z^0$-mesons in the same $isu$-region of the nuclear matter. Virtually, this exchange is effectuated in the annihilation of these quarks and antiquarks of $d\tilde{d}$ and $u\tilde{u}$ pairs producing a correlated $e^-e^+$ pair, which is in good agreement with the decay of a neutral boson studied in [158-160].

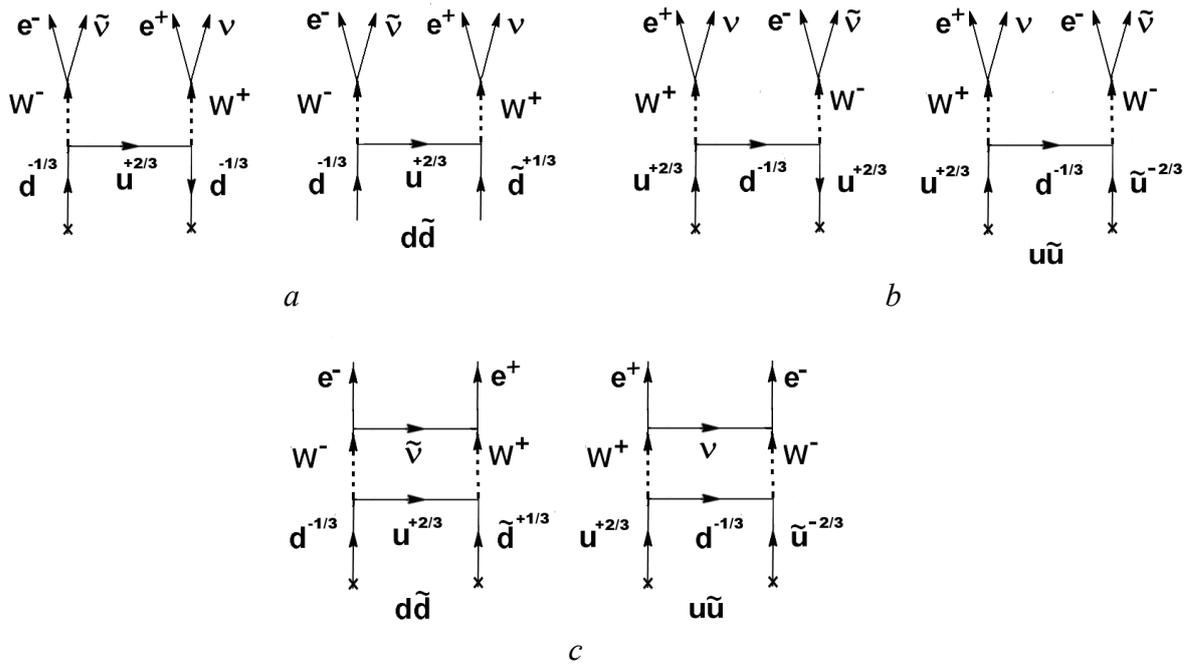

Fig. 4. Feynman diagrams for the formation of correlated $e^-e^+$, $e^-\tilde{\nu}$ и $e^+\nu$ pairs initiated by the interaction between a $\pi^0$-meson (quark-antiquark superposition $d\tilde{d}$ and $u\tilde{u}$) and the nucleons of the excited $^8$Be* nucleus

Possible diagrams for the formation of $e^-e^+$ pairs accompanied by the production of $e^-\tilde{\nu}$ and $e^+\nu$ pairs are plotted in Fig. 4. In contrast to the decays of bosons as particles with a certain set of quantum characteristics analyzed in [158-160], the quark pairs to be annihilated in these diagrams can have different relative orbital moments, as in [162], which does not impose any substantial restrictions on the sets of quantum numbers for the excited states of the nuclei; specifically, $^8_4Be^*$ and $^8_3Li^*_{isu}$. For this reason, the results of [158] were discussed above without referring to the quantum numbers for the excited states of these nuclei.

The developed concept stating the existence of the metastable states of the nuclear matter with a local shake-up of its nucleonic structure makes it possible to qualitatively interpret the formation of correlated $e^-e^+$ pairs in the experiment under discussion [158] without involving the hypothesis of afifth fundamental interaction into the physical science. Admittedly, it is possible so far to speak only about a qualitative understanding of the correlation of $e^-e^+$ pairs in the above process because the exchanges of quarks in the region of nucleus non-nucleonic metastability have not been studied yet.

An additional clarity in discussing the above alternative could be brought by the new experiments proposed in [160] to record the anomalies in the angular correlations between positrons and electrons, like those in [158], that are emitted in the radioactive decay of other excited nuclei. The study [160] deals with the reactions $^7Li(^3He,\gamma)^{10}B^*(19.3MeV)$ [163] and $^7Li(t,\gamma)^{10}Be^*(17.79MeV)$ [164] and assumes that the decay of these excited states of the daughter nuclei can produce $e^+e^-$-pairs with the same type of opening-angle anomaly as in [158].

The developed concept of the radioactive decays of the excited nuclei producing the $^{10}B^*(19.3MeV)$ nucleus implies that, in the first of the above reactions, the $^{10}_{4}Be^*_{isu}$ nucleus rather than the $^{10}_{5}B^*$ nucleus would decay, producing the final products $^7Li$, $^3He$, and an $e^+e^-$-pair. In the second reaction, the final products $^7Li$, $^3H$, and an $e^+e^-$-pair are formed in the decay of the $^{10}_{3}Li^*_{isu}$ nucleus rather than the $^{10}_{4}Be^*$ nucleus. The above differences are significant due to the high difference in the ground-state energies of the nuclei to be decayed when these energies are referred to the unified energy scale. It is this kind of analysis that will enable us to make an unambiguous choice in conducting appropriate experimental studies in favor of the hypothesis of the existence of a fifth fundamental interaction or developed concept of nuclear radioactive decays. The most significant differences are seen for the energy levels of $^{10}_{3}Li^*_{isu}$ and $^{10}_{4}Be^*$ nuclei: the ground state for the lithium-10 nucleus is 20.444 MeV higher than the one for the beryllium-10 nucleus. The corresponding difference between the $^{10}_{4}Be^*_{isu}$ nucleus and $^{10}_{5}B^*$ nucleus is 0.556 MeV [165].

In view of the above differences, the excited state of 19.3 MeV for the boron-10 nucleus should formally be in correspondence with the excited state of 18.74 MeV for the beryllium-10 nucleus, in which the excited-state energy closest to the latter value is 18.55 MeV. When the width of this state is greater than 0.2 MeV, the above correspondence can be true and the anomaly in the angular correlations between positrons and electrons emitted in the radioactive decays of excited $^{10}B^*$ nuclei can, in principle, be recorded. The situation is substantially different when we look for these anomalies in the decays of excited $^{10}Be^*$ nuclei. The excited state of 17.79 MeV for the beryllium-10 nucleus cannot even formally be in correspondence with the ground state for the lithium-10 nucleus because the energy difference between the ground states for the lithium-10 nucleus and beryllium-10 nucleus is higher than the above excited-state energy for the beryllium-10 nucleus. Therefore, the desired correlations in the $^7Li(t,\gamma)^{10}Be^*$ reaction should be sought at kinetic energies $E_t$ of the tritium nuclei higher than those suggested in [160]. For example, the excited states of 1.4, 2.35, and 2.85 MeV for the lithium nucleus, whose decay into $^7Li$ and $^3H$ may be accompanied by the anomaly in the angular correlations of the recorded $e^+e^-$-pair can be achieved using the tritium nucleus energies of 21.8, 22.8, and 23.3 MeV, respectively. This experiment can become an *experimentum crucis* in selecting between the discussed nature alternatives for the correlated opening angle of the $e^+e^-$-pair, as well as in deciding whether it is possible to initiate metastable states with a shaken-up nucleonic structure in the nuclear matter and, hence, validate the new concept of radioactive decays of nuclei.

VII. CONCLUDING REMARKS

The main hypothesis that the author uses to understand the key problems of contemporary fundamental physics – from the well-known cosmological problems to the new class of yet extremely controversial phenomena of nuclear chemical processes, is to introduce the basic reference system associated with the EM vacuum of the expanding Universe, and ideas about the "Casimir" polarization of the EM vacuum in the vicinity of atomic nuclei and electronic subsystems of various objects. It is through Casimir polarization that every elementary particle, every atomic nucleus of macroscopic objects turns out to be inextricably linked with the EM vacuum as an energy supply and all-penetrating world active substance, and this relationship determines the internal dynamics and stability of each atom's nuclei, the specificity of the chemical activity of atoms.

Such a global, all-pervasive medium can be viewed as a kind of "primary material" (materia prima), the existence of which was postulated by Rene Descartes and "which was agreed to be called the ancient name of the ether." It is the introduction of such an all-pervasive base environment, which "serves as a vessel of World energy, ... gives grounds, without illusions, to think that it allows grouping all known facts into one grand synthesis" [166], wrote L. Poincaré in the beginning of the last century, during the time of the famous crisis in physics.

Indeed, ideas about the ether were effectively used by Huygens, Jung and Fresnel in the development of the wave theory of light. Newton and his followers, developing the corpuscular (emission) theory of light, considered the ether as a carrier of light particles, and the refraction and diffraction of light were associated with a change in the "density of the ether near the interphase boundaries" (compare: the Casimir polarization of the EM vacuum in atomic nuclei and electrons). After Maxwell discovered the equations of classical electrodynamics, the ether began to be considered as a common carrier of light, electricity and magnetism, there was a hope to build a physical model of the ether on this foundation.

There have been numerous attempts to link the ether with gravity and on this basis to understand the Newtonian law of the World. However, the Descartes-Huygens ether was moved beyond the framework of physical images at the beginning of the 20th century when creating the special theory of relativity (STR). According to Einstein [167], the concept of the luminiferous ether is incompatible with the principle of relativity, according to which all laws of nature are the same in all inertial reference systems, and the principle of constancy of the speed of light in vacuum, according to which the speed of light is constant and does not depend on the motion of the radiating body.

However, the emerging problems of fundamental physics presented in this article, as well as attempts to understand the physical essence of the basic postulates of special relativity, in particular, the reasons why the material body cannot achieve the speed of light in the EM vacuum, the mass increase at relativistic speeds, the increase in the lifetime of relativistic muons, gave grounds to imagine arguments for a return to the idea of Descartes about the energy-feeding and all-pervading active world substance – ether. Indeed, as shown in this article, the introduction of the EM vacuum of the expanding Universe as the base medium, the "XXI century ether", the absolute frame of reference with a single time $t$ for all objects of the Universe, made it possible to understand, on a phenomenological level, first of all, some key and urgent problems of the contemporary fundamental science – quantum mechanics, gravity, astrophysics.

In particular, on this basis it became possible to substantiate not only the Newton's formula for the Law of Universal Gravitation, but also to understand the physical cause of the unique smallness of the gravitational constant $G$. There was manifested also the gnoseological role of the Planck numbers with the simultaneous awareness of their limitations for solving the problems of gravity. The introduction of yet another, fourth arithmetical relation, a modified Weinberg's relation, allowed moving the collection of all four arithmetical relations to the level of phenomenological relations with the establishment of the physical meaning of the introduced parameters. This became the basis for the understanding of sufficiency of the introduction, besides $\hbar$ and $c$, of another two universal constants $m_Q$ and $R_H$, in order to present at the phenomenological level the main fundamental interconnections in the Universe and to focus on the development of the general theory of gravity and the dynamics of the Universe within not the $\hbar G c$ -, but the $\hbar m_Q R_H c$ -plan. It is necessary to emphasize that the main essence of the $\hbar m_Q R_H c$ -plan under consideration is the postulated dependence of the fundamental parameters $\hbar$, $c$ и $m_Q$ from the global time of the Universe.

      The introduction of the EM vacuum as the base medium of the Universe made it possible to understand at the phenomenological level the complex of problems in nuclear physics arising from the phenomenon of low-energy nuclear reactions. New aspects that determine the possibility of such processes are associated with the concept of an atomic nucleus as an open system, the state and dynamics of which are determined by the conditions at the nucleus boundary with the EM vacuum, which can change under initiate action of electrons of high (chemical scale!) kinetic energies. Extensive research of such processes opens up great prospects associated with the use of the vast hidden energy resources of Nature to solve current problems of modern energy sector. From this point of view, how the prophecy looks like today the statement of D.I. Mendeleev [168]: "The problems of understanding gravity and the problems in energetics cannot be imagined as really solved without a real understanding of the ether as the world medium transmitting energy to distances."

      Indeed, the introduction of a basic reference system associated with the EM vacuum of the expanding Universe, with a single time *t* for all objects, made it possible at the phenomenological level to understand the essence of gravity and, in general, the dynamics of the Universe as an open system. On the same basis, with the establishment of the key role of the EM vacuum in changing the stability of atomic nuclei during the initiating impacts of electrons, possible mechanisms for the implementation of low-energy nuclear-chemical reactions were understood, including the reasons for the absence of dangerous ionizing radiation and the formation of neutrons during such processes. It can be assumed that the understanding of the physical nature of nuclear-chemical reactions, associated with the introduction of new ideas about the initiation of non-nucleon excitations of nuclear matter [94] in the interaction of nuclei with electrons of high (on a chemical scale!) energies, will lead to the expansion of research in this new field of whole energy sector, what dreamed about D.I. Mendeleev.